\documentclass[iop,revtex4]{emulateapj}
\usepackage{apjfonts}
\usepackage{amssymb,appendix,multirow,morefloats,amsmath}
\usepackage[figuresright]{rotating}
\usepackage{threeparttable}
\usepackage{longtable}
\usepackage{subfigure}
\usepackage{epstopdf}
\usepackage{graphicx}
\usepackage{pdflscape}
\usepackage{chngpage}

\newcommand\T{\rule{0pt}{2.6ex}}       
\newcommand\B{\rule[-1.2ex]{0pt}{0pt}} 
\newcommand\M{\rule{0pt}{2.3ex}}    
\newcommand\U{\rule{0pt}{3.2ex}}       
\newcommand\D{\rule[-2.4ex]{0pt}{0pt}} 

\begin{document}

\title{The \textit{NuSTAR} Hard X-ray Survey of the Norma Arm Region}

\shorttitle{The NuSTAR Norma Region Survey}
 
\shortauthors{Fornasini et al.}
\slugcomment{{\sc Accepted to ApJ: } February 19, 2017}

\author{Francesca M. Fornasini\altaffilmark{1,2}, John A. Tomsick\altaffilmark{2}, JaeSub Hong\altaffilmark{3}, Eric V. Gotthelf\altaffilmark{4}, Franz Bauer\altaffilmark{5,6,7}, Farid Rahoui\altaffilmark{8}, Daniel Stern\altaffilmark{9}, Arash Bodaghee\altaffilmark{10}, Jeng-Lun Chiu\altaffilmark{2}, Ma{\"i}ca Clavel\altaffilmark{2}, Jes{\'u}s M. Corral-Santana\altaffilmark{5,11}, Charles J. Hailey\altaffilmark{4}, Roman A. Krivonos\altaffilmark{2,12}, Kaya Mori\altaffilmark{4},
David M.~Alexander\altaffilmark{13}, 
Didier Barret\altaffilmark{14},
Steven E.~Boggs\altaffilmark{2}, 
Finn E. Christensen\altaffilmark{15},
William W. Craig\altaffilmark{2,16},
Karl Forster\altaffilmark{17},
Paolo Giommi\altaffilmark{18},
Brian W.~Grefenstette\altaffilmark{17},
Fiona A.~Harrison\altaffilmark{17},
Allan~Hornstrup\altaffilmark{15},
Takao Kitaguchi\altaffilmark{19},
J. E. Koglin\altaffilmark{20},
Kristin K.~Madsen\altaffilmark{17},
Peter H. Mao\altaffilmark{17},
Hiromasa Miyasaka\altaffilmark{17},
Matteo Perri\altaffilmark{18,21},
Michael J.~Pivovaroff\altaffilmark{16},
Simonetta Puccetti\altaffilmark{18,21}, 
Vikram Rana\altaffilmark{17},
Niels J.~Westergaard\altaffilmark{15},
William W. Zhang\altaffilmark{22}
}

\altaffiltext{1}{Astronomy Department, University of California, 601 Campbell Hall, Berkeley, CA 94720, USA (e-mail: f.fornasini@berkeley.edu)}
\altaffiltext{2}{Space Sciences Laboratory, 7 Gauss Way, University of California, Berkeley, CA 94720, USA}
\altaffiltext{3}{Harvard-Smithsonian Center for Astrophysics, 60 Garden St., Cambridge, MA 02138, USA}
\altaffiltext{4}{Columbia Astrophysics Laboratory, Columbia University, New York, NY 10027, USA}
\altaffiltext{5}{Instituto de Astrof{\'{\i}}sica and Centro de Astroingenier{\'{\i}}a, Facultad de F{\'{i}}sica, Pontificia Universidad Cat{\'{o}}lica de Chile, Casilla 306, Santiago 22, Chile} 
\altaffiltext{6}{Millennium Institute of Astrophysics (MAS), Nuncio Monse{\~{n}}or S{\'{o}}tero Sanz 100, Providencia, Santiago, Chile} 
\altaffiltext{7}{Space Science Institute, 4750 Walnut Street, Suite 205, Boulder, Colorado 80301} 
\altaffiltext{8}{European Southern Observatory, K. Schwarzschild-Strasse 2, D-85748 Garching bei M{\"u}nchen, Germany}
\altaffiltext{9}{Jet Propulsion Laboratory, California Institute of Technology, 4800 Oak Grove Drive, Pasadena, CA 91109, USA}
\altaffiltext{10}{Georgia College, 231 W. Hancock St., Milledgeville, GA 31061, USA}
\altaffiltext{11}{European Southern Observatory, Alonso de C\'ordova 3107, Casilla 19001, Santiago, Chile}
\altaffiltext{12}{Space Research Institute, Russian Academy of Sciences, Profsoyuznaya 84/32, 117997 Moscow, Russia}
\altaffiltext{13}{Centre for Extragalactic Astronomy, Department of Physics, Durham University, Durham, DH1 3LE, U.K.}
\altaffiltext{14}{Universit\'e de Toulouse; UPS-OMP; IRAP; Toulouse, France \& CNRS; Institut de Recherche en Astrophysique et Plan\'etologie; 9 Av. colonel Roche, BP 44346, F-31028 Toulouse cedex 4, France}
\altaffiltext{15}{DTU Space - National Space Institute, Technical University of Denmark, Elektrovej 327, 2800 Lyngby, Denmark}
\altaffiltext{16}{Lawrence Livermore National Laboratory, Livermore, CA 94550, USA} 
\altaffiltext{17}{Cahill Center for Astrophysics, 1216 E. California Blvd, California Institute of Technology, Pasadena, CA 91125, USA} 
\altaffiltext{18}{ASI Science Data Center (ASDC), via del Politecnico, I-00133 Rome, Italy} 
\altaffiltext{19}{Department of Physical Science, Hiroshima University, 1-3-1 Kagamiyama, Higashi-Hiroshima, Hiroshima 739-8526, Japan}
\altaffiltext{20}{Kavli Institute for Particle Astrophysics and Cosmology, SLAC National Accelerator Laboratory, Menlo Park, CA 94025, USA} 
\altaffiltext{21}{INAF - Astronomico di Roma, via di Frascati 33, I-00040 Monteporzio, Italy}
\altaffiltext{22}{NASA Goddard Space Flight Center, Greenbelt, MD 20771, USA} 

\begin{abstract}

\noindent We present a catalog of hard X-ray sources in a square-degree region surveyed by \textit{NuSTAR} in the direction of the Norma spiral arm.  This survey has a total exposure time of 1.7 Ms, and typical and maximum exposure depths of 50 ks and 1 Ms, respectively.  In the area of deepest coverage, sensitivity limits of $5\times10^{-14}$ and $4\times10^{-14}$~erg~s$^{-1}$~cm$^{-2}$ in the 3--10 and 10--20~keV bands, respectively, are reached.  Twenty-eight sources are firmly detected and ten are detected with low significance; eight of the 38 sources are expected to be active galactic nuclei.  The three brightest sources were previously identified as a low-mass X-ray binary, high-mass X-ray binary, and pulsar wind nebula.  Based on their X-ray properties and multi-wavelength counterparts, we identify the likely nature of the other sources as two colliding wind binaries, three pulsar wind nebulae, a black hole binary, and a plurality of cataclysmic variables (CVs).  The CV candidates in the Norma region have plasma temperatures of $\approx$10--20~keV, consistent with the Galactic Ridge X-ray emission spectrum but lower than temperatures of CVs near the Galactic Center.  This temperature difference may indicate that the Norma region has a lower fraction of intermediate polars relative to other types of CVs compared to the Galactic Center.  The \textit{NuSTAR} log$N$-log$S$ distribution in the 10--20~keV band is consistent with the distribution measured by \textit{Chandra} at 2--10~keV if the average source spectrum is assumed to be a thermal model with $kT\approx15$~keV, as observed for the CV candidates. 
\end{abstract}


\section{Introduction}
\label{sec:intro}.pdf

Hard X-ray observations of the Galaxy can be used to identify compact stellar remnants $-$ white dwarfs, neutron stars, and black holes $-$ and probe stellar evolution in different environments.  While a number of sensitive surveys of Galactic regions (e.g. \citealt{muno09}; \citealt{townsley11}; \citealt{fornasini14}) have been performed by the \textit{Chandra X-ray Observatory}, its soft X-ray band (0.5--10~keV) is often insufficient for differentiating between different types of compact objects.  The \textit{Nuclear Spectroscopic Telescope Array} (\textit{NuSTAR}; \citealt{harrison13}), with its unprecedented sensitivity and angular resolution at hard X-ray energies above 10~keV, provides a unique opportunity to study the X-ray populations in the Galaxy.  During the first two years of its science mission, \textit{NuSTAR} performed surveys of the Galactic Center and the Norma spiral arm in order to compare the X-ray populations in these regions of the Galaxy, which differ with regard to their star formation history and stellar density.  The \textit{NuSTAR} sources found among the old, high-density Galactic Center stellar population are described in \citet{hong16}, and, in this paper, we present the results from the \textit{NuSTAR} Norma Arm survey. \par
In 2011, the Norma Arm Region \textit{Chandra} Survey (NARCS) observed a 2$^{\circ} \times 0\fdg8$ region in the direction of the Norma spiral arm (\citealt{fornasini14}, hereafter F14); the near side of the Norma arm is located at a distance of about 4~kpc while the far Norma arm is at a distance of 10-11 kpc \citep{vallee08}.  The Norma region was targeted because its stellar populations are younger than those in the Galactic Center but older than those in the young Carina and Orion star-forming regions observed by \textit{Chandra} (F14 and references therein).  An additional goal of this survey was to identify low-luminosity high-mass X-ray binaries (HMXBs) falling below the sensitivity limits of previous surveys in order to constrain the faint end of the HMXB luminosity function; the evolutionary state of the Norma arm and the large number of OB associations along this line-of-sight \citep{bodaghee12} make it an ideal place to search for HMXBs. \par  About 300 of the 1130 \textit{Chandra} sources detected at $\geq3\sigma$ confidence in the Norma region were found to be spectrally hard in the 0.5--10~keV band, with median energies $>$3~keV. The majority of these sources are expected to be magnetic cataclysmic variables (CVs) and active galactic nuclei (AGN), although some could also be HMXBs, low-mass X-ray binaries (LMXBs), or colliding wind binaries (CWBs).  Distinguishing between these types of sources is not possible based on \textit{Chandra} data alone, especially since most of the Norma X-ray sources have low photon statistics. \par
Since \textit{Chandra}'s resolution enables the identification of unique optical/infrared counterparts, spectral identification of the counterparts has helped shed light on the physical nature of some of the Norma X-ray sources \citep{rahoui14}.  However, not even this information is necessarily sufficient; for example, HMXBs and CWBs both have massive stellar counterparts in the optical/infrared and it can be difficult to differentiate them spectrally in the \textit{Chandra} band with $<100$ photon counts, as is the case for most  NARCS sources.  \textit{NuSTAR} observations, due to their superior sensitivity above 10 keV and in the energy range of the iron K$\alpha$ and K$\beta$ lines, provide critical information to differentiate hard X-ray sources.  For example, CWBs can be distinguished from HMXBs because they have thermal spectra that fall off steeply above 10~keV and strong 6.7~keV Fe emission (\citealt{mikles06} and references therein), and magnetic CVs can be distinguished from non-magnetic CVs by their harder spectra, lower equivalent widths of the 6.7~keV line, and higher line ratios of 7.0/6.7~keV Fe emission (e.g. \citealt{xu16}).
\par
The first set of observations of the \textit{NuSTAR} Norma Arm survey were carried out in February 2013 and improved the identification of three NARCS sources (\citealt{bodaghee14}, hereafter B14), discovered one transient (\citealt{tomsick14}, hereafter T14), and permitted the study of the disk wind of the LMXB 4U~1630-472 \citep{king14}.  In this paper, we present a catalog of all point sources detected in the \textit{NuSTAR} Norma Arm survey.  The \textit{NuSTAR} observations and basic data processing are described in \S~\ref{sec:obs} and \S~\ref{sec:processing}.  Descriptions of our source detection technique, aperture photometry, and spectral analysis are found in \S~\ref{sec:detection}, \S~\ref{sec:phot}, and \S~\ref{sec:spectral} respectively.  In \S~\ref{sec:discussion}, we discuss the physical nature of \textit{NuSTAR} detected sources and compare the Norma X-ray populations to those seen in the Galactic Center region.

\section{Observations}
\label{sec:obs}
\subsection{\textit{NuSTAR}}
\begin{figure*}[t]
\makebox[\textwidth]{ %
	\centering
	\subfigure{\includegraphics[width=1.0\textwidth]{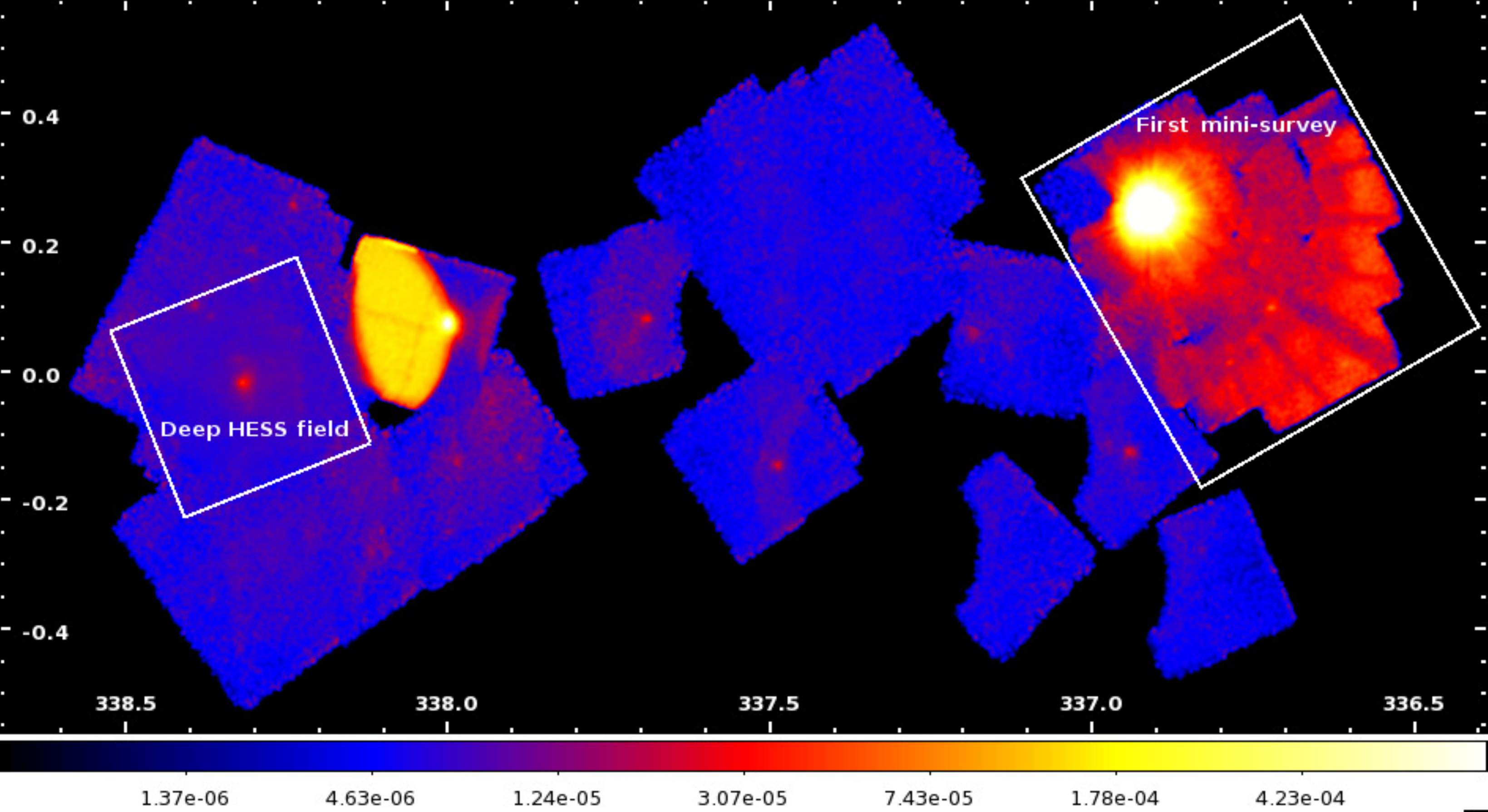}}}	
	\centering	
	\subfigure{\includegraphics[width=1.0\textwidth]{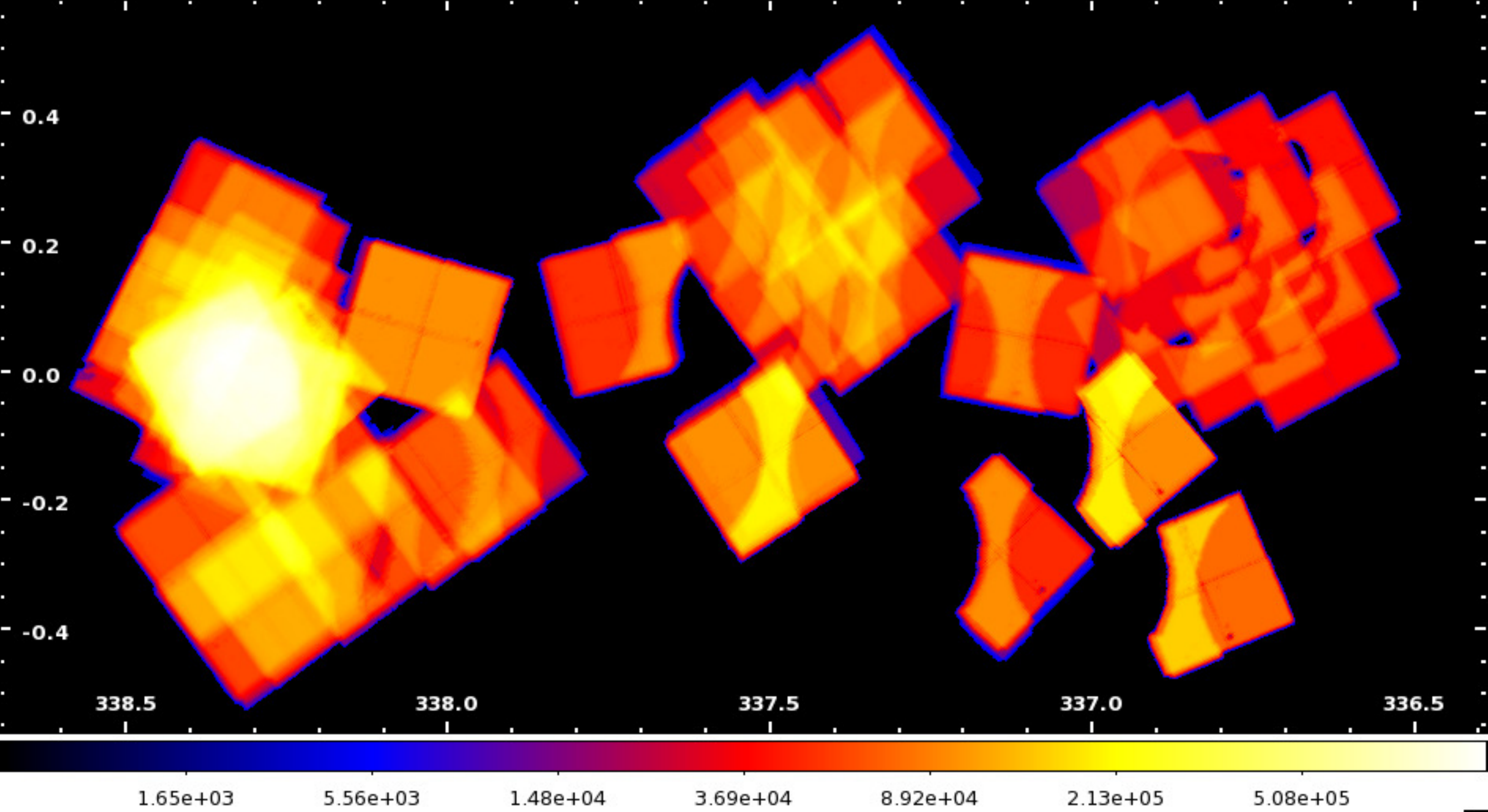}}
\caption{The top panel shows the smoothed 3--40~keV count rate mosaic (units of counts per second) and the bottom shows the 3--40~keV exposure map without vignetting correction (units of seconds).  The mosaics have been cleaned of most contamination from ghost rays and stray light; some residual ghost ray contamination can be seen in the first mini survey (upper right of the mosaic) while one wedge of stray light around ($\ell$, $b$) = (338$^{\circ}$, 0.08$^{\circ}$), which is due to GX~340+0, is not removed since a bright source, IGR~J16393-4643, is embedded in it.} 
\label{fig:mosaic}
\end{figure*}

\textit{NuSTAR} observations of the Norma arm region began in February 2013, and were completed in June 2015.  During this period, \textit{NuSTAR} performed 61 observations in the Norma region, shown in Figure~\ref{fig:mosaic}; every pointing consists of data from two co-aligned focal plane modules (FPM), A and B, each of which has a field-of-view (FOV) of $13^{\prime}\times13^{\prime}$.  \par
The \textit{NuSTAR} observations were planned to minimize contamination from stray light and ghost rays.  Stray light is the result of zero-bounce photons reaching the detector from bright sources within a few degrees of the FOV, while ghost rays are single-bounce photons from bright sources within about 1$^{\circ}$ of the FOV.  The pattern of stray light contamination is well-understood and can be carefully predicted\footnotemark\footnotetext{Stray light constraints for new observations can be checked with the stray light simulation tool at http://www.srl.caltech.edu/NuSTAR$\_$Public/NuSTAROperationSite/CheckConstraint.php}, while the patterns of ghost rays are more challenging to model (\citealt{koglin11}; \citealt{harrison13}; \citealt{wik14}; \citealt{mori15}; \citealt{madsen15}).  \par
Therefore, rather than observing the whole region surveyed by \textit{Chandra}, we performed simulations of stray light contamination and focused our observations on three areas of the sky that would be least affected by stray light.  Even in these ``cleaner'' areas, at least one of the focal plane modules was often affected by stray light, so exposure times for more contaminated observations were lengthened to compensate for the fact that we would not be able to combine data from both modules.  Seven additional pointings were specifically made at the locations of some of the brightest NARCS sources found to be hard in the \textit{Chandra} band and for which optical or infrared spectra have been obtained (\citealt{rahoui14}, Corral-Santana et al., in prep).  Unfortunately, despite this adopted strategy, the first mini-survey of the Norma region was highly contaminated by ghost rays because a black hole binary in the region, 4U~1630-472, serendipitously went into outburst while the \textit{NuSTAR} observations were taking place (B14).  Having learned about the spatial extent of ghost ray contamination, later observations in proximity of 4U~1630-472 were timed to occur only when it was in quiescence.  \par
Finally, in addition to the observations dedicated to the Norma survey either as part of the baseline \textit{NuSTAR} science program or the \textit{NuSTAR} legacy program, a series of observations were made to regularly monitor the pulsar associated with HESS~J1640-465 (\citealt{gotthelf14}, hereafter G14), a very luminous TeV source which resides within the Norma survey area.  When combining all such observations taken prior to March 2015, they yield a total exposure of 1~Ms over a 100~arcmin$^2$ field, which we call the ``deep HESS field''.  While the detailed analysis of the pulsar's braking index is discussed in \citet{archibald16}, here we present the other \textit{NuSTAR} sources detected in the deep HESS field. \par
Table~\ref{tab:obs} lists all the \textit{NuSTAR} observations included in our analysis.  Although the sources in the first mini-survey (\citealt{king14}; B14; T14), HESS~J1640-465 (G14), and IGR~J16393-4643 (\citealt{bodaghee16}, hereafter B16) have been analyzed separately and in more detail by others, we include these sources in our analysis to measure the photometric properties of all sources in a consistent way, allowing us to calculate the number-flux (log$N$-log$S$) distribution of \textit{NuSTAR} Norma Region (NNR) sources.  

\subsection{Chandra}

In this study, we make extensive use of information from the Norma Arm Region \textit{Chandra} Survey (NARCS) catalog as well as the soft ($<10$~keV) X-ray spectra of some of the NARCS sources.  The analysis of these \textit{Chandra} observations and the details of the spectral extraction are provided in F14.  We also use two other archival \textit{Chandra} observations that cover part of the area surveyed by \textit{NuSTAR}: ObsID 7591 provides an additional epoch for a transient source (NuSTAR~J164116-4632.2, discussed in \S~\ref{sec:chandrafollowup}), and ObsID 11008 provides spatially resolved observations of NARCS sources 1278 and 1279 \citep{rahoui14}, which are blended in the NARCS and \textit{NuSTAR} Norma observations.  For reference, we provide information about all these relevant archival \textit{Chandra} observations in Table~\ref{tab:archivalchandra}.  \par
Furthermore, in this study we make use of \textit{Chandra} observations which were triggered to follow-up four transient sources discovered by \textit{NuSTAR}.  These \textit{Chandra} observations were used to constrain their soft X-ray spectra and better localize their positions so as to be able to search for optical and infrared counterparts.  The follow-up observations of one of these transients, NuSTAR~J163433-4738.7, are discussed in T14, and the others are presented in \S~\ref{sec:chandrafollowup} and listed in Table~\ref{tab:chandra}.

\section{NuSTAR Data Processing and Mosaicking}
\label{sec:processing}
The raw data of each observation was processed using CALDB v20150612 and the standard \textit{NuSTAR} pipeline v1.3.1 provided under HEASOFT v6.15.1 to produce event files and exposure maps for both focal plane modules.  We made exposure maps with and without vignetting corrections to be used in different parts of our analysis. \par
Next, we cleaned the event files of stray light contamination by filtering out X-ray events in stray light affected regions.  Table~\ref{tab:obs} indicates whether stray light removal occurred in either FPMA or FPMB as well as the source responsible for the stray light.  In one exceptional case, we did not remove stray light seen in FPMA and FPMB of observation 30001008002, since a bright source, IGR~J16393-4643, is located within the stray light regions caused by GX~340+0 and 4U~1624-49.  We also excised the most significant ghost rays from observations from the first mini-survey, defining the ghost ray pattern regions in the same way as B14.  One observation, 30001012002, was performed to follow-up NuSTAR~J163433-4738.7, a transient source discovered in the first mini-survey; this observation helped to characterize the outburst duration of this transient (T14), but it was so extensively contaminated by ghost rays that it was not included in our analysis. Finally, a few observations show additional contamination features such as sharp streaks, listed in Table~\ref{tab:obs}, which were also removed.  \par
To improve the astrometric accuracy of the \textit{NuSTAR} observations, we calculated the shifts between the positions of bright \textit{NuSTAR} sources and their \textit{Chandra} counterparts in NARCS observations which were astrometrically registered using infrared counterparts in the VISTA Variables in the Via Lactea (VVV; \citealt{minniti10}) survey \citep{fornasini14}.  The positions of bright sources, which could be easily identified in raw images, were determined using the IDL \texttt{gcntrd} tool, which makes use of the DAOPHOT ``FIND'' centroid algorithm.  This source localization was done independently for each FPM of each observation and was used to apply translational shifts to event files and exposure maps.  In performing astrometric corrections, we limited ourselves to using sources with $>100$ net counts in each individual observation and FPM and located on-axis.  For on-axis sources with this number of counts, we expect the statistical error on the centroid to be $<6^{\prime\prime}$ based on simulations (Brian Grefenstette, personal communication, May 7, 2014).  NARCS 999 is very bright, with $>10,000$ net counts, and therefore the statistical uncertainties of the astrometric corrections derived from this source are $<2^{\prime\prime}$ at 90\% confidence; the other sources used for astrometric corrections have $100-300$ net counts, and their associated statistical uncertainties are expected to be $5-6^{\prime\prime}$ at 90\% confidence.  Table~\ref{tab:shift} lists the applied boresight shifts and the bright sources used for astrometric correction.  We were only able to apply these astrometric corrections to 23 out of 60 observations (43 out of 117 modules) due to the dearth of bright X-ray sources in our survey.  Our inability to astrometrically correct all the observations does not significantly impact the results of our photometric and spectral analysis since the radii of the source regions we use are significantly larger than the expected shifts.  The boresight shifts range from 1$^{\prime\prime}$ to 14$^{\prime\prime}$; 20\% of the shifts are larger than 8$^{\prime\prime}$, which is more than expected based on \textit{NuSTAR}'s nominal accuracy of $\pm$8$^{\prime\prime}$ at 90\% confidence \citep{harrison13}, but is not unexpected given that the statistical errors on the source positions may be as high as 6$^{\prime\prime}$.  Checking each shifted and un-shifted image by eye and comparing the locations of \textit{NuSTAR} sources with their \textit{Chandra} counterparts in shifted and un-shifted mosaic images, we confirm that these boresight shifts constitute an improvement over the original \textit{NuSTAR} positions.  \par
We re-projected the event files of each observation onto a common tangent point and merged all the observations and both FPM together to maximize photon statistics.  We then generated mosaic images on the common sky grid in the 3--78, 3--10, 3--40, 10--20, 10--40, 20--40, and 40--78~keV bands.  To create mosaic exposure maps, we combined the individual exposure maps by adding exposure values at the location of each sky pixel in the mosaic image; we made exposure maps both without vignetting corrections and with vignetting corrections evaluated at 8, 10, and 20~keV.  We used the exposure maps without vignetting corrections when we calculated the source significance and net counts, since these calculations require comparing the exposure depth in the source and background region apertures and the background is dominated by non-focused emission.  Instead, when calculating sensitivity curves (\S~\ref{sec:sensitivity}), we used exposure maps with vignetting corrections since the source emission is focused by the telescope mirrors.  When calculating the source fluxes, vignetting corrections are taken into account through the ancillary response file (ARF).  An exposure-corrected \textit{NuSTAR} mosaic image in the 3--40~keV band and exposure map without vignetting correction are shown in Figure~\ref{fig:mosaic}.  As can be seen, the typical exposure depth of the Norma survey is 30--100~ks while the exposure of the deep field is 1~Ms.

\section{Source Detection}
\label{sec:detection}

\subsection{Generating trial maps}
To identify sources in the \textit{NuSTAR} Norma survey, we employed a technique that was specifically developed for the \textit{NuSTAR} surveys.  This technique, which we refer to as the ``trial map'' technique, is described in detail by \citet{hong16}, so we only provide a brief explanation here.  The \textit{NuSTAR} Galactic Center region survey \citep{hong16}, and the \textit{NuSTAR} extragalactic surveys (\citealt{civano15}; \citealt{mullaney15}; Lansbury et al., submitted) all use this technique as the basis for their detection method.    
As a result of \textit{NuSTAR}'s point spread function (PSF) being larger and its background being higher and more complex compared to other focusing X-ray telescopes such as \textit{Chandra} and \textit{XMM-Newton}, the utility of typical detection algorithms, such as \texttt{wavdetect} \citep{freeman02}, is limited when applied to \textit{NuSTAR} data.  One way of dealing with this problem is to add an additional level of screening to the results of conventional algorithms, calculating the significance of detections by independent means and setting a significance detection threshold.  The trial map technique is more direct, skipping over the initial step of using a detection algorithm such as \texttt{wavdetect}. \par
To make a trial map, for each sky pixel, we calculate the probability of acquiring more than the total observed counts within a source region due to a random background fluctuation.  For each pixel, the source and background regions are defined as a circle and an annulus, respectively, centered on that pixel.  The mean background counts expected within the source region are estimated from the counts in the background region scaled by the ratio of the areas and exposure values of the source and background regions.  Using background regions that are symmetric around the central pixel helps to account for spatial variations of the background.  In making trial maps, we plot the inverse of the random chance probability, which is the number of random trials required to produce the observed counts simply by random background fluctuations, such that brighter sources with higher significance have higher values in the maps.  \par
\begin{figure*}[t]
\makebox[\textwidth]{ %
\centering
\includegraphics[width=1.0\linewidth]{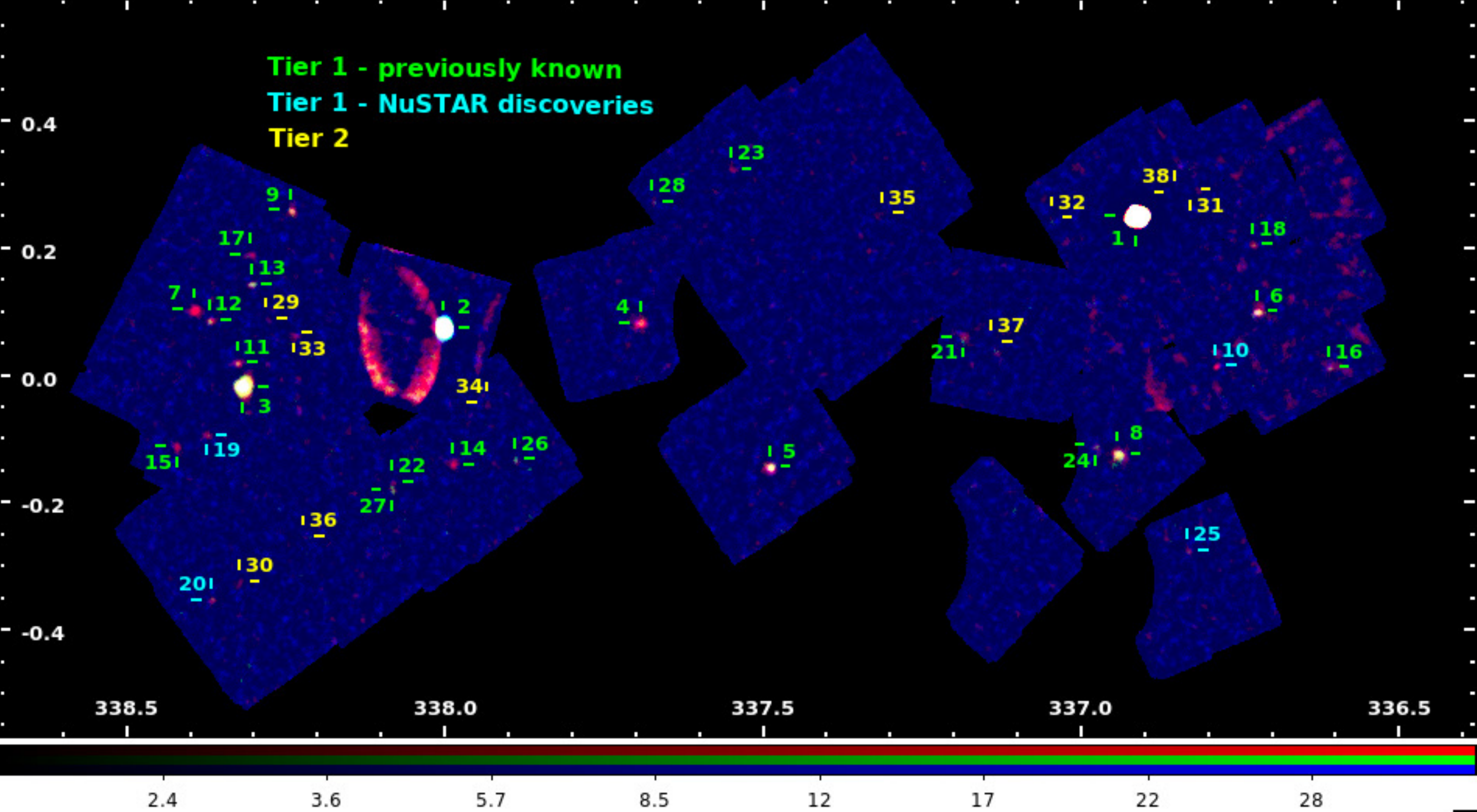}}
\caption{Composite trial map showing the 3--10~keV band in red, 10--20~keV band in green, and 20--40~keV band in blue.  The colors are scaled by the logarithmic trial map values.  Tier~1 sources are labeled in green, if they were observed by NARCS or were previously well-studied, or cyan, if they were discovered by the \textit{NuSTAR} Norma survey.  Tier~2 sources are labeled in yellow.  The streaks in the vicinity of NNR~2 are due to stray light which has not been removed because NNR~2 is partially embedded in it.  The small streaks seen in the area covered by the first mini-survey are due to ghost rays from NNR~1. }
\label{fig:trial}
\end{figure*}

We generated trial maps using three different source region sizes with radii of $9\farcs5$, 12$^{\prime\prime}$, and 17$^{\prime\prime}$ (corresponding to 15, 22, and 30\% enclosures of the PSF, respectively) and six different energy bands (3--78, 3--10, 10--40, 40--78, 10--20, 20--40~keV).  The source region sizes we used are slightly larger than those used in the analysis of the \textit{NuSTAR} Galactic Center survey since the smaller sizes are especially suited for picking out relatively bright sources in areas of diffuse emission, but in the Norma region there is no evident diffuse emission apart from stray light and ghost rays.  The inner and outer radii of the background regions are 51$^{\prime\prime}$ (corresponding to 70\% of the PSF) and 85$^{\prime\prime}$ (equal to 5/3 of the inner radius), respectively, in all cases .  Figure~\ref{fig:trial} shows trial maps made using the 22\% PSF enclosure and the 3--10, 10--20, and 20--40~keV bands; the three energy bands are combined into a three-color image so that spectral differences between sources can be seen. 

\subsection{Detection Thresholds and Source Selection}
\label{sec:threshold}

\begin{figure}
\includegraphics[width=0.47\textwidth]{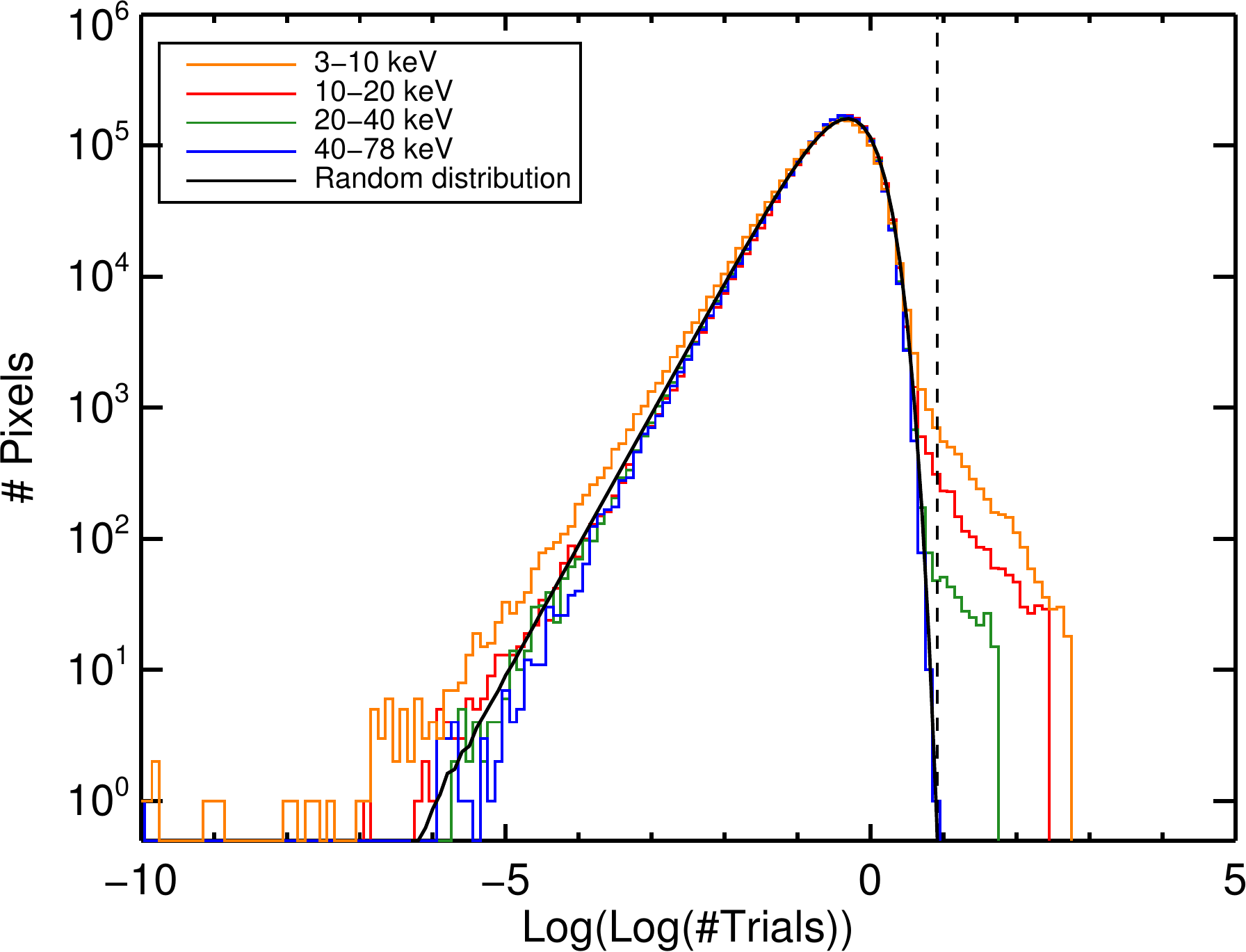}
\caption{Distribution of trial map values in different energy bands for 22\% PSF enclosures.  The x-axis is shown in a double logarithmic scale.  The 40--78~keV distribution closely matches the random distribution expected due to Poissonian fluctuations of the background; this is consistent with the fact that among the ``clean'' observations included in creating this plot, only one source is detected in the 40--78~keV band.  The vertical dashed line shows the detection threshold set for the 3--10~keV band trial map.  The excess of high trial map values relative to the 40--78~keV band distribution is due to the presence of sources, stray light, and ghost rays; the excess of low trial map values result from the vicinity of bright sources, which effectively increase the local background.}
\label{fig:tdist}
\end{figure}

When considering how to set detection thresholds for our trial maps, we excluded the observations from the first mini-survey and observation 30001008002 since they have significantly higher levels of stray light and ghost ray contamination than the rest of the survey; in the remainder of this paper, we will refer to this subset of observations as the ``clean'' sample.  Figure~\ref{fig:tdist} shows the fractional distributions of the values from the ``clean'' trial maps using source region sizes of 22\% PSF enclosures.  As can be seen, the distribution for the 40--78~keV band is very close to that expected for a Poissonian distribution of random background fluctuations, and in fact no sources are clearly visible in the ``clean'' trial maps.  \par
\begin{figure}
\includegraphics[angle=90,width=0.47\textwidth]{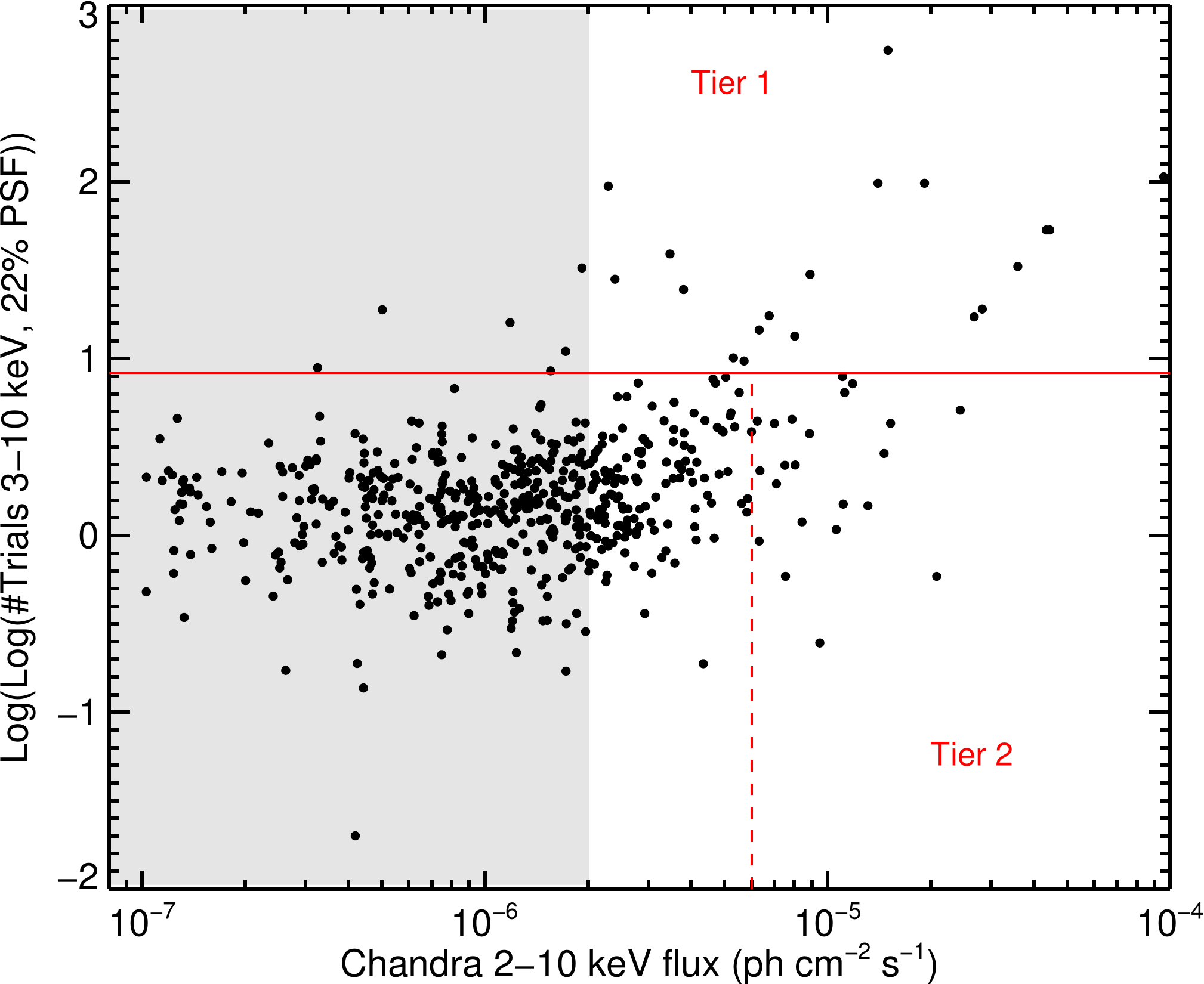}
\caption{Trial map value in the 3--10~keV band using 22\% PSF enclosures versus \textit{Chandra} 2--10~keV photon flux for NARCS sources in the surveyed \textit{NuSTAR} area.  Fluxes of sources in the gray region are uncorrelated with the trial map values and used to set the detection threshold, which is shown by the red horizontal line.  Sources above the horizontal line in at least two trial maps are tier~1 sources, while bright sources below that line but to the right of the vertical dashed line are tier~2 candidates.}
\label{fig:tflux}
\end{figure}

Following the procedure described in \citet{hong16} to establish detection thresholds, we began by cross-correlating each trial map with the NARCS source catalog.  Figure~\ref{fig:tflux} shows the maximum trial map value within 10$^{\prime\prime}$ of the locations of NARCS sources detected at $>3\sigma$ in the 2--10~keV band as a function of \textit{Chandra} photon flux.  Above \textit{Chandra} fluxes of $6\times10^{-6}$~cm$^{-2}$~s$^{-1}$, more than 1/3 of NARCS sources have trial map values which are significantly higher than the bulk of NARCS sources clustered between trial map values of 10$^{0.3}$ to 10$^{3}$.  For \textit{Chandra} fluxes lower than $2\times10^{-6}$~cm$^{-2}$~s$^{-1}$, the distribution of trial map values are uncorrelated with source flux, having a linear Pearson correlation coefficient $|p|<0.04$ for all trial maps.  \par  
For a source to be considered for the final catalog, we require that it exceed the detection threshold in at least two trial maps.  If all 18 trial maps were independent of each other, the expected number of false sources ($N_{\mathrm{F}}$) would be equal to $N_{\mathrm{can}}C(18,2)p^{16}(1-p)^2$, where $N_{\mathrm{can}}$ is the number of NARCS sources included in a \textit{NuSTAR} counterpart search, $C(i,j)$ is a binomial coefficient, and $p$ is the fraction of false sources to be rejected in each map \citep{hong16}.  However, the trial maps are not completely independent given that their energy ranges overlap.  Thus, to at least partly account for the fact that some of the trial maps are correlated, we set a stringent limit on the expected number of false sources, setting $N_{\mathrm{F}}$=0.5.  Since the long-term variability of NARCS sources is unknown, we search for \textit{NuSTAR} detections among all NARCS sources.  Thus, in the ``clean'' map regions, $N_{\mathrm{can}}$=579; limiting $N_{\mathrm{F}}$ to 0.5 requires a rejection percentage $p = 99.76$\%.  Making a cumulative distribution function of the trial map values of uncorrelated NARCS sources lying in the gray area of Figure~\ref{fig:tflux}, we determine the corresponding trial value threshold for each trial map; the detection thresholds range from 10$^{5.2}$ in the 20--40~keV band with 15\% PSF enclosures to 10$^{10.3}$ in the 3--10~keV band for 30\% PSF enclosures.  \par
Having established detection thresholds for each trial map, we first searched for any \textit{Chandra} sources  detected by \textit{NuSTAR}.  We cross-correlated all NARCS sources detected at $>3\sigma$ in the 2--10~keV \textit{Chandra} band with the trial maps of the full set of observations, including those with significant background contamination.  We considered all NARCS sources that exceed the detection threshold in at least two trial maps as tier 1 candidate sources.  All sources with 2--10~keV \textit{Chandra} flux $>6\times10^{-6}$~cm$^{-2}$~s$^{-1}$ that are not tier~1 sources are considered tier~2 candidate sources, regardless of their trial map values.  Although for tier~2 sources, we do not expect to be able to retrieve significant spectral information, we can at least check for significant variability between the \textit{Chandra} and \textit{NuSTAR} observations and place upper limits on the flux above 10~keV.  We also performed a blind search for \textit{NuSTAR} sources that were not detected in NARCS; we consider any clusters of pixels that exceed the detection threshold in at least three trial maps as additional tier~1 candidate sources.  \par
We then inspected all the candidate sources.  First, we checked whether \textit{NuSTAR} sources matched to \textit{Chandra} counterparts are unique matches.  We find 13 cases in which multiple NARCS sources were associated with a single \textit{NuSTAR} detection due to \textit{NuSTAR}'s much larger PSF; however, in all these cases, one NARCS source was more clearly centered on the \textit{NuSTAR} position and was also significantly brighter, demonstrating the more likely association.  We then visually inspect all tier~1 candidate sources without NARCS associations to ensure they are not associated with artifacts due to stray light, ghost rays, or the edges of the fields of view (FoVs).  Based on this visual inspection, we exclude three candidate sources located at the edges of the FoVs, the stray light region near NNR~2, and 21 candidates without a clear point-like morphology that are located in the first mini-survey area contaminated by ghost rays.  In addition, since tier~2 candidate sources do not exceed the trial map detection thresholds, in order for them to be included in our final catalog, we require that their aperture photometry have a signal-to-noise ratio (S/N)$>3\sigma$ in at least one of the 3--10, 3--40, or 10--20~keV energy bands (see \S~\ref{sec:counts} for details).  In total, after these different screenings, 28 tier~1 candidates and 10 tier~2 candidates are included in our final source list, shown in Table~\ref{tab:srclist}.  \par
To determine the best position of tier~1 \textit{NuSTAR} sources, we applied the DAOPHOT ``FIND'' algorithm in the proximity of each source in the 3--10~keV trial map with 22\% PSF enclosure; we found that using the centroid algorithm on the trial maps rather than the mosaic images yielded better results, allowing the algorithm to converge for all tier 1 sources with lower statistical errors.  When applying the centroid algorithm, we used the 3--10~keV, 22\% PSF trial map since all the tier~1 sources are clearly discernible in this map.  The tier~2 sources are not bright enough for the centroid algorithm to yield reliable results, so we simply adopt the \textit{Chandra} positions for these sources.  The offsets between tier~1 sources and their \textit{Chandra} counterparts vary from $0\farcs9$ to 14$^{\prime\prime}$, excluding two extended sources (NNR~8 and 21) whose \textit{Chandra} positions were determined subjectively by eye.  The offsets of four \textit{NuSTAR} point sources from their \textit{Chandra} counterparts are larger than the 90\% \textit{NuSTAR} positional uncertainties.  We estimated the \textit{NuSTAR} positional uncertainty for each tier~1 source as the quadrature sum of statistical and systematic uncertainties.  We calculated the statistical error by performing Gaussian fits to histograms of the spatial count distributions in the X and Y directions in a $25\times25$ pixel image cutout centered on the source position.  These statistical errors are approximate since the \textit{NuSTAR} PSF has non-Gaussian wings, but comparison of the errors derived using the Gaussian approximation to those derived from the accurate PSF simulations performed for some of the brighter sources (see \S\ref{sec:processing}) indicates this approximation is accurate to 10\%.  For the systematic uncertainty, we assumed the nominal $8^{\prime\prime}$ astrometric accuracy \citep{harrison13} for sources located in observations which were not astrometrically corrected and the uncertainties calculated in \S\ref{sec:processing} for sources in astrometrically-corrected observations.  Looking carefully at the four sources with the largest offsets, the similarity between their fluxes and/or spectral properties in the 2--10~keV band between \textit{Chandra} and \textit{NuSTAR} suggests that they are true counterparts despite the large positional offsets.  The fact that 17\% of the \textit{NuSTAR} offsets exceed the 90\% positional uncertainties suggests that the \textit{NuSTAR} positional uncertainty is slightly underestimated.  Large offsets between \textit{NuSTAR} positions and soft X-ray counterparts are also seen in the \textit{NuSTAR} serendipitous survey, where \citet{lansbury16} find that the 90\% positional accuracy of \textit{NuSTAR} varies from 12$^{\prime\prime}$ for the most significant detections to 20$^{\prime\prime}$ for the least significant detections.  The large \textit{NuSTAR} offsets in the serendipitous survey suggest that the 90\% \textit{NuSTAR} systematic uncertainty is larger than 8$^{\prime\prime}$, which would help to explain some of the large offsets seen for sources in the Norma survey. \par
Table~\ref{tab:srclist} provides information about the detection, position, and \textit{Chandra} counterparts of all \textit{NuSTAR} Norma region (NNR) sources.  The tier~1 sources include five sources not detected in NARCS; one of them is the well-known LMXB 4U~1630-472 \citep{kuulkers97} while the others are new transient sources discussed in \S~\ref{sec:chandrafollowup}. 

\section{Aperture Photometry}
\label{sec:phot}

\subsection{Defining source and background regions}
\label{sec:regions}
For photometry and spectral extraction, we used circular source regions and, whenever possible, annular background regions centered on the source positions provided in Table~\ref{tab:srclist}.  At energies below 20~keV, the \textit{NuSTAR} background is not uniform as it is dominated by non-focused emission, which exhibits spatial variations due to shadowing of the focal plane \citep{harrison13}.  Using aperture regions that are symmetric about the source position helps to compensate for this non-uniformity.  We performed our photometric analysis with two different source extraction regions with 30$^{\prime\prime}$ and 40$^{\prime\prime}$ radii (corresponding to roughly 50\% and 60\% PSF enclosures, respectively), to assess possible systematic errors associated with aperture selection.  The default background regions are annuli with 60$^{\prime\prime}$ inner radii and 90$^{\prime\prime}$ outer radii.  For NNR~8 and 21, which appear extended and are not fully contained within the default source regions, we adopted radii of 45$^{\prime\prime}$  and 60$^{\prime\prime}$ for the small and large circular source regions, respectively, and annular background regions with 80$^{\prime\prime}$ inner radii and 110$^{\prime\prime}$ outer radii.  We adjusted the centers of the aperture regions for NNR~8 and 21 by 8$^{\prime\prime}$ and $5^{\prime\prime}$, respectively, so that they were more centered with respect to the full extended emission rather than the peak of the emission \footnotemark\footnotetext{The adjusted locations of the aperture regions for NNR~8 and 21 are ($\alpha$, $\delta$) = (248.9468, -47.6238) and (248.9875, -47.3200), respectively}. \par
For about 1/3 of sources, it was necessary to modify the background aperture regions.  In order to prevent contamination to the background from other sources, it is preferable for background regions not to extend within 60$^{\prime\prime}$ of any tier 1 source.  In addition, above 20~keV, as the relative contribution of the internal background becomes more significant, the background is fairly uniform across any given detector but differs between detectors (\citealt{harrison13}; \citealt{wik14}), so it is advantageous for the background region to be located on the same detector as the source region.  Furthermore, when a source is located close to the edge of the field of view, using an annular background region may not sample a statistically large enough number of background counts.  Finally, although we removed the most significant patches of stray light and ghost ray contamination from \textit{NuSTAR} observations, non-uniform low-level contamination remains.  Thus, we modified the background region in situations where the default background region comes within 60$^{\prime\prime}$ of any tier~1 source, the low-level contamination from stray light or ghost rays appears to differ significantly between the source and default background regions, or $>50\%$ of the annular background region falls outside the observation area or on a detector different from the one where the source is located.  In these cases, we adopted a circle with a 70$^{\prime\prime}$ radius for the background region and placed it in as ideal a location as possible following these criteria: 
\begin{list}{}{%
\setlength{\topsep}{0pt}%
\setlength{\leftmargin}{0.3in}%
\setlength{\listparindent}{0.0in}%
\setlength{\itemindent}{0.0in}%
\setlength{\parsep}{\parskip}%
\setlength{\itemsep}{0pt}
}%
\item[]
i. Keeping the region as close to the source as possible to minimize variations due to background inhomogeneities, but at least 60$^{\prime\prime}$ away from the source and any tier~1 sources.  \\
ii. Maximizing the fraction of the background region area that falls on the same detector as the source region. \\
iii. Placing the background region at a location that exhibits a similar level of low-level stray light or ghost ray contamination as the source region.  
\end{list}
\par
For a given source, background aperture regions were defined for each observation and FPM individually since stray light and ghost ray contamination as well as the fraction of the default annular background that lies on a given detector varies depending on the observation and the module.  Furthermore, if a source fell close to the edge of an observation, such that $>$50\% of the area of a 40$^{\prime\prime}$ radius source region was outside the observation area, that observation was not used to extract photometric or spectral information for the source.  Thus, the exposure value at the location of a source in the mosaicked exposure map may be higher than the effective exposure for the source based only on observations used for photometric analysis; the latter effective exposure is the value reported in Table~\ref{tab:srclist}.  Table~\ref{tab:phot} provides the results of our aperture photometry and includes flags that indicate which sources required modified background regions.  \par
The only exceptions to this method of defining background regions are NNR~22 and 27.  These sources are only separated by 47$^{\prime\prime}$ and thus contaminate each other's default background regions although they do not suffer from any additional background problems.  Therefore, since annular background regions are preferable for minimizing vignetting effect, we simply redefined their background regions as an annulus with an 80$^{\prime\prime}$ inner radius and 110$^{\prime\prime}$ outer radius centered in between the two sources.  Due to their proximity, the photometric and spectral properties of these sources as derived from 40$^{\prime\prime}$ radius circular apertures are less reliable than those from the 30$^{\prime\prime}$ radius apertures. 

\subsection{Net counts and source significance} 
\label{sec:counts}
Having defined aperture regions, we extracted the source and background counts for each source in each observation.  We then calculated the expected number of background counts ($\langle c_{\mathrm{bkg}} \rangle$) in each source region by multiplying the counts in the background region by the ratio $(A_{\mathrm{src}} E_{\mathrm{src}})/(A_{\mathrm{bkg}} E_{\mathrm{bkg}})$, where $A_{\mathrm{src}}$ and $A_{\mathrm{bkg}}$ are the areas, in units of pixels, and $E_{\mathrm{src}}$ and $E_{\mathrm{bkg}}$ are the exposures (without vignetting corrections) of the source and background regions, respectively.  Then for each source, we summed the source counts ($C_{\mathrm{src}}$), total background counts ($C_{\mathrm{bkg}}$), background counts expected in the source region ($\langle C_{\mathrm{bkg}} \rangle$), and exposures across all observations and modules in 7 different energy bands: 3--78, 3--40, 40--78, 3--10, 10--40, 10--20, and 20--40~keV.   The 1$\sigma$ errors in the total counts were calculated using the recommended approximations for upper and lower limits in \citet{gehr}.  Then, the net source counts ($C_{\mathrm{net}}$) were calculated by subtracting the total expected background counts in the source region from the total source counts.  \par
In each energy band, we then calculated the signal-to-noise ratio (S/N) of the photometric measurements from the probability that the source could be generated by a noise fluctuation of the local background using the following equation from \citet{weiss07}:
\begin{multline}
P(\geq C_{\mathrm{src}}|C_{\mathrm{bkg}};C_{\mathrm{net}}=0) = \\ \sum_{c = C_{\mathrm{src}}}^{C_{\mathrm{bkg}}+C_{\mathrm{src}}}\frac{(C_{\mathrm{bkg}}+C_{\mathrm{src}})!}{c!(C_{\mathrm{bkg}}+C_{\mathrm{src}}-c)!}\left(\frac{f}{1+f}\right)^c\left( 1-\frac{f}{1+f}\right)^{C_{\mathrm{bkg}}+C_{\mathrm{src}}-c} 
\label{eq:probnoise}
\end{multline}
where $f = \langle C_{\mathrm{bkg}} \rangle/C_{\mathrm{bkg}}$.  Using this probability, we define the S/N as the equivalent Gaussian significance in units of the standard deviation (e.g., $P=0.0013$ corresponds to S/N$=3\sigma$).  These S/N measurements are used to select which tier 2 sources to include in our catalog, but not to set detection thresholds for tier 1 sources, which are determined by the trial maps.  Only five sources have photometric measurements with S/N$\geq3\sigma$ above 20 keV.  Therefore, we focus the remainder of our analysis on the 3--40, 3--10, and 10--20~keV energy bands.  Of the tier 2 source candidates, we only included those with S/N$\geq3\sigma$ in at least one of these three energy bands, using either of the two source aperture regions, in our final source list.   Table~\ref{tab:phot} provides the significance of each source in our final catalog in these three energy bands, the net counts in the 3--40~keV band, and additional photometric properties described in the following sections.  We estimate that local spatial variations of the background could affect the S/N values reported in this table by $\pm0.4\sigma$, and change the measured net counts and fluxes by $\pm5$\%, variations which are smaller than the statistical uncertainties of the photometric measurements. 

\subsection{Photon and energy fluxes}

In \S~\ref{sec:spectral}, we describe how we derived fluxes from spectral modeling, but for all sources we also derived fluxes in a model-independent way since the spectral fitting of faint sources is prone to significant uncertainty.  For each source and background region in each observation and module, we used \texttt{nuproducts} to extract a list of photon counts as a function of energy and generate both an ARF and a response matrix file (RMF); the ARFs are scaled by the PSF energy fraction enclosed by the aperture region.  We first calculated the source photon flux within each observation and module in the 3--10 and 10--20~keV bands by dividing the counts in each channel by the corresponding ARF, summing all these values within the given energy band, and then dividing by the source region exposure; the estimated background contribution, scaled from the photon flux measured in the background region, was subtracted.  These photon flux measurements assume a quantum efficiency of 1, which is a decent approximation for the \textit{NuSTAR} CdZnTe detectors, which have a quantum efficiency of 0.98 over the vast majority of the \textit{NuSTAR} energy range \citep{bhalerao12}.  If the significance of a source in a particular observation was $<1\sigma$, then we calculated a 90\% confidence upper limit to its photon flux by converting the probability distribution of true source counts (from Equation A21 in \citealt{weiss07}) to a photon flux distribution using the source region effective area.  \par
For the five transient sources which were detected by \textit{NuSTAR} but not by NARCS, we looked at lightcurves of their 3--10~keV photon fluxes to check whether they are detected at $>2\sigma$ confidence in individual \textit{NuSTAR} observations.  We found that NNR~1 is only detected in ObsIDs 40014008002 and 40014009001, NNR~10 is only detected in ObsID 40014007001 (which is consistent with T14), and NNR~19 is only detected in ObsIDs 30002021005, 30002021007, 30002021009, 30002021011, and 30002021013.  Excluding the observations in which the transient sources are not detected, we re-evaluated their 3--40~keV net counts and source significance as described in \S~\ref{sec:counts}, and continued to exclude these observations for these sources when determining their other average photometric and spectral properties.  Thus, the photometric and spectral properties derived for NNR~1, 10, 19, and 25 should be considered as their average properties during high flux states.  \par
For each source, we then computed average 3--10 and 10--20~keV photon fluxes by combining the count lists and ARFs from different observations and modules.  These measurements are presented in Table~\ref{tab:phot}.  We also calculate the average 3--10 and 10--20~keV energy flux for each source using the same model-independent method but with the additional step of multiplying the source counts in each channel by the channel energy.  Fluxes derived using the two different source region sizes are in $1\sigma$ agreement with one another, except for three sources which are located in regions of diffuse emission or ghost rays and thus do not appear as exactly point-like.  Comparing the model-independent fluxes with those we derived from spectral modeling (see \S~\ref{sec:spectral}) for tier~1 sources, we find they are in good agreement when using the smaller aperture regions, but show a significant number of discrepancies at $>2\sigma$ confidence when using the larger aperture regions.  In the larger aperture regions, while the net number of source counts is higher, so is the background/source count ratio, which is why in most cases the source significance derived from the larger aperture regions is slightly lower; as a result, accurate background subtraction is more important when using the larger aperture regions and it is not surprising that our crude subtraction method, which assumes a spectrally flat background, for the model-independent fluxes leads to discrepancies with the spectral fluxes.  

\subsection{X-ray variability}
\label{sec:var}
\textit{NuSTAR}'s high time resolution allows us to characterize the timing properties of detected sources over a range of timescales. \textit{NuSTAR}'s time resolution is good to $\sim 2$~ms rms, after being corrected for thermal drift of the on-board clock, and the
absolute accuracy is known to be better than $< 3$~ms (\citealt{mori14}; \citealt{madsen15}). 
For our timing studies, all photon arrival times were converted to barycentric dynamical time (TDB) using the \textit{NuSTAR} coordinates of each point source. \par
To characterize the source variability on $\sim\mathrm{hourly}$ timescales we
used the Kolmogorov-Smirnov (KS) statistic to compare the temporal
distributions of X-ray events extracted from source and background
apertures in the $3-20$~keV energy band.  The background light curve acts
as a model for the count rate variations expected in the source region due to the background. The
maximal difference between the two cumulative normalized light curves gives
the probability that they are drawn from the same distribution, i.e. that the light curve in the source region is consistent with that expected from the background plus a source with constant flux.  Any source with a KS statistic
lower than $0.05\%$ in any observation is flagged as short-term variable by an ``s'' in Table~\ref{tab:phot}.  For each source, we ran the KS test independently for each of the observations in which it was covered.
 Since the KS test is applied 160 times in total, the adopted threshold corresponds to $\lesssim1$ spurious detection.  We identify two sources as variable using the KS test.  An examination of the lightcuves of these sources, NNR~2 (presented in B16) and NNR~15 (Figure 5), shows clear variability on $\sim$hourly timescales.\par

\begin{figure}
\includegraphics[width=0.5\textwidth]{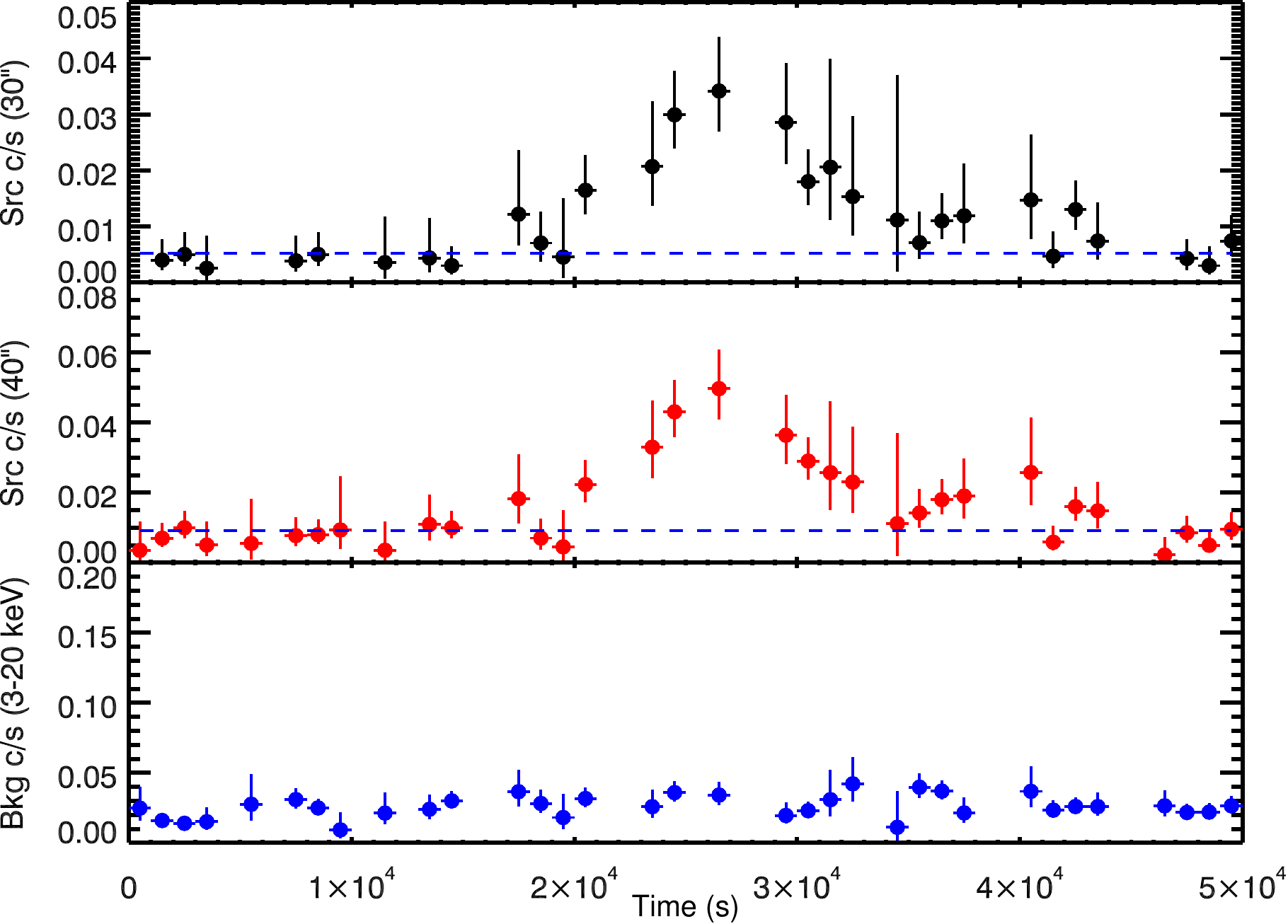}
\caption{Light curve of NNR~15 in the \textit{NuSTAR} 3--20~keV band from ObsID 40014016001, FPMA and FPMB combined, as measured from an aperture region with a 30$^{\prime\prime}$ radius (top) and a 40$^{\prime\prime}$ radius (middle).  The light curve exhibits evident short-term variability.  The bottom panel displays the light curve extracted from the background aperture region; the blue dashed lines in the top two panels show the mean background count rate scaled by the source region area.  The light curves display the average count rate in each 1~ks time interval; note that during some of these time intervals, the effective exposure time is less than 1~ks due to Earth occultations or periods of poor data quality.}
\label{fig:lightcurves}
\end{figure}
We checked for variability of the NNR sources on week to year
timescales by comparing the flux detected between repeated \textit{NuSTAR}
observations. Sources were flagged as long-term variable with an ``l'' in Table~\ref{tab:phot} if their $3-10$~keV photon flux differed by $>3\sigma$ based on their flux measured uncertainties; given the number of flux comparisons performed, this $3\sigma$ threshold should result in $\lesssim1$ spurious detection.  NNR~1, 10, 11, 19, and 29 were found to be variable using this criterion.  In addition, we compared \textit{Chandra} and \textit{NuSTAR} fluxes to check for variability on year timescales.  For all sources with sufficient photon statistics, we compared the joint spectral fits to \textit{Chandra} and \textit{NuSTAR} data (see \S~\ref{sec:spectral} for details), and identified sources with normalizations that differed at the $>90$\% confidence level.  Since we performed these joint fits for 24 sources, we would expect as many as two spurious detections of variability, but we made the criterion more stringent by requiring that for a source to be considered variable between the \textit{Chandra} and \textit{NuSTAR} observations, its \textit{Chandra} and \textit{NuSTAR} normalizations must be inconsistent regardless of which of three different spectral models is adopted.  This more selective criterion is only met by NNR~4, 11, and 27.  For fainter sources (NNR~29-38), we considered a range of spectral models that would be consistent with their quantile values, and assessed whether their 2--10~keV \textit{Chandra} flux was incompatible with their average 3--10~keV \textit{NuSTAR} flux at $>90$\% confidence, regardless of the spectral model assumed.  NNR~28, 35, and 36 are found to be variable by this criterion.  In Table~~\ref{tab:var}, we provide maximum photon fluxes and the ratio of maximum and minimum fluxes for all \textit{NuSTAR} sources that demonstrate X-ray variability; the transient sources, NNR~1, 10, 19, 20, and 25, which are detected by \textit{NuSTAR} but not detected in NARCS, are flagged as long-term variable and included in this table as well. \par
We searched for a periodic signal from those \textit{NuSTAR} sources with
sufficient counts to detect a coherent timing signal, determined as
follows. The ability to detect pulsations depends strongly on the
source and background counts and number of search trials.  For a
sinusoidal signal, the aperture counts (source plus
background) necessary to detect a signal of pulsed fraction $f_p$ is
$N=2S/f^2_p$, where $S$ is the power associated with the single trial false
detection probability of a test signal $\wp = e^{-S/2}$; $S$ is
distributed as $\chi^2$ with two degrees of freedom \citep{vanderklis89}.  In practice, for
a blind search, we need to take into account the number of frequencies
tested $N_{trials} = T_{span}/f_{Nyq}$, when $T_{span}$ is the data
span and $f_{Nyq} = 250$~Hz, the effective \textit{NuSTAR} Nyquist
frequency. In computing $N$ we must allow for the reduced
sensitivity of the search due to background contamination in the
source aperture ($N_b$); the minimum detectable pulse fraction
$f_p({\mathrm{min}})$ is then increased by $(N_s+N_b)/N_s$.\par
We computed the detectability in individual observations for each source in our sample and considered those suitable for a pulsar
search, with $f_p({\mathrm{min}}) > 50\%$ at the $3\sigma$ level.  For the
three brightest sources in the Norma survey, their timing properties
are already presented elsewhere: i) the quasi-periodic oscillations of
the black hole binary 4U~1630- 472 (NNR~1) were extensively studied
using the \textit{Rossi X-ray Timing Explorer} (\citealt{tomsick00}; \citealt{dieters00}; \citealt{seifina14}), ii) the high mass X-ray binary pulsar HMXB IGR~J16393-4643 (NNR~2) with a period of 904 seconds, whose spin-up rate
was determined from recent \textit{NuSTAR} observations (B16),
and iii) the \textit{NuSTAR}-discovered 206~ms pulsar PSR~J1640$-$4631 (NNR~3)
associated with the TeV source HESS~J1640$-$465 (G14;\citealt{archibald16}).\par

\begin{figure*}[t]
	\centering
	\subfigure[NNR 19, \textit{Chandra} ObsID 16170]{\includegraphics[width=2.3in]{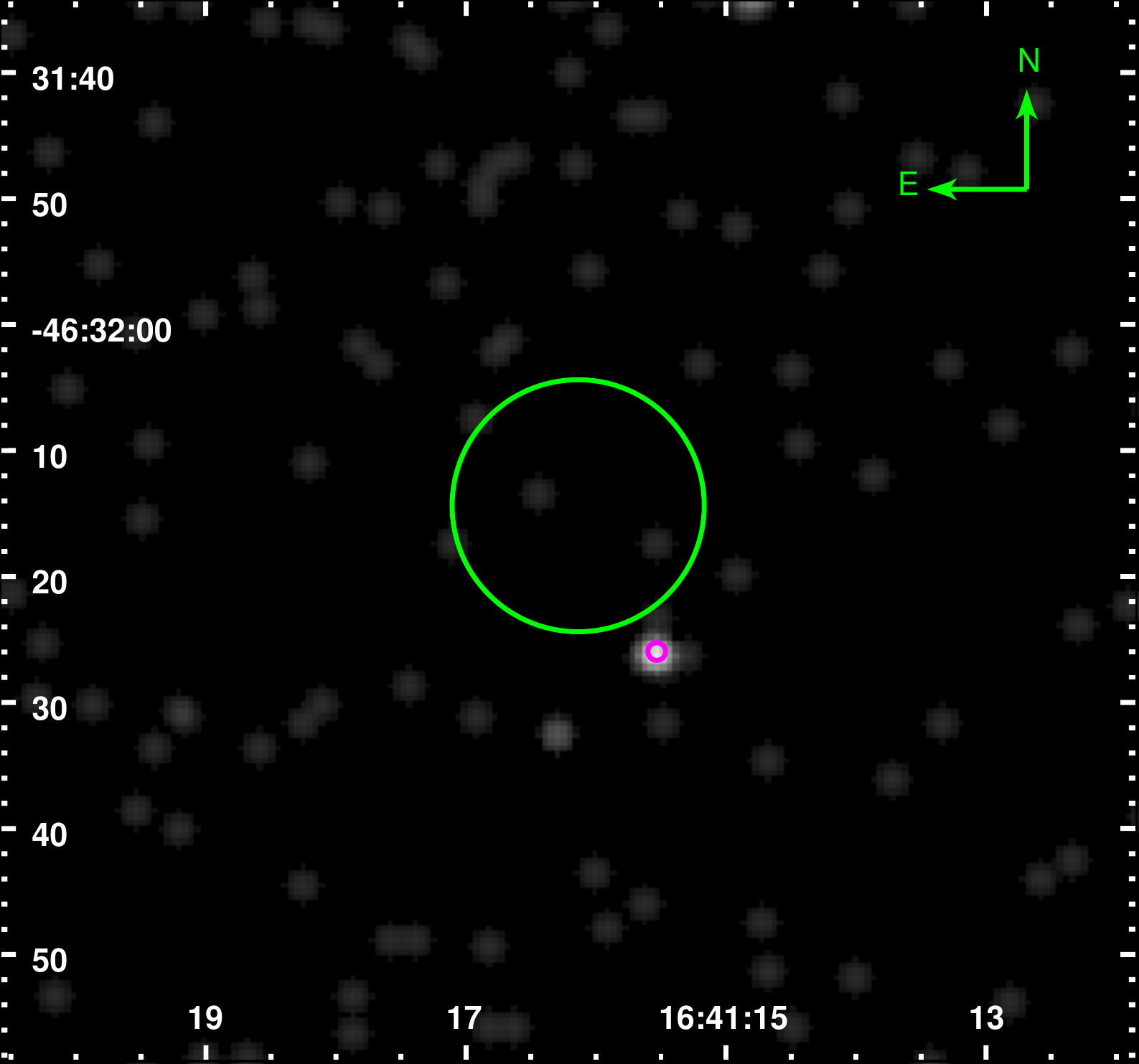}
	\label{fig:pos1}}
	\subfigure[NNR 20, \textit{Chandra} ObsID 16171]{\includegraphics[width=2.3in]{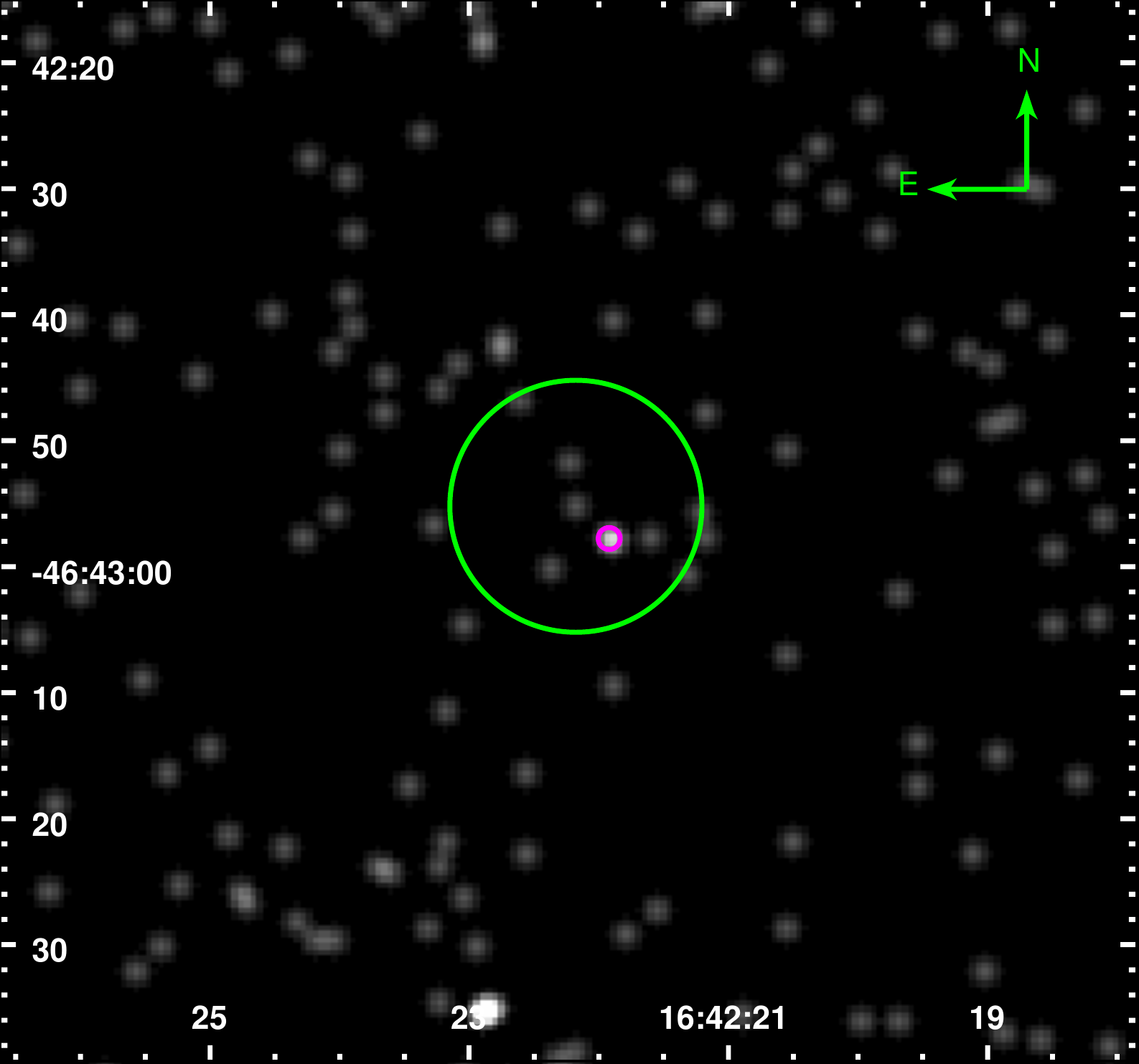}
	\label{fig:pos2}}
	\subfigure[NNR 25, \textit{Chandra} ObsID 17242]{\includegraphics[width=2.3in]{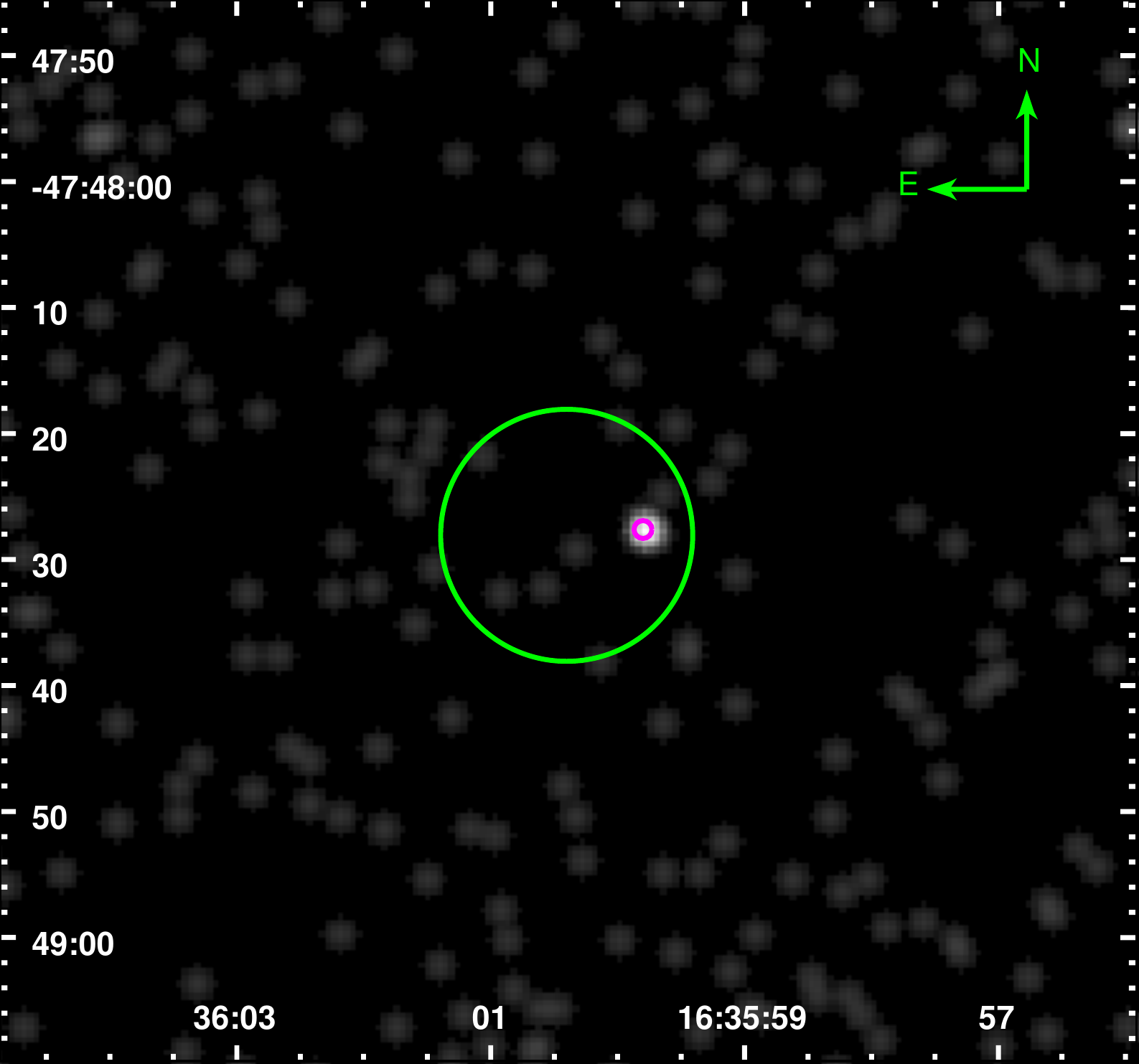}
	\label{fig:pos3}}
	\caption{\textit{Chandra} follow-up observations of \textit{NuSTAR} transients in the 0.5--10 keV band (see Table~\ref{tab:chandra}).  \textit{NuSTAR} source positions are shown with 90\% confidence error circles in green, and the locations of the nearest \textit{Chandra} sources are indicated with 90\% confidence error circles in magenta.  The \textit{NuSTAR} and \textit{Chandra} positional uncertainties are provided in Tables \ref{tab:srclist} and \ref{tab:chandrasrc}, and are approximately 10$^{\prime\prime}$ and 0\farcs7, respectively, for all three sources. }
\label{fig:chandrapos}
\end{figure*}

For NNR~4, 5, 8, and 21, we extracted event lists in the
$3-20$~keV band from $r=40^{\prime\prime}$ radius apertures and
searched for periodic signals between 4~ms and 100~seconds. For each
source, we evaluated the power at each frequency (oversampling by a
factor of two) using the unbinned $Z^2_n$ test statistic \citep{buccheri83} summed over $n =1,2,3,5$ harmonics, to be sensitive to both
broad and narrow pulse profiles.  We repeated our search for an
additional combination of energy ranges $3<E<25$~keV, $3<E<10$~keV,
$10<E<25$~keV, $10<E<40$ keV, and aperture size $r< 20^{\prime\prime}$
and $r< 30^{\prime\prime}$. For all these searches, no significant
signals were detected. For NNR 5 and 8, we can constrain the pulsed
fraction of X-ray emission to be $< 45\%$ and $< 48\%$, respectively,
at the $3\sigma$ confidence.  We also performed periodic searches for
longer periods, with special attention to NNR~4 for which \textit{Chandra}
detected a 7150~second period, but we were unable to pick out any
signals that could clearly be attributed to the \textit{NuSTAR} sources due to
the artifacts introduced by \textit{NuSTAR}'s orbital occultations to the
Fourier power spectrum.

\subsection{\textit{Chandra} follow-up of \textit{NuSTAR} discoveries}
\label{sec:chandrafollowup}

As discussed in \S~\ref{sec:obs}, we triggered \textit{Chandra} follow-up observations for the four sources discovered by \textit{NuSTAR}, NNR~10, 19, 20, and 25.  NNR~10, 19, and 25 were not detected by NARCS despite its much higher sensitivity compared to the \textit{NuSTAR} Norma survey, indicating these are transient sources.  NNR~20 falls outside the area surveyed by \textit{Chandra}, but our follow-up \textit{Chandra} observations show that its flux is also highly variable.  \par
The analysis of the \textit{Chandra} follow-up of NNR~10 is presented in T14, while the analysis of the other three observations, which are listed in Table~\ref{tab:chandra}, is described here.  The archival \textit{Chandra} observation 7591 (see Table~\ref{tab:archivalchandra}, which provides additional coverage of NNR 19) was also subjected to the same analysis.  The \textit{Chandra} observations were processed using CIAO version~4.7 adopting standard procedures.  Then we used \texttt{wavdetect} to determine the positions of \textit{Chandra} sources in the vicinity of the \textit{NuSTAR} sources.  The statistical uncertainties of the \textit{Chandra} positions were calculated using the parametrization in Equation~5 of \citet{hong05}; the 90\% statistical uncertainty was then combined with \textit{Chandra}'s $0\farcs64$ systematic uncertainty\footnotemark\footnotetext{See http:/cxc.harvard.edu/cal/ASPECT/celmon.} in quadrature.  Since NNR 19 was also detected in an archival \textit{Chandra} observation, we averaged the positions determined from ObsIDs 7591 and 16170.  The \textit{Chandra} positions and uncertainties are reported in Table~\ref{tab:chandrasrc}.  The \textit{Chandra} follow-up observations of NNR~9, 20, and 25 are shown in Figure~\ref{fig:chandrapos}, where green circles indicate the \textit{NuSTAR} source positions and magenta circles show the locations of the nearest \textit{Chandra} sources.\par
\par
The closest \textit{Chandra} source to NNR~19 is located at a distance of $13\farcs2$, which is outside of the 90\% confidence \textit{NuSTAR} error circle.  However, as noted in Table~\ref{tab:srclist}, a few of the NARCS counterparts have similarly large offsets, suggesting that in some cases the systematic \textit{NuSTAR} positional uncertainties may be underestimated.  The fact that only three days elapsed between the \textit{NuSTAR} and \textit{Chandra} observations of NNR~19 strengthens the case that these sources are indeed associated.  Furthermore, this \textit{Chandra} source was detected in 2007 in \textit{Chandra} ObsID 7591, but undetected in 2011 in ObsID 12508; the fact that this \textit{Chandra} source is a transient boosts the probability that it is the counterpart of NNR 19.\par
The only \textit{Chandra} source in the vicinity of NNR 20 lies within the \textit{NuSTAR} error circle but is only detected at 2.9$\sigma$ confidence.  NNR 20 was not covered by previous \textit{Chandra} observations, including NARCS, so before our follow-up observation (ObsID 16171), we did not know whether this source was a transient or not; based on its \textit{NuSTAR} 3--10~keV flux, we would have expected to detect at least 10 counts from its \textit{Chandra} counterpart if it was persistent.  Thus, even if it is not definite that the weak \textit{Chandra} detection is truly the counterpart of NNR~20, the lack of any brighter \textit{Chandra} sources proves NNR~20 is a variable source.  \par
Follow-up observations of NNR~25 were performed 34 days after the \textit{NuSTAR} observations, and a \textit{Chandra} source is clearly detected within the \textit{NuSTAR} error circle.  This \textit{Chandra} source was not detected during the 2011 NARCS observations; its transient nature boosts the probability that it is the true counterpart of the transient NNR~25.  As was done by F14 for all the NARCS sources, we searched for infrared counterparts to the \textit{NuSTAR}-discovered sources in the VVV survey.  We did not find any infrared counterparts to NNR~19, 20, or 25 within the 95\% uncertainty of the \textit{Chandra}-derived positions. \par
In order to extract photometric and spectral information for each \textit{Chandra} counterpart, we defined source aperture regions as circles with $2\farcs5$ radii and background regions as annuli with 15$^{\prime\prime}$ inner radii and 44$^{\prime\prime}$ outer radii.  As the counterpart of NNR~19 was at a larger angular offset from the \textit{Chandra} aimpoint in ObsID 7591, and the \textit{Chandra} PSF increases in size with angular offset, the circular source region used for this observation had a 5$^{\prime\prime}$ radius.  For each source in each \textit{Chandra} observation, we calculate the net 0.5--10~keV counts, detection significance, and quantile values (see \S~\ref{sec:quantile}), which are provided in Table~\ref{tab:chandrasrc}. 


\subsection{Hardness ratio and quantile analysis}
\label{sec:quantile}

Since spectral fitting can be unreliable or impractical for faint sources, we use hardness ratios and quantile values \citep{hong04} to probe and compare the spectral properties of \textit{NuSTAR} sources.  In order to reduce the level of background contamination and prevent the hardness ratios and quantile values from being skewed towards the values of the \textit{NuSTAR} background, we opted to use the aperture regions with smaller radii to derive these spectral parameters.  The hardness ratio for each source is calculated as $(H-S)/(H+S)$, where $H$ is the counts in the hard (10--20~keV) band and $S$ is the counts in the soft (3--10~keV) band.  The \textit{NuSTAR} hardness ratios are listed in Table~\ref{tab:phot}.  \par
\begin{figure*}[t]
\makebox[\textwidth]{ %
	\centering		
	\subfigure[\textit{NuSTAR} quantile diagram.]{\includegraphics[width=0.7\textwidth]{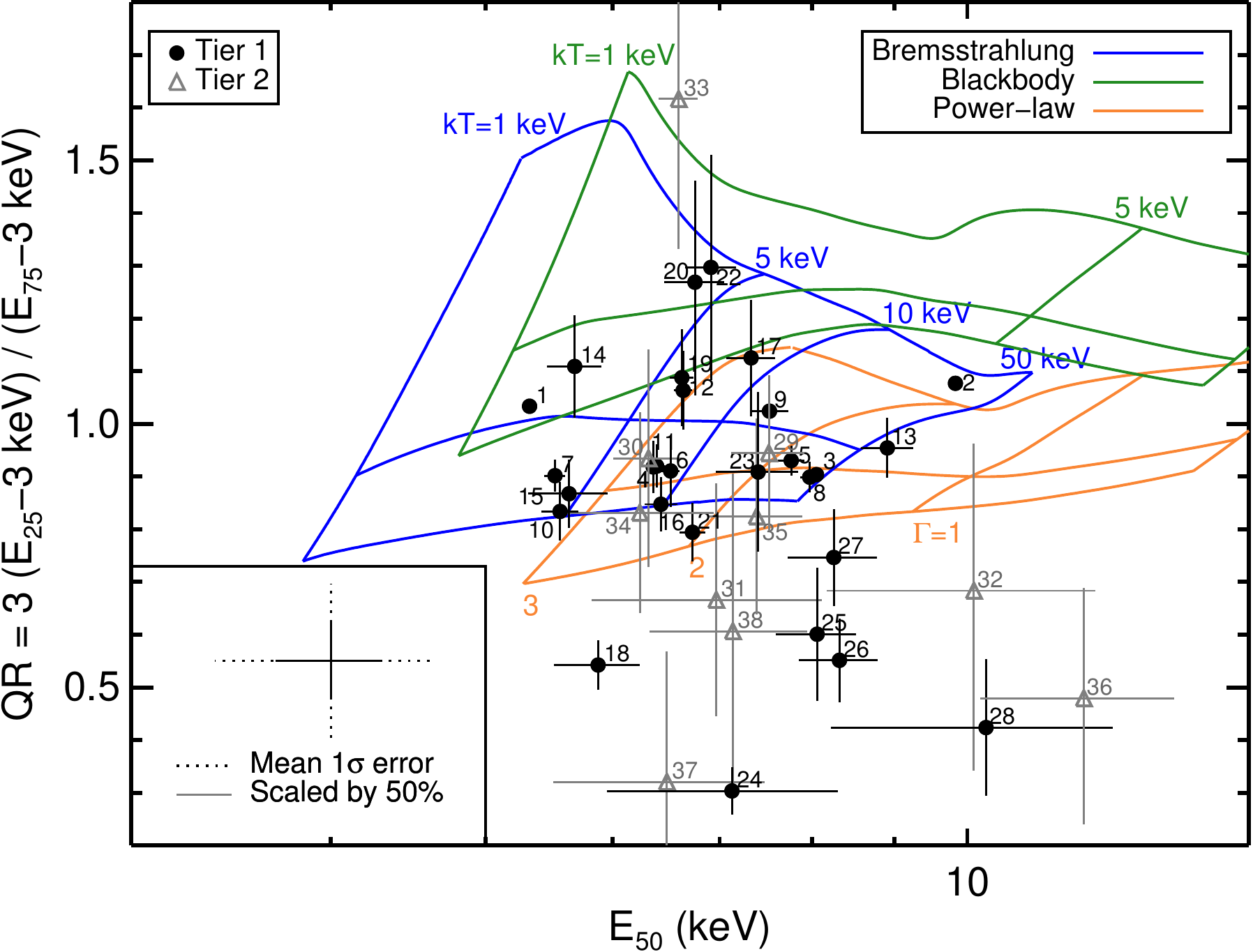}
	\label{fig:quant1}}}
	\centering
	\subfigure[\textit{Chandra} quantile diagram.]{\includegraphics[angle=90,width=0.7\textwidth]{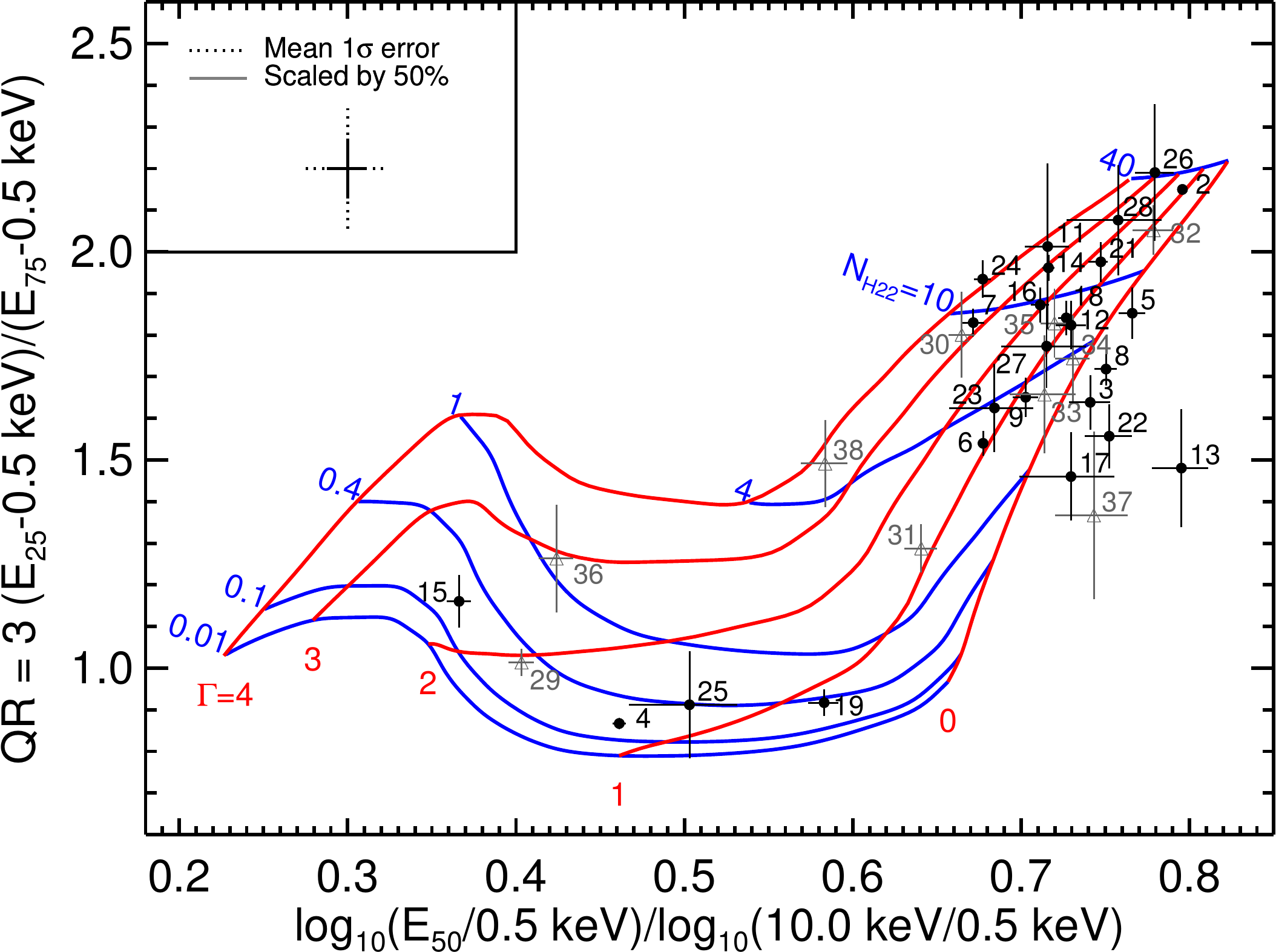}
	\label{fig:quant2}}
	\caption{Quantile diagrams showing the quantile ratio on the y-axis and the median energy on the x-axis (or median energy ``normalized'' by the \textit{Chandra} 0.5--10~keV band for the lower panel).  Quantile values of tier~1 sources are shown with black circles and those of tier~2 sources are shown with gray trinagles.   Comparing the positions of sources in the quantile diagrams to the spectral model gridlines provides a rough measurement of their spectral parameters.  The \textit{Chandra} quantiles are very sensitive to the amount of absorption suffered by a source, while the \textit{NuSTAR} quantiles are more useful for separating sources with different spectral slopes.  To improve the legibility of the plots, 1$\sigma$ error bars have been scaled down by 50\%.  As a visual aide, the corner boxes in each plot show the mean 1$\sigma$ uncertainty for the tier 1 sources and the same mean error scaled by 50\%. \textit{(a)} The \textit{NuSTAR} 3--40~keV background has $E_{50} =$ 10--15~keV and $QR$ = 0.4--0.6, which is why several tier~2 sources, which are most affected by the background, are found near that position in the diagram. Grids representing absorbed bremsstrahlung, blackbody, and power-law models are shown in blue, green, and orange, respectively.  Roughly vertical grid lines represent different values of the temperature ($kT$) or photon index ($\Gamma$).  Primarily horizontal grid lines represent $N_{\mathrm{H}} = 10^{22}, 10^{23}, 5\times10^{23}$~cm$^{-2}$ from bottom to top.  \textit{(b)} A grid of a power-law spectral model attenuated by interstellar absorption is overlaid.  Red (primarily vertical) lines represent values of the photon index $\Gamma = 0, 1, 2, 3,$ and 4 from right to left.  Blue (primarily horizontal) lines represent values of the hydrogen column density $N_{\mathrm{H}} = 0.01$, 0.1, 0.4, 1, 4, 10, and 40 in units of $10^{22}$~cm$^{-2}$ from bottom to top.}
\label{fig:quantile}
\end{figure*}

While hardness ratios are the most widely used proxy for spectral hardness of faint X-ray sources, they are subject to selection effects associated with having to choose two particular energy bands and they do not yield meaningful information for sources which have zero net counts in one of the two energy bands.  Therefore, we also calculated quantile values for each source in the 3--40~keV band; these values are the median energy $E_{50}$, $E_{25}$ and $E_{75}$, the energies below which 25\% and 75\% of the source counts reside, respectively.  The latter energies were combined into a single quantile ratio ($QR$) which is a measure of how broad or peaked the spectrum is and is defined as $QR = 3 (E_{25}-E_{\mathrm{min}})/(E_{75}-E_{\mathrm{min}})$, where $E_{\mathrm{min}}$ is the lower bound of the energy band, 3 keV for \textit{NuSTAR} and 0.5 keV for \textit{Chandra}.  The \textit{NuSTAR} median energy and $QR$ value of each source is provided in Table~\ref{tab:phot} and shown in Figure~\ref{fig:quant1}.  The gridlines in this figure indicate where a source with a particular blackbody, bremsstrahlung, or power-law spectrum would fall in the \textit{NuSTAR} quantile space; gridlines which are roughly vertical represent different temperatures ($kT$) or photon indices ($\Gamma$) while roughly horizontal gridlines represent different values of the absorbing column density along the line-of-sight to the source ($N_{\mathrm{H}}$).  \par
Figure~\ref{fig:quant2} shows the quantile values of the \textit{Chandra} counterparts of the \textit{NuSTAR} sources in the 0.5--10~keV band.  Most of these values are taken from the NARCS catalog (F14).  The quantile values for \textit{Chandra} counterparts of NNR~19 and 25 were derived using the aperture regions described in \S~\ref{sec:chandrafollowup}; the values for NNR~19 derived from ObsIDs 7591 and 16170 were combined in a weighted average.  The \textit{Chandra} counterpart of NNR~20 only has 3 counts, which are too few for quantile analysis; however, all three photons have energies $>4$ keV, indicating that this source is subject to significant absorption since \textit{Chandra}'s effective area peaks below 2~keV.  Finally, we did not adopt the NARCS catalog quantile values for extended sources, because they were derived using aperture regions whose position and extent were determined by eye and which removed embedded point sources not distinguishable with \textit{NuSTAR}.  Therefore, we recalculated the quantile values for extended sources using circular aperture regions with 45$^{\prime\prime}$-radii centered on the \textit{NuSTAR}-determined positions of NNR~8 and 21; these \textit{Chandra} quantiles are weighted averages of values derived from ObsIDs 12528 and 12529\footnotemark\footnotetext{The \textit{Chandra} counterpart of NNR~8 is also observed in ObsID 12525.  However, in this observation, a nearby transient point source which falls within the aperture region is visible.  Comparing the 3--10~keV photon fluxes of NNR~8 in \textit{Chandra} and \textit{NuSTAR}, it does not appear that this nearby transient was present during the \textit{NuSTAR} observation, and therefore we decided not to include ObsID 12525 in our \textit{Chandra} analysis.} for the counterpart of NNR~8 and ObsIDs 12523 and 12526 for the counterpart of NNR~21.  \par
As can be seen in Figure~\ref{fig:quantile}, the \textit{Chandra} quantiles can easily differentiate between foreground sources and those subject to high levels of absorption due to gas along the line-of-sight.  The integrated column density of neutral and molecular hydrogen due to the interstellar medium along the line-of-sight in the Norma region varies from $4-9 \times 10^{22}$~cm$^{-2}$, as derived from the sum of $N_{\mathrm{HI}}$ measured by the Leiden/Argentine/Bonn survey \citep{kalberla05} and $N_{\mathrm{H_2}}$ estimated from the MWA CO survey \citep{bronfman89} using the $N_{\mathrm{H_2}}/I_{\mathrm{CO}}$ factor from \citet{dame01}; since these surveys have 0.5$^{\circ}$ resolution, the interstellar $N_{\mathrm{HI+H_2}}$ values we derive are averages over 0.25~deg$^2$ regions, so it is possible that the interstellar absorption is actually higher or lower along particular lines-of-sight due to the clumpy nature of molecular clouds.  Thus, the sources whose X-ray spectra show column densities in excess of these values may be located behind dense molecular clouds or suffer from additional absorption due to gas or dust local to the X-ray source.  The \textit{NuSTAR} quantiles are not particularly sensitive to $N_{\mathrm{H}}$, but instead are able to separate sources with intrinsically soft and hard spectra, regardless of their level of absorption.  Thus, the combination of quantile values in the \textit{Chandra} and \textit{NuSTAR} bands allows us to learn a fair amount about the spectral properties of sources which are too faint for spectral fitting and provide a check on spectral fitting results which can depend on the choice of binning for low photon statistics.  

\subsection{Spectral Analysis}
\label{sec:spectral}

For all tier 1 sources with $>$100 net counts in the 40$^{\prime\prime}$ radius aperture in the 3--40~keV band, we perform spectral analysis using XSPEC version~12.8.2 \citep{arnaud96}, jointly fitting the \textit{NuSTAR} and \textit{Chandra} data when it is available.  All spectral parameters were tied together for these joint fits, except for a cross-normalization factor between the \textit{Chandra} and \textit{NuSTAR} observations which was left as a free parameter to account for source variability and differences in instrumental calibrations (measured to be consistent to 10\% precision, \citealt{madsen15}).  We also included a cross-normalization constant between \textit{NuSTAR} FPMA and FPMB in our models; for most sources, due to limited photon statistics, the errors on this normalization constant are large and the constant is consistent with 1.0 to better than 90\% confidence.  Thus, for the \textit{NuSTAR} sources detected with lowest significance (i.e., with trial map values $<10^{15}$), we fixed the FPMA/B normalization constant to 1. To maximize the number of counts per spectral bin, we used the larger aperture source regions to extract information for spectral fitting; however, for NNR~22 and 27, which are only separated by 47$^{\prime\prime}$, we extracted spectral information from 30$^{\prime\prime}$ source regions to limit the blending of the two sources.  The spectra of the \textit{Chandra} counterparts were extracted as described in F14 for NARCS sources and \S~\ref{sec:chandrafollowup} for the counterparts of \textit{NuSTAR} discoveries; however, for the extended counterparts of NNR~8 and 21, we defined aperture regions as 60$^{\prime\prime}$-radius circles centered on the \textit{NuSTAR}-derived position in order to match the \textit{NuSTAR} extraction region.  \par
The \textit{Chandra} and \textit{NuSTAR} spectra were grouped into bins of $>2-10\sigma$ confidence, depending on the net counts of each source.  For the three brightest sources which have been carefully analyzed in other papers, we adopt simplified versions of the best-fitting models found in \citet{king14}, B16, and G14, in order to easily measure their observed and unabsorbed fluxes in the 3--10 and 10--20~keV bands which we use to calculate the log$N$-log$S$ distribution of our survey (\S~\ref{sec:lognlogs}).  For other tier 1 sources, we fit absorbed power-law, bremsstrahlung, and collisionally-ionized models; we employed the \texttt{tbabs} absorption model with solar abundances from \citet{wilms00} and photoionization cross-sections from \citet{verner96}. When Fe line emission was clearly visible between 6.4 and 7.1~keV, we also included a Gaussian line in the spectral models.  Due to \textit{NuSTAR}'s 0.4 keV resolution at 6--7~keV enegies, multiple Fe lines would appear blended in our spectra, especially given the low photon statistics.  Thus, measurements of the Fe line parameters should be interpreted as the average energy of the Fe line complex and the combined equivalent width of the Fe lines.  If Fe line emission was not evident, the source spectrum was first fit without a Gaussian component.  Then, having determined which of the three spectral models best fit the spectrum, a Gaussian component was added in order to place constraints on the strength of Fe line emission that may not be visible due to poor photon statistics.  The central energy of this Gaussian component was constrained to be between 6.3 and 7.1~keV and its width was fixed to zero; we tested the effect of fixing the width to values as high as 0.1~keV, but the impact on the results was negligible.  Then the 90\% upper limit on the line normalization was used to calculate the 90\% upper limit on the Fe line equivalent width.  In addition, when significant residuals remained at soft energies, we introduced a partial covering model (\texttt{pcfabs}) to test if it provides a significant improvement of the chi-squared statistic.  Including this component substantially improved $\chi^2$ for NNR~4 and 6, but for NNR~6 the $N_{\mathrm{H}}$ of the partial absorber could not be well constrained and the covering fraction was found to be consistent with 1.0 to 90\% confidence.  Thus, since the spectral quality of NNR~6 was not good enough to constrain the additional \texttt{pcfabs} component, we did not include it in our final model fit for NNR~6.  \par
\begin{figure*}[t]
\makebox[\textwidth]{ %
	\centering
	\subfigure{
		\includegraphics[angle=270,width=0.47\textwidth]{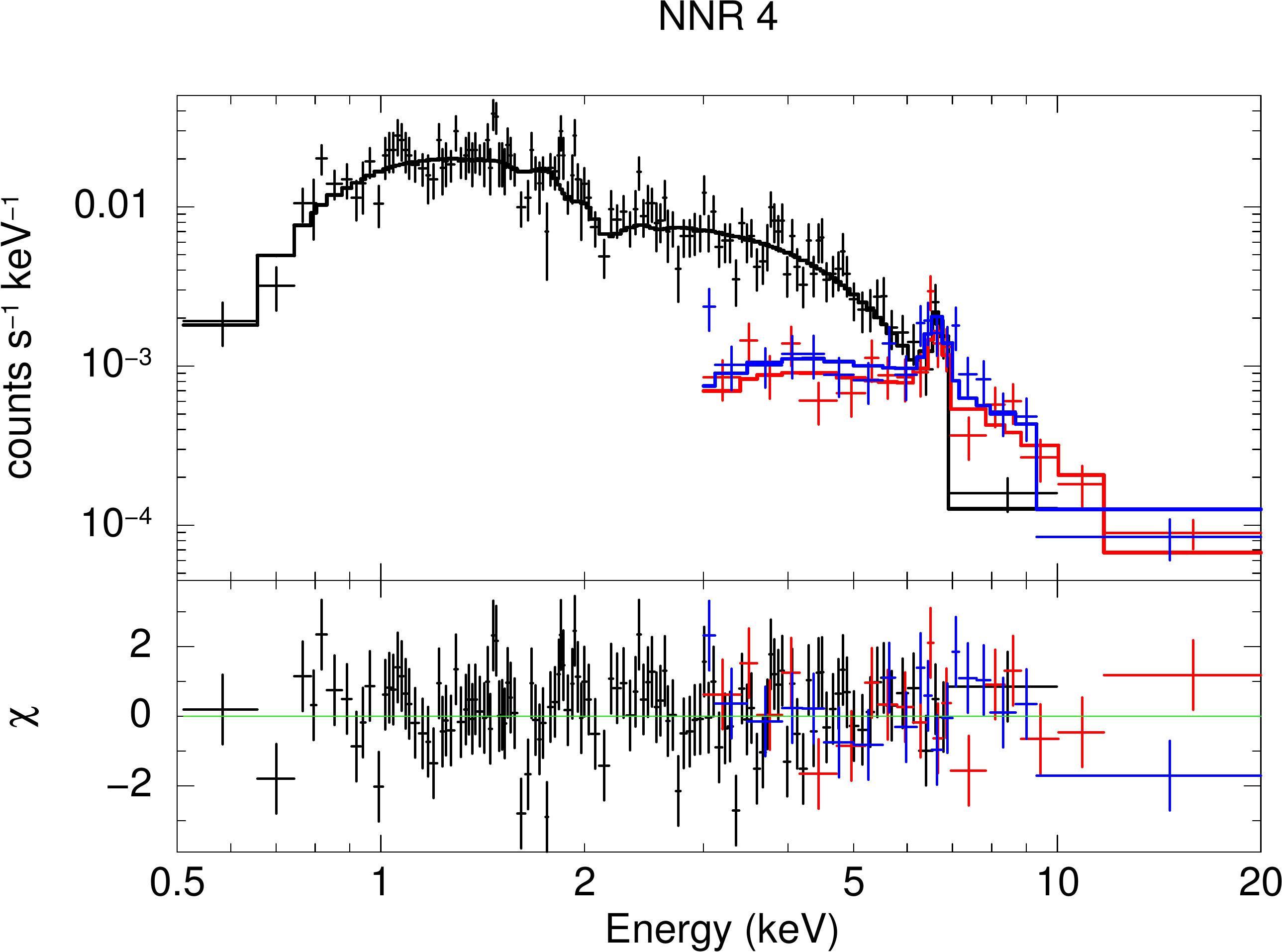}}
	\subfigure{
		\includegraphics[angle=270,width=0.47\textwidth]{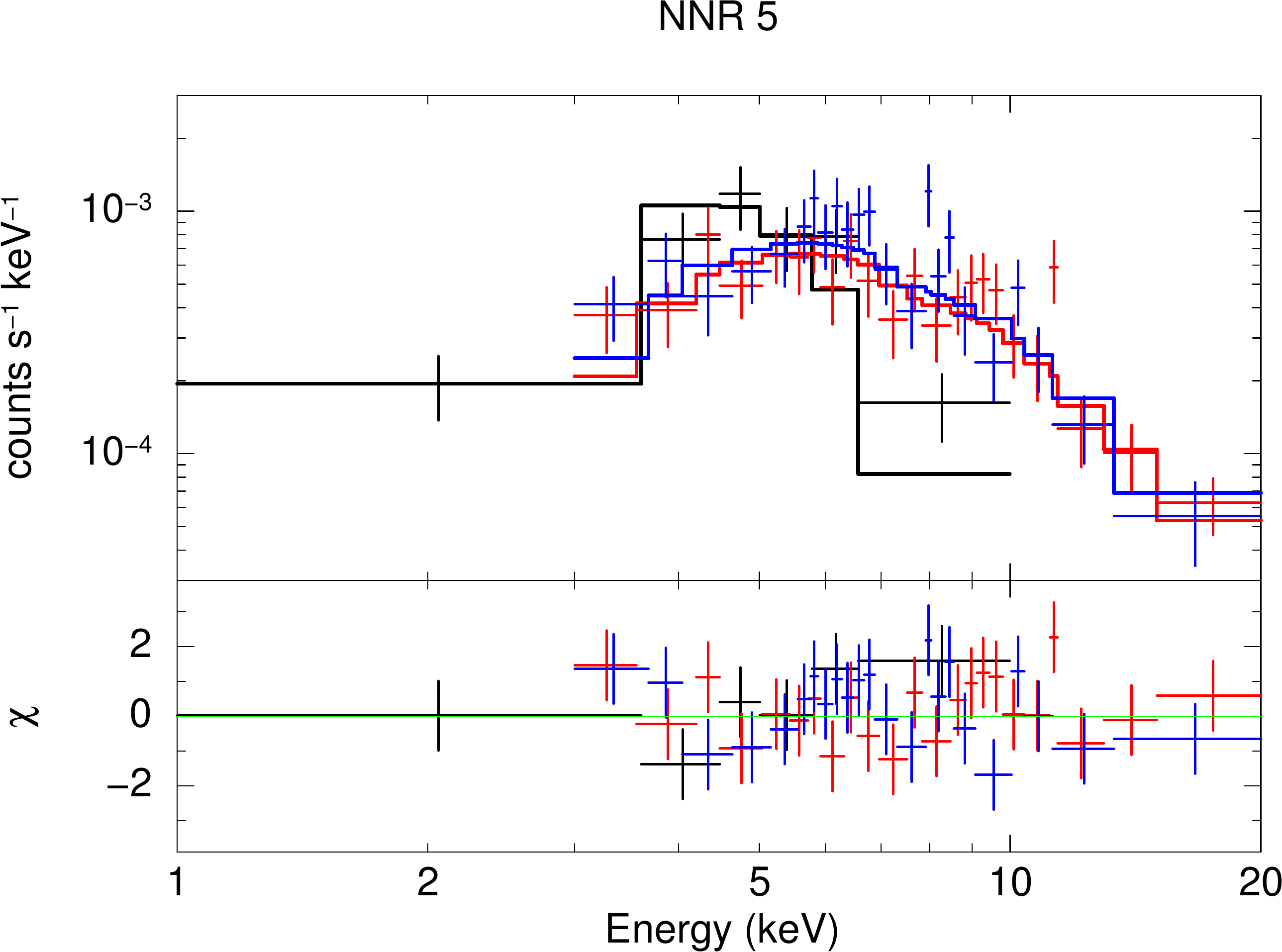}}}

	\centering
	\subfigure{
		\includegraphics[angle=270,width=0.47\textwidth]{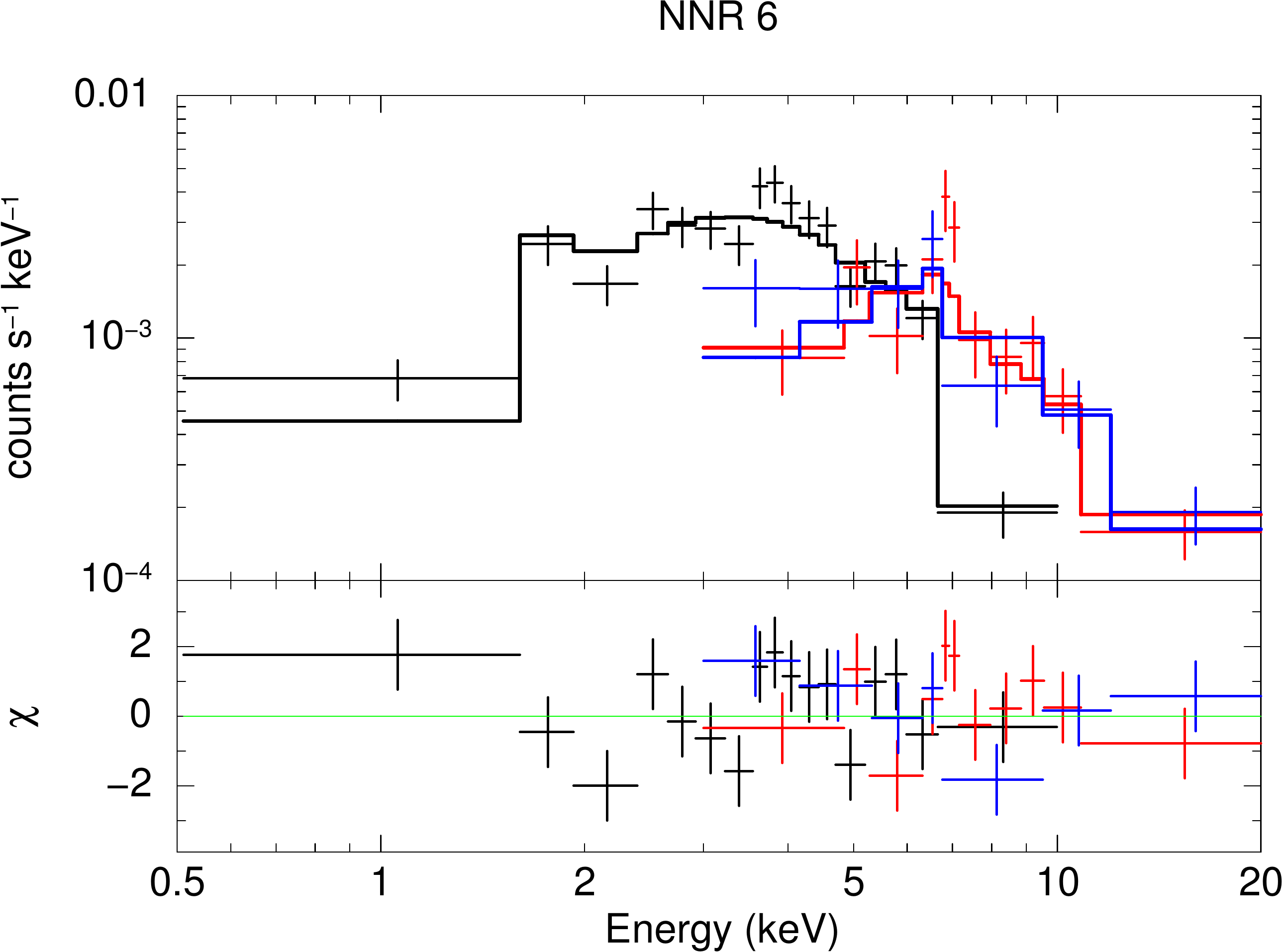}}
	\subfigure{
		\includegraphics[angle=270,width=0.47\textwidth]{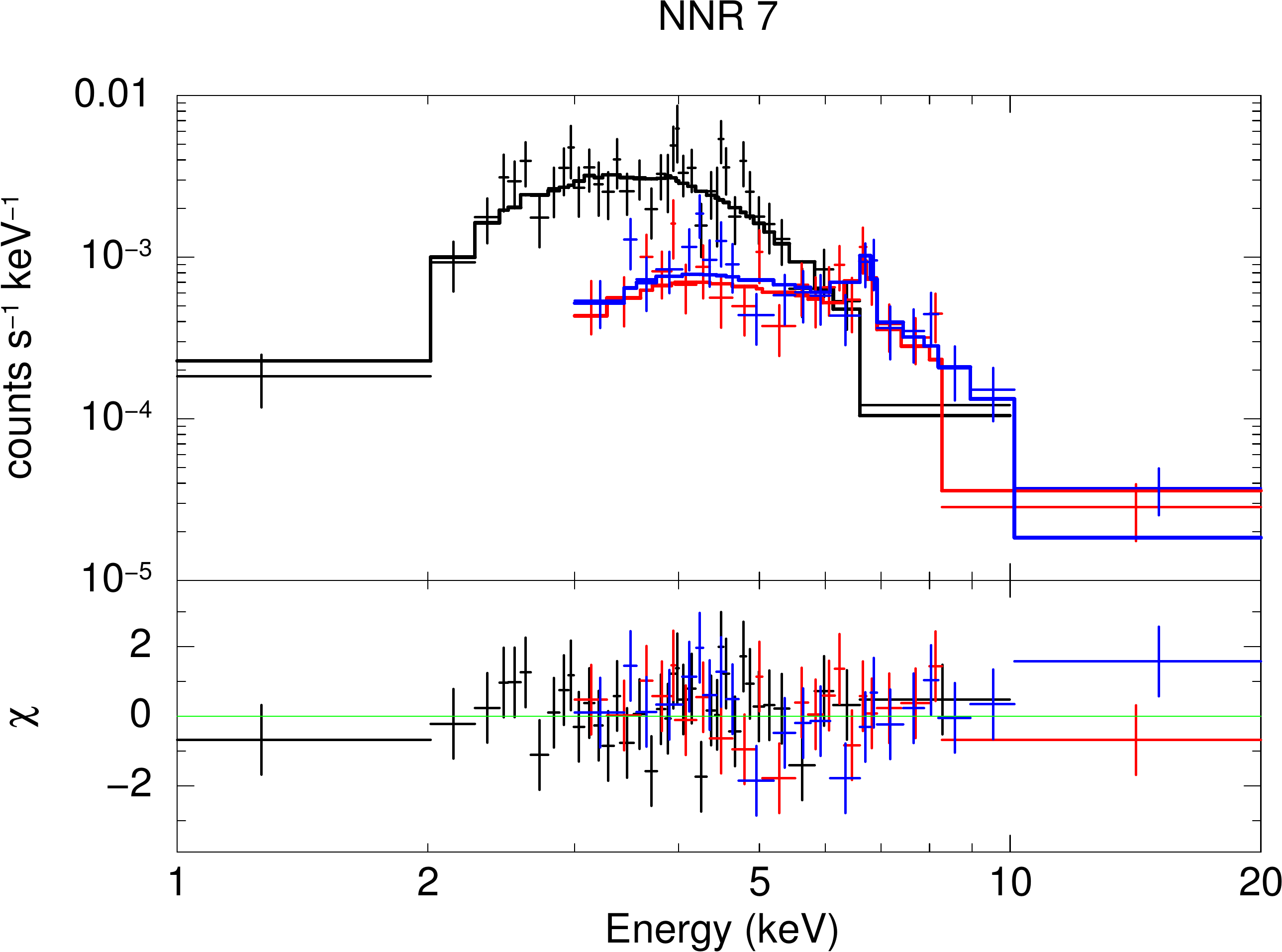}}
		
	\centering
	\subfigure{
		\includegraphics[angle=270,width=0.47\textwidth]{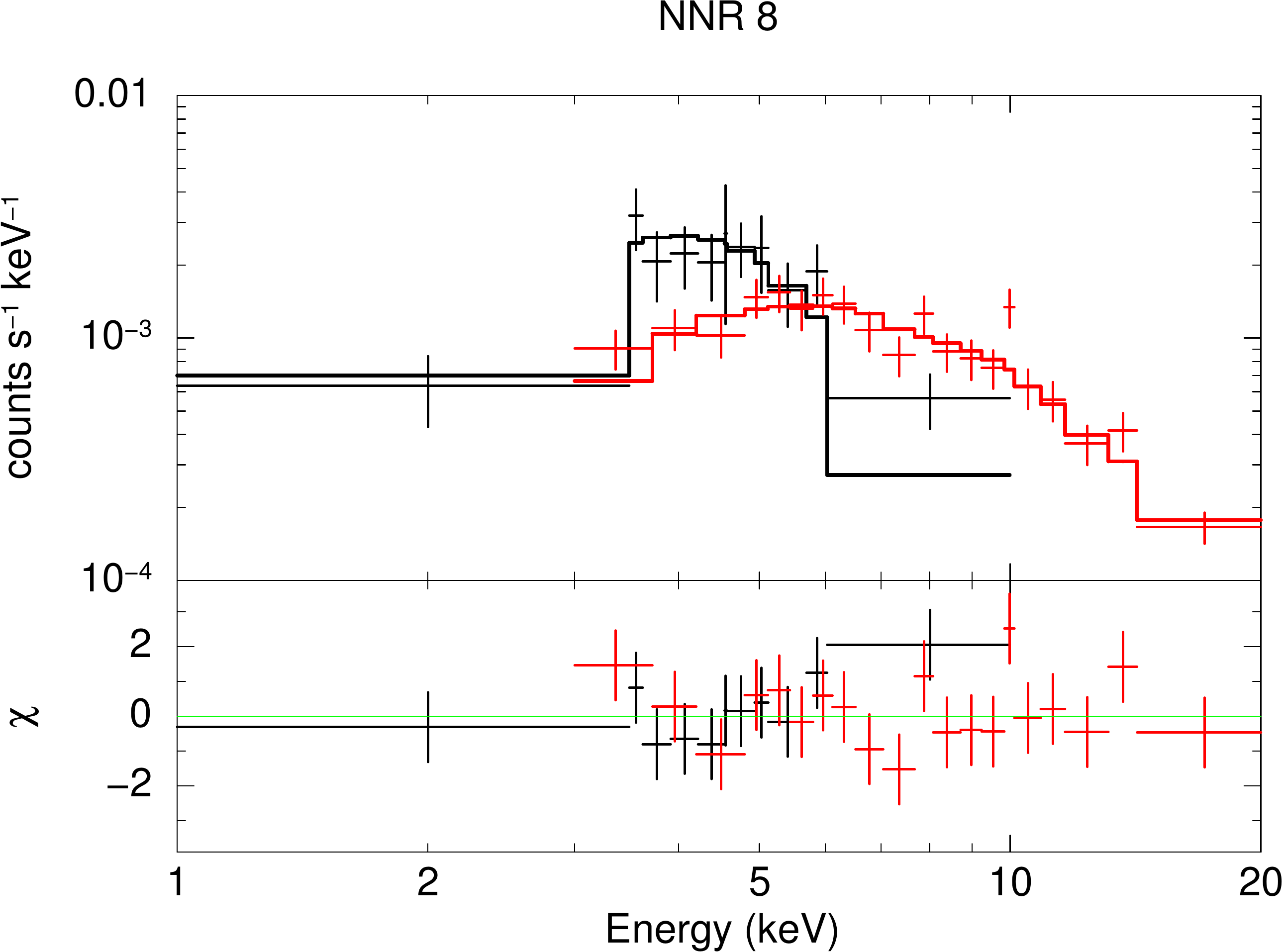}}
	\subfigure{
		\includegraphics[angle=270,width=0.47\textwidth]{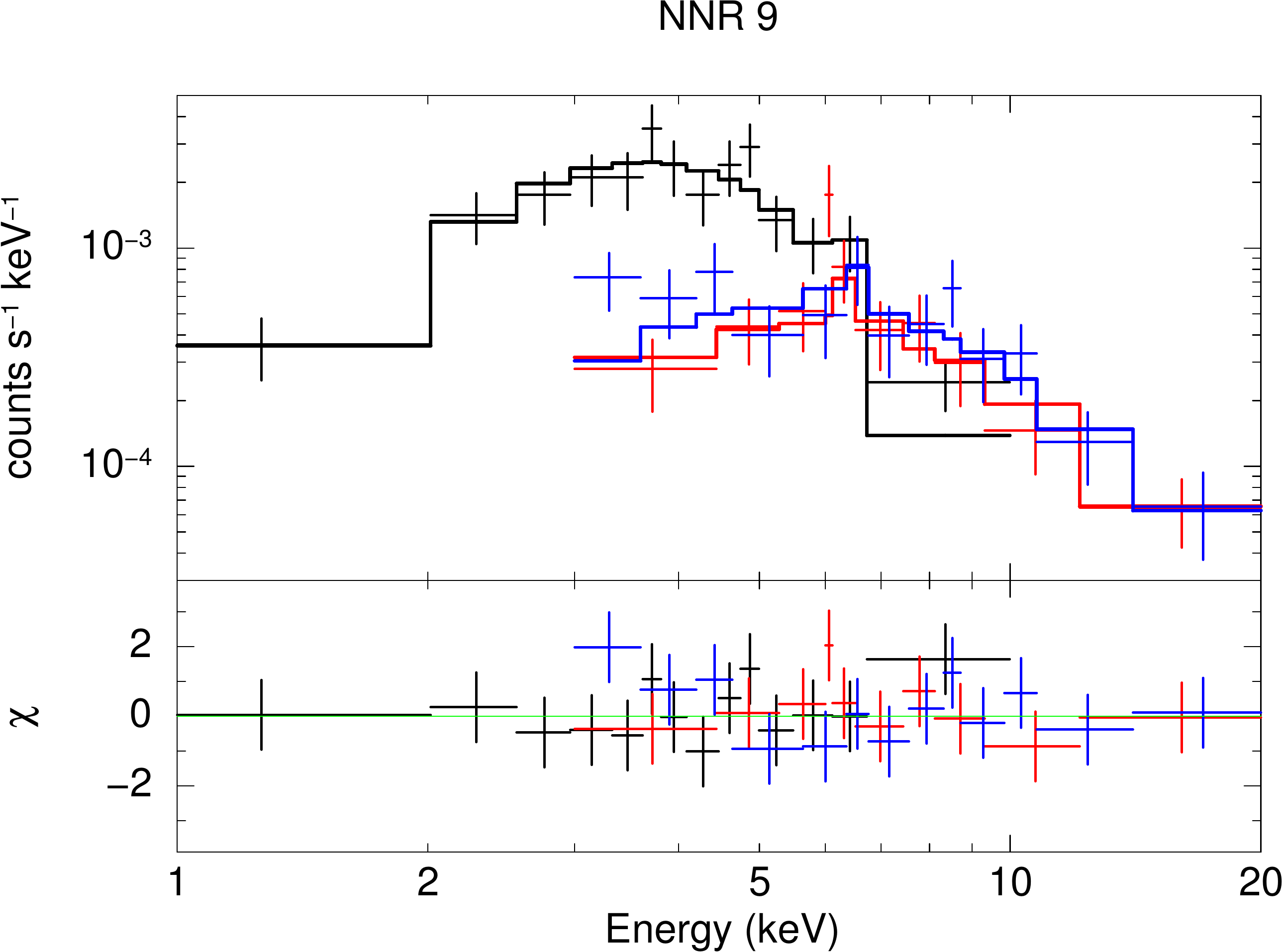}}
\caption{Example \textit{Chandra} and \textit{NuSTAR} spectra with residuals of best-fitting model.  \textit{Chandra} data is shown in black, \textit{NuSTAR} FPMA data is shown in red, and FPMB data is shown in blue.  Additional spectra are shown in the appendix at the end of the paper. Spectral analysis results can be found in Table~\ref{tab:spectra}.}
\label{fig:examplespec}
\end{figure*}

 
The results of our spectral analysis can be found in Table~\ref{tab:spectra}, and the spectra and fit residuals are shown in Figure~\ref{fig:examplespec} and the appendix.  As can be seen, spectra with $<300$ \textit{NuSTAR} counts cannot place strong constraints on the spectral parameters.  However, we nonetheless include these results to be able to compare non-parametric fluxes with spectrally derived fluxes, and as a reference to aid the design of future \textit{NuSTAR} surveys.  \par
We used the model fit with the best reduced chi-square statistic to determine observed energy fluxes for each source in the 2--10, 3--10, and 10--20~keV bands and conversion factors from photon fluxes to unabsorbed energy fluxes, which are listed in Table~\ref{tab:convfactors}.  These conversion factors are used to calculate the log$N$-log$S$ distribution for unabsorbed fluxes (see \S~\ref{sec:lognlogs}).  The faintest tier~1 source, NNR~28, does not have enough counts to permit spectral fitting; based on its quantile values, it has $N_{\mathrm{H}}\approx10^{23}$~cm$^{-2}$ and $\Gamma\approx1.8$.  Fixing the parameters of an absorbed power-law model to these values while allowing the \textit{Chandra} and \textit{NuSTAR} normalizations to vary independently, we fit the unbinned spectra of NNR~28 using the C-statistic \citep{cash79} and find a goodness of fit lower than 28\%.  The observed and unabsorbed fluxes of NNR~28 measured from these fits are included in Table~\ref{tab:convfactors}.  \par

To ensure that these results were not significantly dependent on the binning that was chosen, we compared the best-fitting parameters with those derived by fitting unbinned spectra using the C-statistic and the locations of sources in the quantile diagrams; no significant discrepancies were found except for sources with strong Fe lines, which is to be expected since the quantile grids do not account for the presence of Fe lines.  However, for NNR~17, our analysis yields a harder spectrum than is found by B14.  This source lies in the ghost ray pattern of 4U~1630-472, making background subtraction particularly challenging.  The background region we selected contains higher ghost ray contamination than the background chosen by B14; we consider our selection more appropriate given that this source resides in a region of high ghost ray contamination.  Since the spectrum of 4U~1630-472 is dominated by a blackbody component with $kT\approx1.4$ keV, the fact that B14 measured a softer spectrum for NNR 17 than we do, with $\Gamma=3.7\pm0.5$ rather than 2.0$^{+1.0}_{-0.8}$, suggests that the background contribution from ghost rays may have been underestimated by B14.  The photon index we measure is also more consistent with the hard photon index indicated by the \textit{Chandra} quantiles (see Figure~\ref{fig:quant2}).

\section{Discussion}
\label{sec:discussion}

\subsection{Classification of \textit{NuSTAR} Sources}
\label{sec:classification}
The X-ray spectral and timing properties of the \textit{NuSTAR} sources, as well as information about their optical and infrared counterparts, can help identify their physical nature.  The three brightest sources in the \textit{NuSTAR} Norma survey are well-studied and classified; 4U~1630-472 (NNR~1) is a black hole LMXB (e.g. \citealt{barret96}; \citealt{klein04}), IGR~J16393-4643 (NNR~2) is a neutron star HMXB (\citealt{bodaghee06}; B16), and HESS~J1640-465 (NNR~3) is a pulsar and associated pulsar wind nebula (G14;\citealt{archibald16}).  Here we present the most likely classifications of the fainter \textit{NuSTAR} sources and their hard X-ray properties.

\subsubsection{Colliding wind binaries}
\label{sec:cwb}
Two of the \textit{NuSTAR} sources in the Norma region are likely colliding wind binaries (CWBs), NNR~7 and 14. \par
NNR~7 actually consists of two \textit{Chandra} sources blended together due to \textit{NuSTAR}'s PSF.  In \textit{Chandra} ObsID 11008, where these two sources are resolved, they exhibit very similar spectral properties ($N_{\mathrm{H}}$ and $kT$ values are consistent at <1$\sigma$ level), but the 0.5--10~keV flux of NARCS~1279 is 2 times higher than the flux of NARCS~1278.  These sources are blended in \textit{Chandra} ObsIDs 12508 and 12509 because they are far off-axis, and the combined flux of the two sources is a factor of 3 higher in these later observations.  Spectroscopic follow-up of the near-IR counterparts of both of these \textit{Chandra} sources revealed they are Wolf-Rayet stars of spectral type WN8 \citep{rahoui14}. These stars belong to the young massive cluster Mercer~81 \citep{mercer05} located at a distance of 11$\pm$2~kpc \citep{davies12}.  The \textit{Chandra} spectra of these sources were better fit by thermal plasma models than power-law models, suggesting that these sources were more likely to be CWBs than HMXBs with compact objects accreting from the powerful Wolf-Rayet stellar winds.  \par 
The \textit{NuSTAR} data provides even stronger support for the CWB hypothesis for NNR~7.  Joint fitting of the \textit{Chandra} (from NARCS) and \textit{NuSTAR} spectra of these blended sources reveal that they fall off steeply above 1~ keV and show prominent Fe line emission, primarily due to Fe~XXV based on its 6.76$\pm$0.1~ keV line energy \citep{house69}.  The spectra are best fit by an \texttt{apec} thermal model with $kT=3.2^{+0.8}_{-0.5}$~keV and a metal abundance of 0.5$\pm$0.3 solar, or a steep power-law model with $\Gamma=3.4^{+0.4}_{-0.3}$ and Fe line emission with $650\pm20$~eV equivalent width.  These spectral properties rule out the possibility that NNR~7 could be an accreting HMXB, since accreting HMXBs have harder power-law spectra and Fe~I~K$\alpha$ emission at 6.4~keV, typically with equivalent widths $<100$~eV \citep{torrejon10}.  \citet{elshamouty16} found that, in quiescence, one neutron star HMXB, V0332+53 exhibits a soft spectrum ($\Gamma\approx4$ or $kT_{\mathrm{BB}}\approx0.4$~keV) without prominent Fe lines; if this spectrum is typical of quiescent HMXBs, then we can also rule out the possibility that NNR~7 is a quiescent HMXB given its hard spectrum and prominent Fe emission.  The unabsorbed 0.5--10~keV flux of NNR~7 based on the combined NARCS and \textit{NuSTAR} spectrum\footnotemark\footnotetext{The unabsorbed 0.5--10~keV flux reported here for NARCS~1278 and 1279 combined is higher than that reported in \citet{rahoui14} because we account for the absorption due to the X-ray derived $N_{\mathrm{H}}$ while in \citet{rahoui14} only absorption attributed to the ISM is removed.} is $1.20^{+0.04}_{-0.12}\times10^{-12}$~erg~cm$^{-2}$~s$^{-1}$.  Adopting the 0.5--10~keV flux ratio for NARCS~1278 and 1279 and the bolometric luminosities of their Wolf-Rayet counterparts calculated by \citet{rahoui14}, we find that their respective X-ray luminosities are $5\times10^{33}$~erg~s$^{-1}$ and $1.2\times10^{34}$~erg~s$^{-1}$, and they have $L_X/L_{\mathrm{bol}} = 1.3\times10^{-6}$ and $8\times10^{-7}$, respectively. \par
Isolated high-mass stars are known to be X-ray emitters, but their spectra typically have $kT\sim0.5$~keV and their 0.5--10~keV luminosities follow the scaling relation $L_X/L_{\mathrm{bol}}\approx10^{-7}$ (e.g. \citealt{berghoefer97}; \citealt{sana06}).  The harder X-ray emission and higher $L_X/L_{\mathrm{bol}}$ exhibited by NNR~7 have been observed from the wind-wind shocks in CWBs (\citealt{zhekov00}; \citealt{portegies02}) and the magnetically channeled shocks of high-mass stars with $\sim$kG fields (\citealt{gagne05}; \citealt{petit13}).  For NNR~7, a CWB nature is more likely given the strength of the Fe line at 6.7~keV; magnetic high-mass stars tend to exhibit weak Fe~XXV line emission (\citealt{schulz00}; \citealt{schulz03}), while the Fe XXV lines in CWB spectra can have equivalent widths as large as $\sim1-2$ keV (\citealt{viotti04}; \citealt{mikles06}).  The X-ray spectrum of NNR~7 exhibits substantial absorption corresponding to $N_{\mathrm{H}} = 1.1\pm0.2\times10^{23}$~cm$^{-2}$, which is in excess of the integrated interstellar absorption along the line-of-sight ($N_{\mathrm{HI+H_2}}=7.8\times10^{22}$~cm$^{-2}$).  The excess absorption measured in the X-ray spectrum of NNR~7 could either be due to inhomogeneities in the ISM or local absorption, which is observed in some CWBs, such as $\eta$ Carinae \citep{hamaguchi07}.  Finally, X-ray variability is more common in CWBs than isolated high-mass stars \citep{corcoran96}.  The X-ray flux variations displayed by CWBs are primarily associated with the orbital period of the binary and can be as large as a factor of $\approx20$ (\citealt{pittard98}; \citealt{corcoran05}).  Thus, the X-ray variability exhibited by NNR~7 provides further evidence of its CWB origin.\par
NNR~14 shares many similarities with NNR~7 and is also likely to be a CWB.  The near-IR spectrum of the counterpart of NNR~14 shows emission lines typical of a Wolf-Rayet star of spectral type WN7 in the $K$-band, but the $H$-band spectrum lacks the emission lines expected for this spectral type.  Overall, the near-IR spectrum may be consistent with an O3I star (Corral-Santana et al., in prep).  Its X-ray spectrum is well fit by an \texttt{apec} thermal model with $kT=2.1^{+0.9}_{-0.5}$~keV or a power-law with $\Gamma=4.1^{+1.2}_{-0.9}$ and Fe line emission centered at 6.59$^{+0.08}_{-0.06}$~keV (consistent with Fe~XXV 6.7 ~keV emission) with a very high equivalent width of $1.8\pm0.5$~keV, making it very similar to the CWB candidate CXO~J174536.1-285638 \citep{mikles06}.  Furthermore, NNR~14 exhibits a very high X-ray absorbing column ($N_{\mathrm{H}} = 2.9^{+0.9}_{-0.7}\times10^{23}$~cm$^{-2}$) that is well in excess of the integrated interstellar column density along the line-of-sight ($N_{\mathrm{HI+H_2}}=8\times10^{22}$~cm$^{-2}$); this amount of absorption local to the X-ray source is larger than for NNR 7 but still within the range observed in CWBs \citep{hamaguchi07}.  NNR~14 is coincident with G338.0-0.1, an H{\small II} region most likely located at a distance of 14.1 kpc (\citealt{wilson70}; \citealt{kuchar97}; \citealt{jones12}).  It would not be surprising for NNR 14 and G338.0-0.1 to be physically associated since H{\small II} regions are photoionized by high-mass stars and the extreme $N_{\mathrm{H}}$ along the line-of-sight to NNR~14 indicates it is likely located in the far Norma arm or beyond.  Thus, adopting a distance of 14~kpc for NNR~14, its unabsorbed 3--10~keV luminosity is $10^{34}$~erg~s$^{-1}$, which is within the typical range for CWBs. \par


\subsubsection{Supernova remnants and pulsar wind nebulae}

In addition to HESS~J1640-465, there are three other extended sources in the \textit{NuSTAR} Norma survey, NNR~5, 8, and 21.  \par
\citet{jakobsen14} identified the \textit{Chandra} counterpart of NNR~5 as a pulsar wind nebula (PWN) candidate due to its bow-shock, cometary morphology and hard power-law spectrum.  Although an AGN or LMXB origin cannot be ruled out, these possibilities were disfavored due to the lack of significant X-ray variability, both on short-term timescales during the \textit{NuSTAR} observation and on long-term timescales between the \textit{Chandra} and \textit{NuSTAR} observations, separated by three years.  Our search for pulsations in the \textit{NuSTAR} data did not yield a detection that would have secured a PWN origin, but our search was only sensitive to high pulsed fractions $>45$\%. A joint spectral fit to the \textit{NuSTAR} and \textit{Chandra} data, covering the point source and extended emission in both data sets, yielded a higher $N_{\mathrm{H}}$ value and steeper photon index than measured by \citet{jakobsen14}.  Our best fit photon index of $\Gamma=2.3\pm0.3$ for a power-law model is possible for a pulsar/PWN ($\Gamma\sim1-2$; \citealt{kargaltsev08}), which is consistent with the earlier results, derived using \textit{Chandra} and \textit{XMM-Newton} data.  However, the $N_{\mathrm{H}}$ value we measure (2.7$^{+1.0}_{-0.8}\times10^{23}$~cm$^{-2}$) is higher than the integrated interstellar absorption along the line-of-sight ($N_{\mathrm{H}} = 8\times10^{22}$~cm$^{-2}$), indicating that NNR 5 is likely on the far side of the Galaxy and may be associated with the star-forming complexes located at $\sim10$~kpc; this source may be subject to additional local absorption or lie within or behind molecular clouds. \par
\begin{figure}
\includegraphics[width=0.47\textwidth]{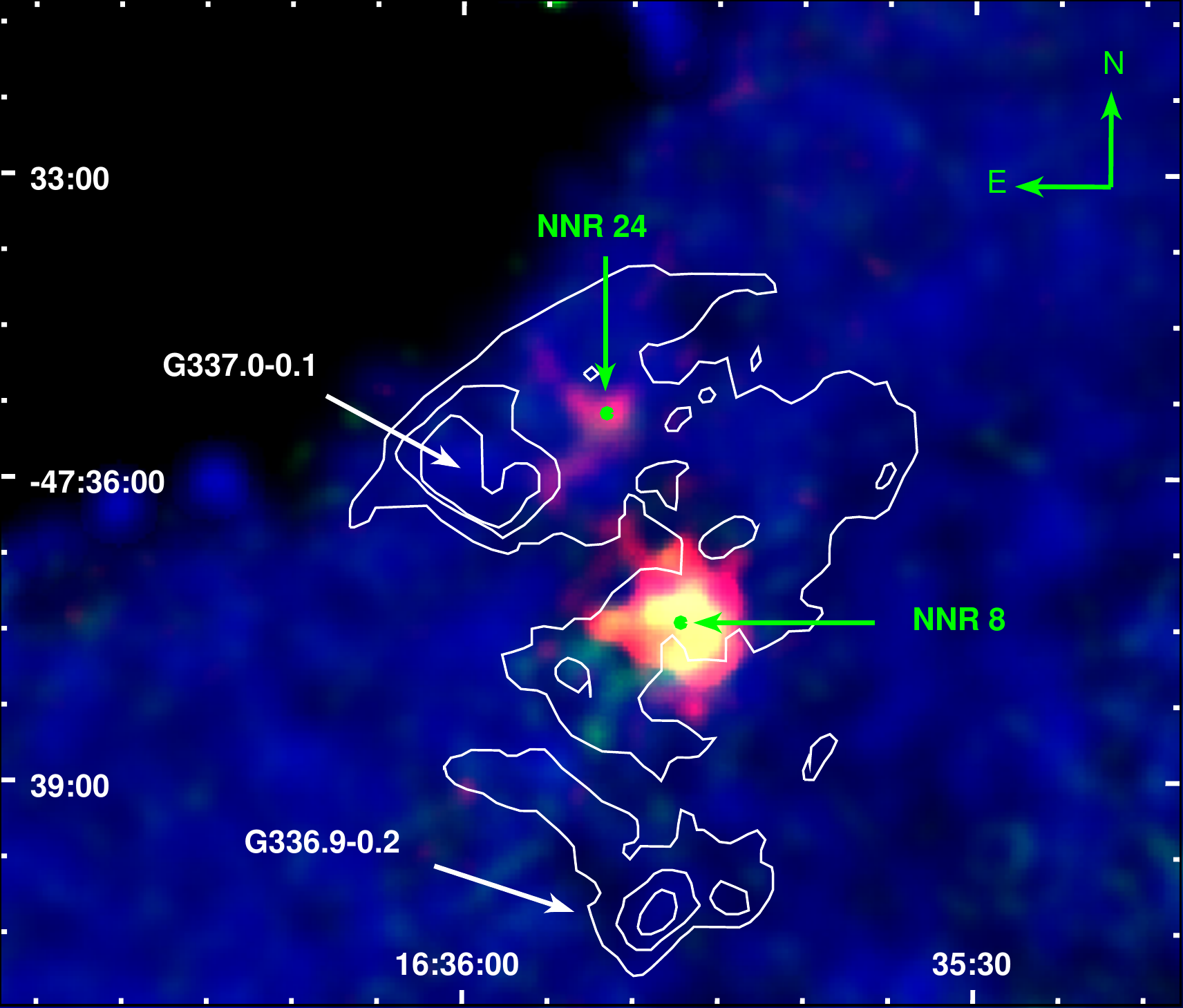}
\caption{\textit{NuSTAR} image of the region around NNR~8.  The 3--10~keV band is shown in red, 10--20~keV in green, and 20--40~keV in blue.  White contours show the radio continuum emission of the CTB~33 complex from \citet{sarma97}.  Green points denote the positions of \textit{NuSTAR} sources. G337.0-0.1 is a confirmed supernova remnant while G336.9-0.2 is an H{\small II} region.  It has been suggested that the magnetar, NNR~24, is associated with this SNR \citep{brogan00}.  However, the extended emission of NNR~8 is clearly not coincident with G337.0-0.1 and its origin may be an unassociated PWN. }
\label{fig:ctb33}
\end{figure}

The source NNR~8 is a region of extended emission with a centrally peaked morphology coincident with the CTB~33 supernova remnant (SNR) and H{\small II} complex located at a distance of $\sim$11~kpc and visible at radio wavelengths \citep{sarma97}.  While NNR~8 may be associated with this complex, it notably does not overlap nearby SNR G337.0-0.1, as shown in Figure~\ref{fig:ctb33}. This hard X-ray diffuse emission was discovered in an \textit{XMM-Newton} field containing the soft gamma-ray repeater (SGR)~1627-41 (here NNR~24) and is attributed by \citet{esposito09} to either a galaxy cluster or a PWN. 
The joint \textit{Chandra} and \textit{NuSTAR} spectrum of NNR~8 is well-fit by an absorbed power-law model with a typical pulsar/PWN index of $\Gamma=1.8\pm0.2$.  In contrast, an absorbed
bremsstrahlung model yields a temperature of $kT = 25^{+22}_{-9}$~keV in the 0.5--20~keV band, which is higher than expected for most galaxy clusters \citep{maughan12}.  No pulsations were detected from NNR~8, but our search was only sensitive to periodic signals with very high pulsed fractions ($>48$\%), leaving open the possibility of a pulsar embedded in diffuse PWN emission.\par
Assuming NNR~8 is a PWN, we can estimate the spin down energy loss of the pulsar from correlations based on the PWN X-ray luminosity and photon index.  Since the high $N_{\mathrm{H}}$ (1.4$^{+0.7}_{-0.5}\times10^{23}$~cm$^{-2}$) measured from the X-ray spectrum of NNR~8 indicates that it lies on the far side of the Galaxy and it is reasonable to expect a PWN to be in the vicinity of star-forming regions, we adopt the 11~kpc distance of the far Norma arm and CTB~33 for NNR~8 and calculate its unabsorbed 2--10~keV luminosity to be $1.0\times10^{34}$~erg~s$^{-1}$.  Using the correlation between 2--10~keV luminosity and spin down energy loss from \citet{possenti02}, we estimate the pulsar $\dot{E}\approx7\times10^{36}$~erg~s$^{-1}$.  The pulsar spin-down luminosity can also be estimated from the PWN photon index using correlations derived by \citet{gotthelf03}; the photon index of NNR~8 yields $\dot{E}\approx1.4\times10^{37}$~erg~s$^{-1}$, which is consistent with the value determined from the correlation of $L_X$ and $\dot{E}$ given the statistical uncertainties of the X-ray luminosity and photon index of NNR~8.  The fact that these estimates of $\dot{E}$ are consistent provides additional support in favor of a PWN origin for this source. \par

The extended emission of NNR~21 is associated with SNR G337.2+0.1, located at a distance of $\sim14$~kpc. Using \textit{Chandra} observations, \citet{jakobsen13} found that the radial profile of the SNR exhibits a central compact source, suggesting a pulsar powering a PWN, as well as excess emission at a radius of $\approx 1\farcm8$, attributable to the SNR shell.  The dearth of \textit{NuSTAR} photons from the central point source does not allow for a significant detection of a pulsar signal, so we cannot confirm the PWN origin of NNR~21.  \textit{XMM-Newton} observations of this SNR revealed that it has a non-thermal spectrum which steepens further from the central core (\citealt{combi06}, hereafter C06), as is seen in many plerionic SNRs (e.g. IC~443, 3C~58, G21.5-0.9; \citealt{bocchino01} and references therein).  Spectral fitting of the \textit{NuSTAR} and \textit{Chandra} data results in a higher column density ($N_{\mathrm{H}} = 2.6^{+0.9}_{-0.7}\times10^{23}$~cm$^{-2}$) and steeper photon index ($\Gamma = 2.6^{+0.5}_{-0.4}$) than measured by C06 for the pulsar/PWN (central source and extended emission combined).  The \textit{Chandra/NuSTAR}-derived photon index, while consistent at the 90\% confidence level with the \textit{XMM} measured value ($\Gamma = 1.82\pm0.45$), is steeper than expected for a pulsar/PWN.  We find that the unabsorbed 2--10 keV luminosity of NNR 21 is $3\times10^{34}$~erg~s$^{-1}$, and thus the $L_X-\dot{E}$ correlation from \citet{possenti02} yields a spin-down luminosity estimate of $\dot{E}\approx1.5\times10^{37}$~erg~s$^{-1}$.  The spin-down luminosity that is estimated using the $\Gamma-\dot{E}$ correlation from \citet{gotthelf03} is in good agreement if it is based on the \textit{XMM}-derived $\Gamma=1.8$ ($\dot{E}\approx1.4\times10^{37}$~erg~s$^{-1}$), but it is at odds if the \textit{Chandra/NuSTAR}-derived $\Gamma$ is adopted ($\dot{E}>1.7\times10^{38}$~erg~s$^{-1}$).\footnotemark\footnotetext{The $\Gamma-\dot{E}$ correlation is only valid for $\Gamma<2.36$, so we can only provide a lower bound on $\dot{E}$ for the \textit{Chandra/NuSTAR}-derived $\Gamma = 2.6^{+0.5}_{-0.4}$.}  \par
Comparing our power-law fits of NNR~21 with the results of C06, the \textit{Chandra/NuSTAR}-derived $N_{\mathrm{H}}$ is statistically higher than the $N_{\mathrm{H}} = 1.15\pm0.27\times10^{23}$~cm$^{-2}$ measured by C06 for the whole PWN, but it is consistent at better than 90\% confidence with the value C06 measure for the outer region of the PWN ($N_{\mathrm{H}} = 1.62\pm0.56\times10^{23}$~cm$^{-2}$), which excludes the central 12$^{\prime\prime}$-radius region; this central region has a much lower column density of $5.9\pm1.5\times10^{22}$~cm$^{-2}$.  Even if we compare the results of our \texttt{apec} model fits with C06, the \textit{Chandra/NuSTAR}-derived $N_{\mathrm{H}}$ is more consistent with the $N_{\mathrm{H}}$ value that C06 measure for the outer region rather than the whole PWN.  One possible explanation for these spatial and temporal $N_{\mathrm{H}}$ variations is that the outer region of the PWN is interacting with a molecular cloud.  This scenario would naturally explain the higher $N_{\mathrm{H}}$ measured in the outer region of the PWN compared to the central region by C06, and the increase in the average $N_{\mathrm{H}}$ measured for the whole PWN between the 2004 \textit{XMM} observation and the 2011 \textit{Chandra} observation could be attributed to a larger fraction of the PWN interacting with the dense interstellar medium as the PWN expands.  Additional X-ray observations to obtain spatially resolved spectroscopy of NNR~21 are required to better understand the origin of the spectral variations exhibited by this SNR.\par 


\subsubsection{Magnetars}
\label{sec:magnetar}
A known magnetar and a magnetar candidate are present in the \textit{NuSTAR} Norma survey.  NNR~24 is a known soft gamma-ray repeater, SGR~1627-41, which was discovered by the Burst and Transient Source Experiment (BATSE) when the source went into outburst in 1998 June \citep{woods99}.  It has been suggested that this SGR is associated with the young SNR G337.0-0.1 in the CTB~33 complex \citep{hurley99}, shown in Figure~\ref{fig:ctb33}.  SGR~1627-41 last went into outburst in 2008 \citep{esposito08} and it was found to have returned to quiescence by 2011 in NARCS observations \citep{an12}.  The cross-normalization constant from fitting the \textit{NuSTAR} and \textit{Chandra} spectra is consistent with 1.0 at 90\% confidence, indicating that the magnetar persists in quiescence and has not significantly decreased in flux since 2011.  We measure a photon index of $5.0^{+2.2}_{-1.4}$ which is steeper but still consistent with that measured by \citet{an12} at 90\% confidence.  Assuming a distance of 11~kpc, based on the association with the CTB~33 complex, we find that NNR~24 has unabsorbed luminosities of $2.3\times10^{33}$~erg~s$^{-1}$ in the 3--10~keV band and $5.2\times10^{31}$~erg~s$^{-1}$ in the 10--20~keV band. \par
NNR~10, a transient source, may also be a magnetar.  The long-term variability and spectral analysis of this source is described in detail in T14, and our spectral analysis yields consistent results.  The flux of NNR~10 varies by more than a factor of 20 over a three-week period, with the peak of activity lasting between 11~hours and 1.5~days and having a soft spectrum with $\Gamma=4.1^{+0.9}_{-0.8}$ or $kT=3^{+2}_{-1}$~keV for a bremsstrahlung model.  The high $N_{\mathrm{H}}$ measured from the X-ray spectrum of NNR~10 suggests that this source is located at $\gtrsim10$~kpc and thus has a peak $L_X\gtrsim10^{34}$~erg~s$^{-1}$ in the 2--10~keV band.  As argued by T14, NNR~10 is most likely either a shorter than average outburst from a magnetar or an unusually bright flare from a chromospherically active binary.  

\subsubsection{Black hole binary candidate}
\label{sec:bhcandidate}
Among the remaining \textit{NuSTAR} Norma sources not discussed in \S~\ref{sec:cwb}-~\ref{sec:magnetar}, NNR~15 stands out as the only source showing clear short-timescale variability in the \textit{NuSTAR} band and also having the lowest median energy.  As can be seen in Figure~\ref{fig:lightcurves}, NNR~15 displays flaring behavior in the 3--20~keV band; during one flare lasting about 15~ks, the source flux increases by a factor of $>$6, and during a smaller flare lasting about 7~ks, the flux increases by a factor of $>$2.  This source also shows variability on year-long timescales since the 3--10~keV flux measured in 2013 \textit{NuSTAR} observations is a factor of 2 higher than the \textit{Chandra} flux measured from 2011 observations.  The \textit{NuSTAR} and \textit{Chandra} spectra are well-fit by an absorbed power-law model with very low $N_{\mathrm{H}}$, indicating that the source must reside within a few kiloparsecs, and $\Gamma=2.6\pm0.4$ (or $kT=2.9^{+1.0}_{-0.7}$~keV for a bremsstrahlung model).  No Fe line is visible in the spectrum, but due to the limited photon statistics, we can only constrain the equivalent width of a potential Fe line feature to be $<1.7$~keV, a loose constraint that does not help to distinguish between different types of X-ray sources.  Assuming a distance of 2~kpc, NNR~15 has an average unabsorbed 3--20~keV luminosity of $1.5\times10^{32}$~erg~s$^{-1}$. Its optical/infrared counterpart has been identified as a mid-GIII star \citep{rahoui14}.  \par
Based on these properties, we identify NNR 15 as a black hole LMXB candidate in quiescence, although an active binary (AB) or CV origin cannot be entirely ruled out.  In quiescence, ABs typically have $L_X=10^{29-31.5}$~erg~s$^{-1}$ and $kT<2$~keV \citep{dempsey93} but they can exhibit flares with peak luminosities of $\sim10^{32}$~erg~s$^{-1}$ and $kT\approx10$~keV \citep{franciosini01}.  However, AB flares tend to have very short rise times and long decay times \citep{pandey12}, whereas the flares seen in NNR~15 appear to have more symmetric profiles.  CVs have $L_X=10^{29-33}$~erg~s$^{-1}$ and $kT=1-25$~keV (e.g. \citealt{eracleous91}; \citealt{muno04}), with magnetic CVs being more luminous and spectrally harder than non-magnetic CVs (\citealt{barlow06}; \citealt{landi09}), so their properties are consistent with NNR~15.  However, the flaring exhibited by NNR~15 is not typically seen in CVs.  Non-magnetic CVs have outbursts that last several days and have recurrence times of weeks to months; intermediate polars (IPs) have outbursts of similar duration but which are very rare (\citealt{hellier97}; \citealt{szkody02}), and polars exhibit flares with $\sim$hour-long durations but they tend to be very soft ($kT<1$~ keV; \citealt{choi99}; \citealt{still01}; \citealt{traulsen10}).  The properties of NNR~15 are reminiscent of the quiescent state of V404~Cyg, a well-known LMXB hosting a black hole (BH; \citealt{makino89}; \citealt{casares92}; \citealt{shahbaz96}).  Recent \textit{NuSTAR} observations of V404~Cyg in quiescence show that, in the 3--25~keV band, its power-law spectrum has $\Gamma=2.35\pm0.2$ and it exhibits flux variations of up to a factor of 10 over periods of a few hours \citep{rana16}.  Given the similarities between the X-ray spectra and light curves of NNR~15 and V404~Cyg, NNR~15 is most likely a BH LMXB, although it may be a CV or an AB.  To order-of-magnitude accuracy, it is estimated that $\sim1000$ quiescent BH LMXBs reside in the Galaxy \citep{tanaka96}; the primary source of uncertainty in this estimate is our limited knowledge of the typical recurrence timescale of BH transients.  Making the simplifying assumption that quiescent BH binaries trace the stellar mass distribution of Galaxy and using the estimate of the stellar mass enclosed in the Norma survey area by F14, we would expect $\sim4$ BH LMXBs to reside in the survey area.  Thus, it is at least plausible that one BH binary would be detected in the \textit{NuSTAR} Norma survey.


\subsubsection{Cataclysmic variables and active galactic nuclei}
\label{sec:cv}
Based on NARCS results, we expect that the majority of \textit{NuSTAR} Norma sources should be a mixture of CVs and AGN.  CVs typically have thermal spectra with $kT\approx1-30$~keV although IPs can display even higher temperatures ($kT\approx30-50$~keV; \citealt{landi09}), while AGN exhibit power-law spectra with $\Gamma\approx1.5-2$ (\citealt{tozzi06}; \citealt{sazonov08}).  The remaining 17 tier~1 sources (NNR~4, 6, 9, 11--13, 16--20, 22, 23, 25--28) have bremsstrahlung temperatures and photon indices consistent with being either CVs or AGN.  
With the \textit{NuSTAR}, \textit{Chandra}, and infrared data available for these sources, there are three primary ways to distinguish CVs and AGN: 
\begin{list}{}{%
\setlength{\topsep}{0pt}%
\setlength{\leftmargin}{0.3in}%
\setlength{\listparindent}{0.0in}%
\setlength{\itemindent}{0.0in}%
\setlength{\parsep}{\parskip}%
\setlength{\itemsep}{4pt}
}%
\item[]
i.  If the absorbing column density inferred from X-ray spectral fitting or \textit{Chandra} quantiles is significantly lower than the integrated interstellar $N_{\mathrm{H}}$ along the line-of-sight to the source, it is a Galactic source.  
\item[]
ii. If the source does not have a point-like infrared counterpart with $>98$\% reliability in the VVV survey, it may be an AGN or a Galactic source with a K or M main-sequence companion, which would fall below the VVV sensitivity limits ($Ks<18$~mag) if located at $\gtrsim2$~kpc.  Since the energy bands used by the \textit{Wide-Field Infrared Survey Explorer} (WISE; \citealt{wright10}) can be more useful than the $J$, $H$, or $K$ bands for identifying AGN (\citealt{stern12}; \citealt{mateos12}), we also searched for counterparts to the \textit{NuSTAR} sources in the AllWISE catalog \citep{cutri13}.  The BH binary candidate NNR~15 and four tier~2 sources (NNR~29, 30, 31, and 38) have \textit{WISE} matches located within the 95\% positional uncertainty of their \textit{Chandra} counterparts.  The counterparts of NNR~15, 29, and 30 have been idenfitied as low-mass stars through spectroscopic follow-up \citep{rahoui14}, and the other \textit{WISE} counterparts have $W1-W2<0.1$, far below the typical value of $W1-W2\geq0.8$ for X-ray luminous AGN \citep{stern12}; furthermore, the near-IR spectra of the counterparts of NNR~31 and 38 indicate they are Galactic sources (Corral-Santana et al., in prep).  Thus, none of the NNR sources with \textit{WISE} counterparts are AGN, but we cannot rule out the possibility that some AGN are undetected by \textit{WISE}.  For instance, in the \textit{NuSTAR} serendipitous survey, which has comparable sensitivity limits to the \textit{NuSTAR} Norma survey, about 25\% of \textit{NuSTAR} sources at Galactic latitudes $|b|>10^{\circ}$, which are likely to be AGN, do not have a WISE counterpart (Lansbury et al., submitted).    
\item[]
iii. If the source exhibits strong unshifted Fe emission, it is more likely to be a CV than an AGN.  Both magnetic and non-magnetic CVs often exhibit Fe emission; in some sources, individual Fe lines at 6.4, 6.7, and 6.97~keV with equivalent widths of 100--200~eV can be seen, while in others, a broad component centered around 6.7~keV with an equivalent width of up to a few keV is seen, likely resulting from the blending of multiple Fe lines due to low energy resolution (e.g. \citealt{mukai93}; \citealt{ezuka99}; \citealt{baskill05}; \citealt{bernardini12}; \citealt{xu16}).  Both type~I and type~II AGN often exhibit red-shifted Fe emission, with the neutral Fe line typically being strongest, except in some highly ionized AGN where the He-like and H-like Fe lines can rival the neutral Fe line in strength; Fe line emission from AGN typically has equivalent widths $<100$~eV, but they can be higher in Compton-thick AGN (\citealt{page04}; \citealt{iwasawa12}; \citealt{ricci14}).  The Fe lines in X-ray binaries also tend to have equivalent widths $\lesssim100$~eV, so the strength of Fe line emission can also help discriminate between CVs and LMXBs (\citealt{hirano87}; \citealt{nagase89}).  
\end{list}
\par
Seven of the tier~1 sources (NNR~4, 6, 9, 12, 18, 19, 25) fulfill at least one of the three criteria listed above and are most likely CVs.  NNR~4 meets all three criteria and there is strong evidence that it is an IP, a CV in which the white dwarf (WD) magnetic field is strong enough ($B\approx10^{6-7}$~G) to truncate the accretion disk and channel the accreting material onto the magnetic poles.  The X-ray spectrum of NNR 4 shows low absorption ($N_{\mathrm{H}}<4\times10^{21}$~cm$^{-2}$), indicating it is a Galactic source residing at a distance of $\lesssim2$ kpc.  The joint fitting of the \textit{Chandra} and \textit{NuSTAR} spectra provides evidence for partial-covering absorption, which is frequently observed in IPs as some of the X-rays produced in the accretion column pass through the accretion curtain on their way to the observer (\citealt{demartino04}; \citealt{bernardini12}).  The near-IR counterpart of NNR~4 is variable and displays emission lines often produced in the accretion streams of IPs \citep{rahoui14}.  Furthermore, this source also exhibits Fe line emission centered at 6.65$^{+0.10}_{-0.06}$~keV line with high equivalent width ($0.9^{+0.2}_{-0.1}$~keV), and a 7150~second period detected by \textit{Chandra}, both of which are typical for IPs (\citealt{scaringi10}).  NNR 4 exhibits flux variations on month-year timescales, which is more typical for non-magnetic CVs and polars than IPs \citep{ramsay04}, but the flux only varies by a factor $<2$, so the case for this source being an IP remains strong.  Assuming a distance of 2~kpc, the unabsorbed 3--10~keV luminosity of NNR 4 is $2-4\times10^{32}$~erg~s$^{-1}$, which is within the luminosity range of IPs (\citealt{muno04} and references therein).  \par
Sources NNR~6, 9, and 12 all have strong Fe emission centered between 6.4 and 6.8~keV and equivalent widths of $1.3\pm0.4$, $0.4\pm0.2$, and $1.2\pm0.4$~keV, respectively, strongly indicating that these sources are CVs since both AGN and X-ray binaries tend to have much weaker Fe emission and the Fe emission from AGN is likely to be redshifted.  These large equivalent widths are likely due to multiple Fe lines being blended due \textit{NuSTAR}'s low energy resolution.  Both NNR~6 and 9 are best-fit by thermal models with high plasma temperatures ($kT>15$~keV), which are more typical of magnetic rather than non-magnetic CVs (\citealt{landi09}; \citealt{xu16}).  The lack of flux variations for NNR~6 and 9 suggest they are most likely IPs.  In addition, NNR~6 has a low-mass (late GIII) stellar counterpart \citep{rahoui14}, lending further support to a CV origin for this source.  The nature of NNR~12 is less certain, because its softer spectrum ($kT = 6^{+3}_{-1}$~keV for an \texttt{apec} model) is typical for both non-magnetic and magnetic CVs.  NNR~21 is likely located at a distance $>10$~kpc given its high $N_{\mathrm{H}}$, so its 3--10~keV luminosity is likely $\gtrsim2\times10^{33}$~erg~s$^{-1}$; this high luminosity coupled with the lack of flux variability suggests this source is also probably an IP. \par   
Another likely IP candidate is NNR~13.  This source displays one of the hardest spectra of all the \textit{NuSTAR} Norma sources, having $kT>21$~keV or $\Gamma=1.0\pm0.5$.  Its very hard spectrum and constant flux over long timescales is typical of IPs. \par  
The nature of NNR~18 is discussed in B14; our spectral analysis yields consistent results, finding a high $N_{\mathrm{H}}$ of $1.9^{+0.9}_{-0.6}\times10^{23}$~cm$^{-2}$ and $\Gamma=2.6^{+1.0}_{-0.8}$.  Assuming a distance of $>10$~kpc based on the high $N_{\mathrm{H}}$ value, NNR~18 has an unabsorbed 3--10~keV luminosity $\gtrsim5\times10^{33}$~erg~s$^{-1}$.  NNR~18 has an early MIII counterpart and exhibited mild X-ray variability on short timescales in \textit{Chandra} observations.  As discussed by B14, these properties are consistent with an IP or an LMXB.  Another possibility is that this source is a hard-spectrum symbiotic binary (SB) hosting a WD or a symbiotic X-ray binary (SyXB) hosting a NS \citep{luna13}; the compact objects in SBs and SyXBs accrete material from the wind of a red giant companion, which is typically of spectral type M or K \citep{morihana16}.  Hard-spectrum SBs and SyXBs display X-ray luminosities between $10^{32}$ and $10^{34}$~erg~s$^{-1}$ (\citealt{masetti02}; \citealt{smith08}; \citealt{nespoli10}), and variability on short and long timescales (\citealt{luna07}; \citealt{corbet08}).  An IP origin is favored for NNR~18 based on its low levels of variability, while its estimated luminosity and the M giant spectral type of its counterpart favors an SB or SyXB origin.   \par  
NNR~19 and 25 show low absorption in their X-ray spectra, indicating they are Galactic sources and probably located at a distance of a few kpc.  Both sources are transients which were not detected in NARCS, but they are detected in follow-up \textit{Chandra} observations taken 3 and 34 days after the \textit{NuSTAR} observations, respectively.  The flux of NNR~25 increased by a factor of $\geq4$ in the couple of years between the NARCS and \textit{NuSTAR} observations and remained high for at least 34 days.  NNR~19 was detected at a consistent flux level in multiple \textit{NuSTAR} observations that span $\approx$100 days.  About 250 days before it is first detected by \textit{NuSTAR}, the 90\% confidence upper limit for its 3--10~keV photon flux is $2\times10^{-6}$~cm$^{-2}$ s$^{-1}$ (a factor 4 below its peak flux), and about 40 days after it is detected by \textit{NuSTAR}, its flux falls below $4\times10^{-6}$~cm$^{-2}$~s$^{-1}$.  Thus, we find that the flux of NNR~19 increased by a factor of $\geq4$ and remained high for a period between 100 and 400 days.  In addition, NNR~19 was detected in the archival \textit{Chandra} ObsID 7591, demonstrating that this transient experienced an outburst in 2007, during which its flux was a factor of $\geq7$ higher than the upper limit measured by NARCS in 2011.  The spectra of NNR~19 and 25 have $kT=11^{+18}_{-4}$~keV ($\Gamma=1.7^{+0.3}_{-0.4}$) and $kT>6$~keV ($\Gamma=1.8\pm0.7$), respectively.  The temporal and spectral properties of NNR~19 and 25 most closely resemble those of polars, CVs with magnetic fields so strong ($B>10^{7}$~G) that the white dwarf magnetosphere inhibits the formation of an accretion disk.  Thus, compared to other CVs, polar X-ray emission is very sensitive to changes in the mass transfer rate, and they exhibit flux variations of factors $\geq4$  as they transition between low and high accretion states on $\sim$month-year timescales (\citealt{ramsay04}; \citealt{worpel16}), very similar to the behavior of NNR~19 and 25.  No IR counterparts in the VVV survey are found for NNR~19 or 25 within the 90\% positional uncertainty determined from \textit{Chandra}.  While the variability and spectra of NNR~19 and 25 would also be consistent with hard-spectrum SBs or SyXBs, the lack of a counterpart with $Ks<18$~mag rules out the possibility that these sources have red giant companions, which should be visible out to $\gtrsim10$~kpc.  In contrast, it is possible for main-sequence K or M-type stars located at distances of a few kiloparsecs to fall below the VVV survey sensitivity. \par


The nine remaining tier 1 sources (NNR 11, 16, 17, 20, 22, 23, 26, 27, 28) are well-fit either by thermal models with $kT=4-30$ keV or power-law models with $\Gamma\approx2$, consistent with the spectra of CVs, SBs, SyXBs, LMXBs, or AGN; the uncertainties in the spectral parameters for many of these sources are quite large since they are among the faintest in our survey.  All of these sources have high absorption, that is equal to or in excess of the ISM column density through the Galaxy, and they lack IR counterparts, so it is difficult to determine whether they are Galactic or extragalactic.  The lack of counterparts does rule out the possibility that these sources are SBs or SyXBs since their red giant companions should be visible through most of the galaxy given the sensitivity of the VVV survey ($Ks<18$ mag).  Based on the log$N$-log$S$ distribution of AGN measured in the COSMOS survey \citep{cappelluti09} and accounting for Galactic absorption, conversion from the 2--10 keV to the 3--10 keV band, and the sensitivity curve of the \textit{NuSTAR} Norma survey (see \S~\ref{sec:sensitivity}), we estimate that about five AGN are present in this survey.  Therefore, roughly half of the remaining tier 1 sources may be AGN.  The other half are probably CVs since quiescent LMXBs are expected to be relatively rare \citep{tanaka96}.  Additional \textit{NuSTAR} or \textit{XMM} observations are required to distinguish between the possible CV or AGN origin of these nine sources by measuring the strength of Fe line emission and better constraining their spectral hardness.  The 3--10 keV fluxes of NNR 11, 20, and 28 vary by factors of $>5$ between the NARCS and \textit{NuSTAR} observations.  Such long-term variability is common for AGN, polars, and non-magnetic CVs (\citealt{orio01}; \citealt{markowitz04}; \citealt{ramsay04}; \citealt{baskill05}), so it does not help us discriminate betwen Galactic and extragalactic sources but it at least excludes an IP origin for these three sources.\par
The ten tier~2 sources included in our catalog do not have enough \textit{NuSTAR} counts to meaningfully constrain their spectral properties, but their distribution in the \textit{Chandra} quantile diagram is very similar to the distribution of the 17 tier~1 sources described in this subsection; two are foreground sources while the rest are heavily absorbed and have $\Gamma<2$.  Seven of these tier~2 sources (NNR~29, 30, 31, 34, 35, 36, and 38) have reliable IR counterparts, three of which (NNR~29, 30, and 36) have been spectrally identified as low-mass stars \citep{rahoui14}.  These seven sources are likely to be a mixture of CVs, SBs, and SyXBs like the majority of identified tier~1 sources.  Sources NNR~29 and 36 display such low absorption that they are likely located within a few kpc and thus have 3--10~keV luminosities $\lesssim3\times10^{31}$~erg~s$^{-1}$, so they could also be active binaries given their low luminosity \citep{strassmeier93}.  An AGN origin cannot be ruled out for NNR~34 and 35, which have VVV counterparts but are not detected by \textit{WISE}.  NNR~32, 33, and 37 lack IR counterparts and are heavily absorbed, and could be AGN or Galactic sources.  Based on the log$N$-log$S$ derived for AGN and CVs in the Norma region by F14, a 1:2 ratio of AGN to CVs/ABs is expected in the 2--10~keV flux range of these tier~2 sources (4$\times10^{-14} < f_X <1\times10^{-13}$~erg~cm$^{-2}$~s$^{-1}$).  Such a ratio is plausible among tier 2 sources given the current constraints we can place on their physical nature.  However, it is odd that none of the sources which may be AGN are detected by \textit{WISE} with $W1-W2\geq0.8$, since the majority of AGN discovered in the \textit{NuSTAR} serendipitous survey have these properties \citep{lansbury16}.  The fact that our AGN candidates either lack IR counterparts or have only VVV, but not \textit{WISE}, counterparts indicates that, if they truly are AGN, they are likely to have low-luminosities ($L_X\lesssim10^{43}$~erg~s$^{-1}$).  


\subsubsection{On the search for low-luminosity high-mass X-ray binaries}

As discussed in \S\ref{sec:intro}, one of reasons the Norma arm was targeted by \textit{Chandra} and \textit{NuSTAR} was to search for low-luminosity HMXBs with fluxes below the sensitivity limits of previous surveys.  A key criterion for identifying an HMXB candidate is to find a source with a high-mass stellar counterpart, which should be visible in the infrared through the whole Galaxy in the VVV survey.  Of the Norma sources detected by \textit{NuSTAR}, only two, NNR~7 and 14, have high-mass stellar counterparts and their broadband X-ray spectra indicate they are CWBs (see \S\ref{sec:cwb}).  However, there are three \textit{Chandra} sources, NARCS~239, 1168, and 1326, with high-mass stellar counterparts \citep{rahoui14} which were not detected by the \textit{NuSTAR} Norma survey.  The $2-10$~keV \textit{Chandra} fluxes of these sources ($f_X\approx7-8\times10^{-14}$~erg~cm$^{-2}$~s$^{-1}$; F14) are comparable to the sensitivity limit of the \textit{NuSTAR} survey (see \S\ref{sec:sensitivity}), and thus it is not surprising that they are not detected.  In the \textit{Chandra} band, these three sources have harder spectra than NNR~7 and 14, residing in a region of quantile space consistent with $\Gamma\approx2$ power-law spectra (F14), which are typical of accreting HMXBs.  \par
Future spectroscopic observations of the infrared counterparts of these \textit{Chandra} sources will confirm whether these systems are HMXBs and help to estimate better distances to these sources.  Once we determine how many of these HMXB candidates, if any, are truly HMXBs, the constraints provided by the Norma surveys on the faint end of the HMXB luminosity function will be presented in a future paper.  By extrapolating the measured slope of the HMXB luminosity function above $10^{34}$~erg~s$^{-1}$ to lower luminosities, F14 predicted that at least a few HMXBs should be detected in the Norma region with $f_X>7\times10^{-14}$~erg~cm$^{-2}$~s$^{-1}$.  Thus, if the three HMXB candidates are confirmed, the number of HMXBs in the Norma region would be consistent with a continuation of the HMXB luminosity function slope to lower luminosities, but if none of these sources prove to be HMXBs, it would imply that the HMXB luminosity function flattens substantially at $L_X<10^{34}$~erg~s$^{-1}$.  

\subsection{Survey sensitivity}
\label{sec:sensitivity}
To compute the sky coverage for the \textit{NuSTAR} Norma survey, we used the same method employed for NARCS, which is taken from \citet{georgakakis08}.  For a given detection probability threshold, $P_{\mathrm{thresh}}$, we determined the minimum number of total counts required for a detection ($C_{\mathrm{lim}}$) at each position in the image, such that $P(\geq C_{\mathrm{lim}})$ = $P_{\mathrm{thresh}}$.  To this end, we made background maps in the 3--10~keV and 10--20 keV~bands by removing the counts within 60$^{\prime\prime}$ (90$^{\prime\prime}$) radius circular regions centered on the point (extended) source positions listed in Table~\ref{tab:srclist}, and then filling in these regions by randomly distributing the expected background counts determined from the local background.  Using these background maps, we calculated the mean expected background counts ($\langle C_{\mathrm{bkg}}\rangle$) in circular regions centered on each pixel with radii equal to the 15, 22, and 30\% PSF enclosures, which are the cell sizes we used for source detection (see \S~\ref{sec:detection}).  The probability that the observed counts will exceed $C_{\mathrm{lim}}$ within a particular region is
\begin{equation}
P(\geq C_{\mathrm{lim}}) = \gamma(C_{\mathrm{lim}}, \langle C_{\mathrm{bkg}}\rangle)
\label{eq:probnoise2}
\end{equation}
where $\gamma(a,x)$ is the lower incomplete gamma function, defined as
\begin{equation}
\gamma(a,x) = \frac{1}{\Gamma(a)}\int_0^x e^{-t}t^{a-1}dt  
\end{equation}
Calculating $C_{\mathrm{lim}}$ requires setting $P(\geq C_{\mathrm{lim}}) = P_{\mathrm{thresh}}$, and inverting Equation~\ref{eq:probnoise2} numerically.  \par
Then we computed the probability of detecting a source of a given flux $f_X$ at each pixel, given by
\begin{equation}
P_{f_X}(\geq C_{\mathrm{lim}}) = \gamma(C_{\mathrm{lim}},C_{\mathrm{src}}) 
\label{eq:probfx}
\end{equation}
and $C_{\mathrm{src}} = f_X t_{\mathrm{exp}} A_{\mathrm{src}} \epsilon + \langle C_{\mathrm{bkg}}\rangle$, where $t_{\mathrm{exp}}$, $A_{\mathrm{src}}$, and $\epsilon$ are the exposure time, mean effective area, and unabsorbed energy flux to observed photon flux conversion factor, respectively.  For $\epsilon$, we used the mean ratio of photon flux to energy flux measured for tier~1 sources in a given energy band.  To estimate the effective area at each pixel location, we made vignetting-corrected exposure maps, and by comparing the ratio of the vignetting-corrected exposure over the uncorrected exposures to the effective areas of tier~1 sources, we derived a linear relation to convert the exposure ratio at a given location to the effective area for the average source spectrum. These relations were derived using vignetting corrections evaluated at 8~keV for the 3--10~keV band and at 10~keV for the 10--20~keV band; they were also calibrated for the three different cell sizes.  \par
There are a few possible sources of systematic error in our calculation of the sky coverage curves.  Different $\epsilon$ values and effective area to exposure ratio relations were derived based on non-parametric and modeling-derived fluxes to account for systematic errors associated with flux calculation methods; the sky coverage curves derived using these different fluxes vary by about 0.1~dex.  Another potential source of systematic error arises from our use of the average source spectrum as representative for the \textit{NuSTAR} Norma sources; based on the spread of spectral properties they exhibit, a systematic error of roughly 0.1~dex on the sky coverage could result from the choice of a representative source spectrum.  Finally, the calculated sky coverage includes all the observations shown in Figures \ref{fig:mosaic} and \ref{fig:trial}, but, as can be seen, a significant wedge of stray light and some residual ghost rays are present, especially in the 3--10~keV band.  This contamination effectively reduces our sky coverage, because even though there were about 20 clusters of pixels in these regions which exceeded our detection threshold, we ascribed most of them to artifacts associated with stray light and ghost rays rather than true sources; the only exceptions which we included in our final source list (NNR~2, 6, 10, 16, 18) were either very bright sources, had bright ($\gtrsim10^{-5}$~photons~cm$^{-2}$~s$^{-1}$) \textit{Chandra} counterparts, or had a clear point-like morphology.  Since the contaminated areas only make up about 2\% of the total survey area, their inclusion in our sky coverage does not significantly impact our log$N$-log$S$ results.  \par
\begin{figure}[t]
\includegraphics[width=0.47\textwidth]{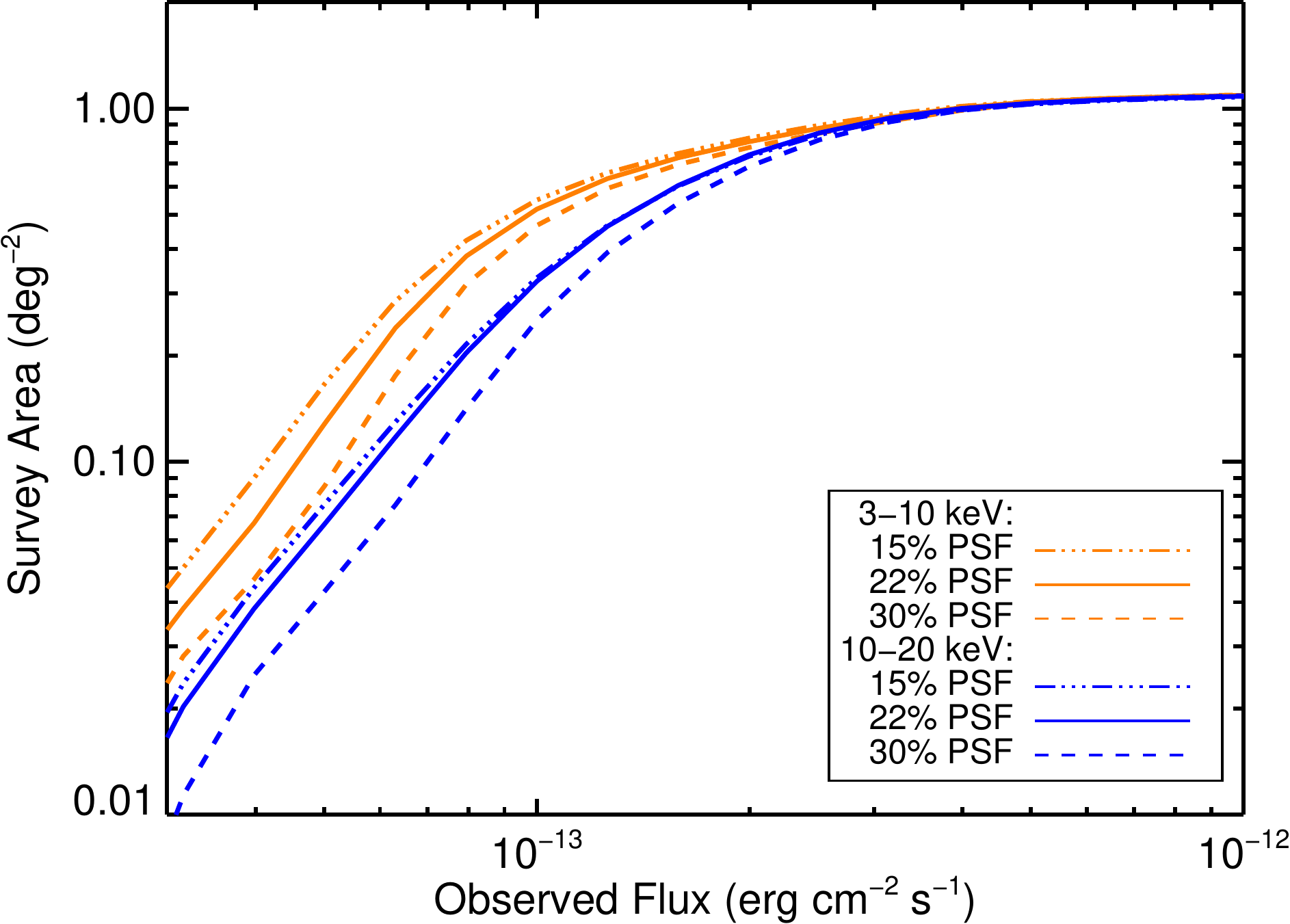}
\caption{Sky coverage of the \textit{NuSTAR} Norma Region survey for different energy bands and PSF enclosure fractions.  The curves in this plot use a photon flux to energy flux conversion factor based the spectral modeling of tier~1 sources.}
\label{fig:sensitivitycurve}
\end{figure}

\begin{figure*}[t]
\makebox[0.49\textwidth]{ %
	\centering
	\hspace{-0.15in}\subfigure{\includegraphics[angle=90,width=0.445\textwidth]{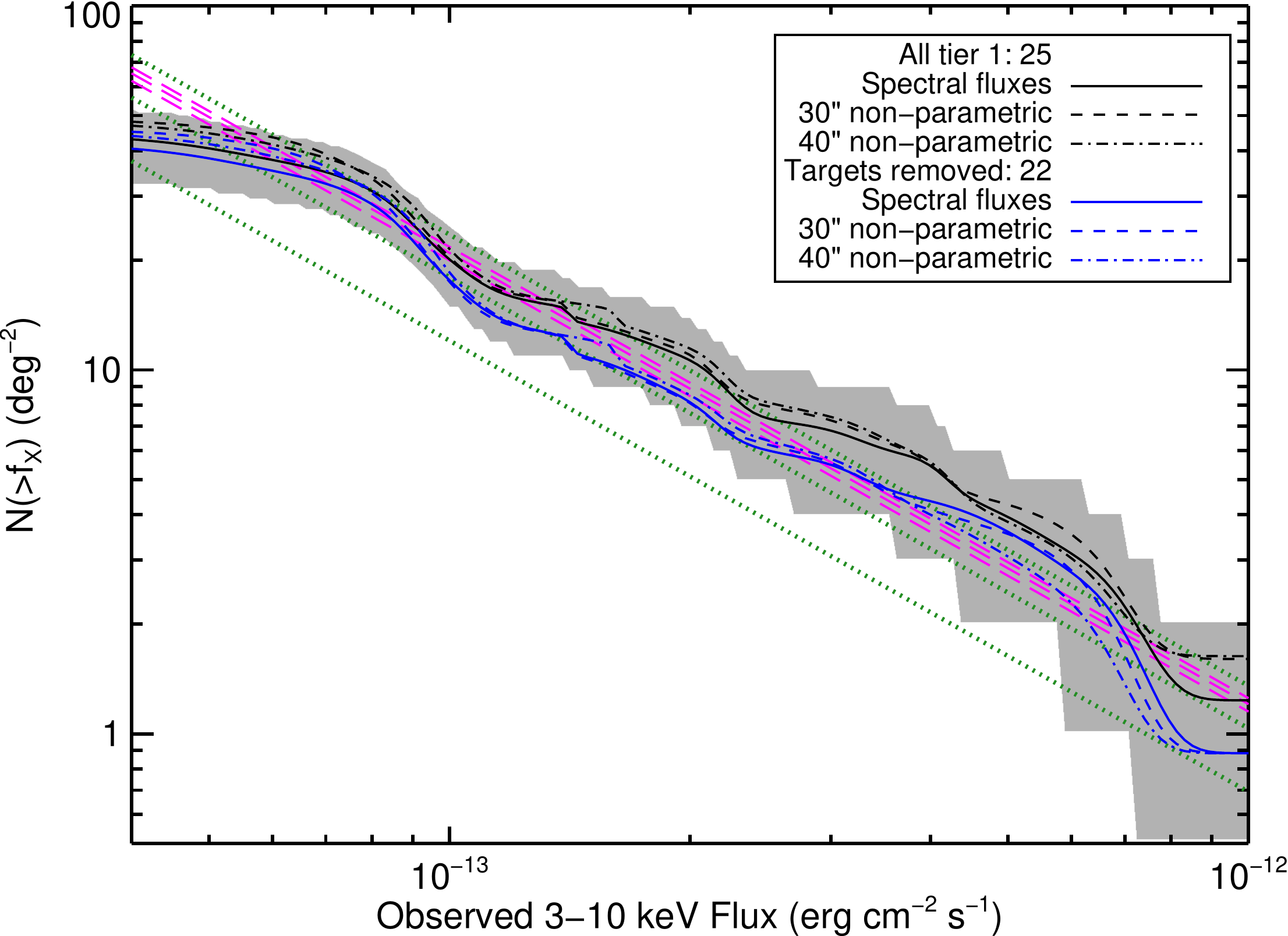}}}
\vspace{-0.08in}
\centering	
	\subfigure{\includegraphics[angle=90,width=0.445\textwidth]{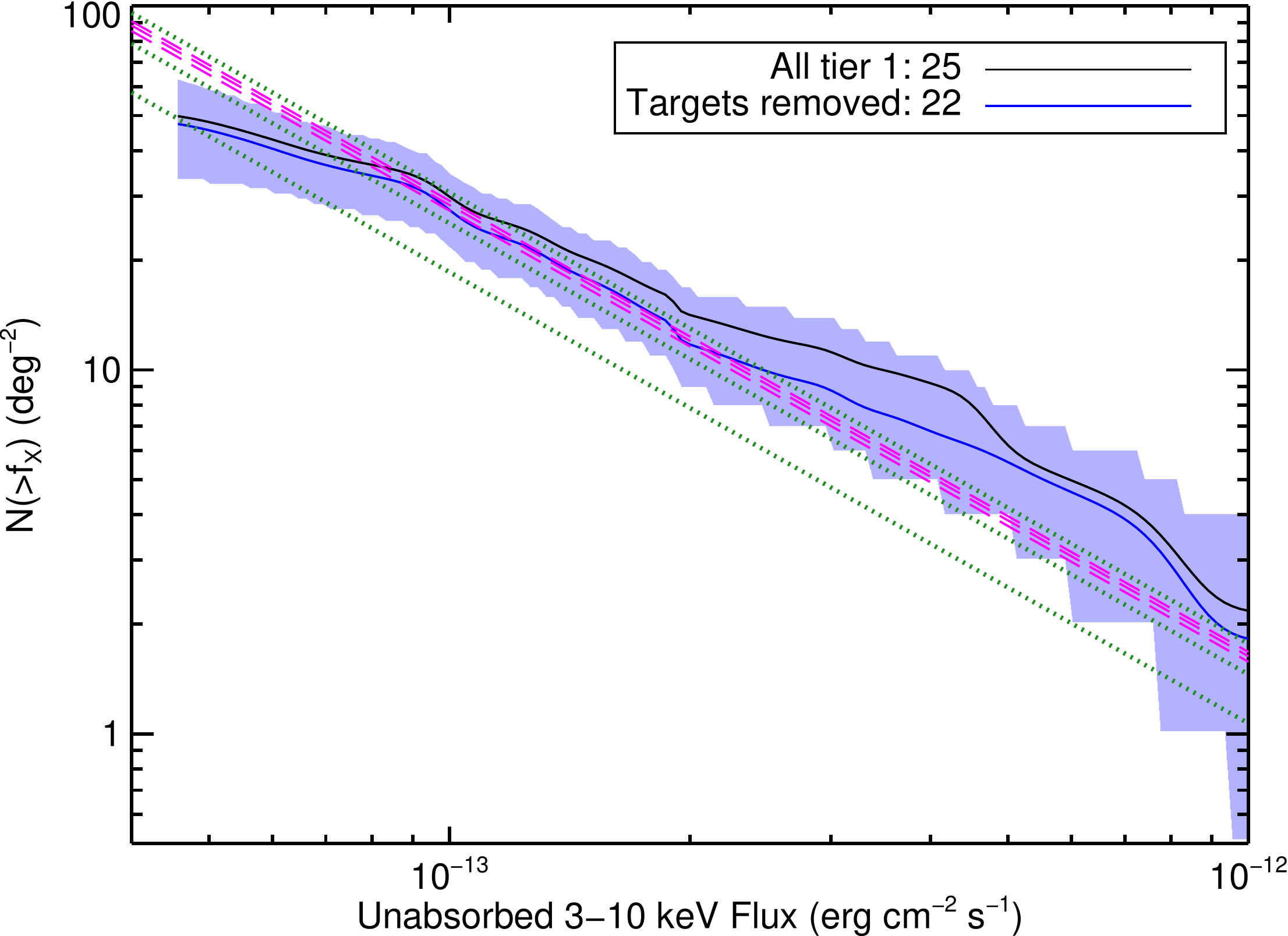}}
\vspace{-0.09in}	
\centering	
	\subfigure{\includegraphics[angle=90,width=0.445\textwidth]{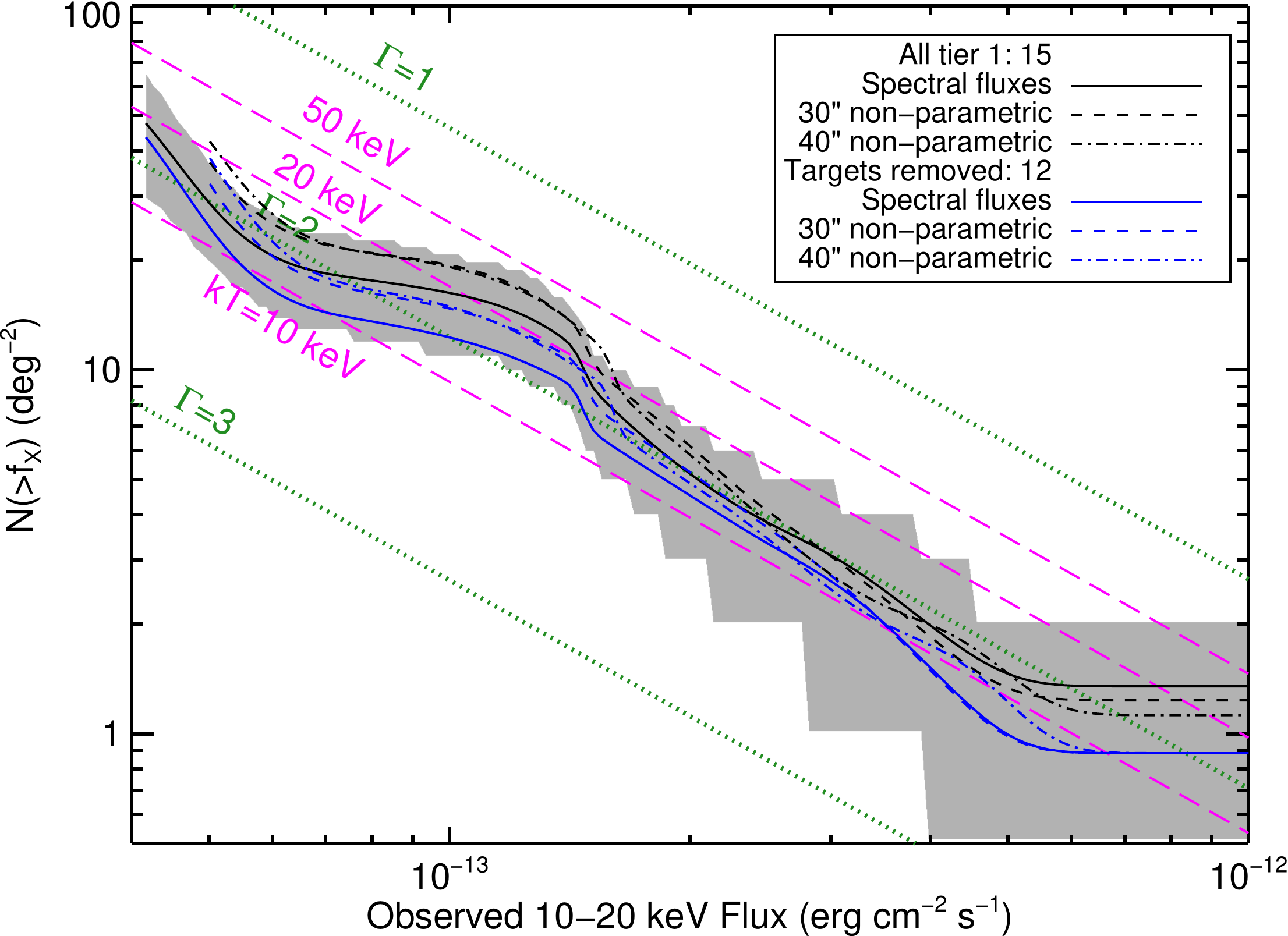}}
\caption{\textit{All}: The log$N$-log$S$ distributions shown in black include all tier~1 sources exceeding the detection threshold in a given energy band; the gray band shows the 1$\sigma$ errors on the log$N$-log$S$ distribution.  The log$N$-log$S$ distributions shown in blue exclude NNR~2, 4, and 5, which were specifically targeted by \textit{NuSTAR}.  The green dotted (magenta dashed) lines show the NARCS log$N$-log$S$ converted from unabsorbed 2--10~keV into the given bands assuming power-law spectral models with $\Gamma= 3$, 2, and 1 (thermal models with $kT$=10, 20 and 50~keV).  When converting the NARCS distribution into the observed 3--10 or 10--20 keV bands, a column density of $N_{\mathrm{H}} = 10^{23}$~cm$^{-2}$ is used, the mean of measured $N_{\mathrm{H}}$ values for the \textit{NuSTAR} sources; varying $N_{\mathrm{H}}$ between $0.7-2.0\times10^{23}$~cm$^{-2}$ does not significantly change the conversion factor. \textit{Top left}: The log$N$-log$S$ distribution in the 3--10~keV band, calculated using observed fluxes derived from spectral fitting, as well as non-parametric fluxes calculated from aperture photometry using 30$^{\prime\prime}$ and 40$^{\prime\prime}$ radius regions. \textit{Top right}: The log$N$-log$S$ distribution in the 3--10~keV band calculated using unabsorbed fluxes derived from spectral fitting.  The blue band shows the $1\sigma$ errors on the distribution shown in blue.  \textit{Bottom}: Same as top except the log$N$-log$S$ distributions are shown as a function of observed 10--20~keV flux.} 
\label{fig:lognlogs}
\end{figure*}

The sky coverage was calculated as the sum of probabilities in Equation~\ref{eq:probfx} over all pixels multiplied by the solid angle per pixel.  We repeated this calculation for a range of fluxes to produce a sensitivity curve for each of the three detection cell sizes in both the 3--10~keV and 10--20~keV bands.  Figure~\ref{fig:sensitivitycurve} shows the sky coverage for different energy bands and cell sizes.  We used the sensitivity curves for the 22\% PSF enclosures to calculate the log$N$-log$S$ distribution and sensitivity limits of our survey in the 3--10~keV and 10--20~keV bands, because all but one of the 26 (17) tier~1 sources which were detected in the 3--10~keV (10--20~keV) band were detected in the 22\% PSF trial maps, and the 22\% PSF sky coverage represents a rough average of the different PSF fraction curves.  The deep field of the \textit{NuSTAR} Norma survey has an area of about 0.04~deg$^2$ and sensitivity limits of $4\times10^{-14}$~erg~cm$^{-2}$~s$^{-1}$ ($5\times10^{-14}$~erg~cm$^{-2}$~s$^{-1}$) in the observed (unabsorbed) 3--10~keV band and $4\times10^{-14}$~erg~cm$^{-2}$~s$^{-1}$ in the 10--20~keV band.  The shallow survey has an area of $\sim$1~deg$^2$ with sensitivity limits of $1\times10^{-13}$~erg~cm$^{-2}$~s$^{-1}$ ($1.5\times10^{-13}$~erg~cm$^{-2}$~s$^{-1}$) in the observed (unabsorbed) 3--10~keV band and $1.5\times10^{-13}$~erg~cm$^{-2}$~s$^{-1}$ in the 10--20~keV band. 

\subsection{Log$N$-log$S$ distribution}
\label{sec:lognlogs}
Since many of the \textit{NuSTAR} Norma sources have fluxes approaching our sensitivity limits, when calculating the number-count distribution for our survey, it is important to consider the effect of Poisson fluctuations of the source and background counts on the measured source flux.  Thus, rather than assigning a single flux value to each source, we determine its flux probability distribution by computing the source count distribution from Equation~A21 in \citet{weiss07} and converting counts to energy fluxes.  The number count distribution is then equal to the sum of the flux probability distributions of individual sources, divided by the sensitivity curve calculated in \S~\ref{sec:sensitivity}.  \par
We compute log$N$-log$S$ distributions in the 3--10~keV and 10--20~keV bands, both for observed and unabsorbed fluxes; in order to check for systematic errors, we perform these calculations using both the modeling-derived and non-parametric fluxes.  When constructing the distribution in a given energy band, we only included the sources that exceed the detection threshold in that particular energy band.  In addition, in order to compare the \textit{NuSTAR} number-count distribution with that derived from NARCS, we excluded extended sources, and for sources that are blended in \textit{NuSTAR} observations but resolved with \textit{Chandra}, we estimate the \textit{NuSTAR} fluxes of individual sources by assuming the ratio of fluxes (see comments in Table~\ref{tab:spectra}) of the two sources is the same in \textit{NuSTAR} as it is in the \textit{Chandra} 2--10~keV band.  Thus, NNR~8 is excluded from the sample of sources used in the number-count distribution, the fluxes of the point sources at the center of the extended sources, NNR~3 and 21, are estimated to be 30\% and 20\% of the total respectively, and the fraction of NNR~7's flux attributed to NARCS~1278 and 1279 is 30\% and 70\%, respectively. \par
We calculated the statistical errors of the log$N$-log$S$ distribution using the bootstrapping method.  We account for the errors associated with our sample size as well as the distribution of fluxes within that sample by generating new samples of sources from our original list used to calculate the log$N$-log$S$ distribution.  For each new sample, we draw the sample size ($N_{\mathrm{sample}}$) from a Poisson distribution with mean equal to the original sample size, and then randomly select $N_{\mathrm{sample}}$ sources from the original list.  We generate 10,000 new samples and calculate the resulting log$N$-log$S$ distribution for each of them.  Then we use the simulated distributions to determine the 1$\sigma$ upper and lower confidence bounds of the measured log$N$-log$S$ distribution.  The 1$\sigma$ statistical errors are comparable in size to the systematic errors associated with the sensitivity curves, which are discussed in \S\ref{sec:sensitivity}. \par
In addition to the possible sources of systematic error already discussed, there is a simplification in our log$N$-log$S$ calculation which could bias our measurements.  As described in \S\ref{sec:threshold}, we require that a source exceed the detection threshold in two or three trial maps in order to be included in our source list, but our log$N$-log$S$ calculation does not explicitly account for this criterion.  This criterion helps to screen out spurious detections which may occur in a given trial map but are unlikely to correlate across different energy bands or aperture sizes; however, it is an easy criterion for a real source to pass, since it is very likely for a real source exceeding the threshold in one trial map to exceed it in at least another trial map with the same energy band but a different PSF enclosure fraction.  Thus, we do not expect our detection criterion to substantially alter the sensitivity curves or, in turn, the log$N$-log$S$ distribution.  To gauge the magnitude of the possible bias due to our choice of sensitivity curve, we produced different versions of the log$N$-log$S$ distribution in the 3--10~keV and 10--20~keV bands, adopting sensitivity curves for different PSF enclosures, and testing the effect of limiting the source subsample to tier~1 sources exceeding the threshold in the trial map with both the same energy band and PSF enclosure as the sensitivity curve rather than just the same energy band.  The resulting variations in our log$N$-log$S$ results are smaller than the statistical uncertainties. \par
Figure~\ref{fig:lognlogs} shows the resulting log$N$-log$S$ distributions for the \textit{NuSTAR} Norma region.  In these panels, magenta and green lines show the log$N$-log$S$ distribution measured by NARCS converted from the unabsorbed 2--10~keV band to the \textit{NuSTAR} bands assuming different spectral models, thermal models with $kT=10$, 20, and 50~keV and power-law models with $\Gamma=1$, 2, and 3; when converting to observed energy fluxes, a typical $N_{\mathrm{H}}$ value of 10$^{23}$~cm$^{-2}$ is used.  The log$N$-log$S$ distributions shown in black include all tier 1 sources that exceed the detection threshold in a given energy band, while the blue distributions exclude the sources that were specifically targeted by \textit{NuSTAR} and detected (NNR~2, 4, and 5), which could unnaturally inflate the log$N$-log$S$ distribution.  As shown in the top panel, there is little difference between the NARCS distributions converted using different spectral models into the observed 3--10~keV band, which is not surprising given its large amount of overlap with the \textit{Chandra} 2--10~keV band.  Regardless of how the source energy fluxes are calculated, the \textit{NuSTAR} distribution is consistent with the NARCS distribution at 1$\sigma$ confidence, exhibiting a similar slope of $\alpha\approx-1.24$. The \textit{NuSTAR} distribution only deviates significantly from the NARCS distribution at low fluxes.  This discrepancy may be due to the Eddington bias or variance in the spatial density of sources, given that the sources with the lowest fluxes are only detected in the deep HESS field, which is only 100~arcmin$^2$ in size. \par
The middle panel of Figure~\ref{fig:lognlogs} shows the log$N$-log$S$ distribution calculated using the unabsorbed 3--10~keV fluxes from spectral fitting.  Although this distribution is still largely consistent with the NARCS distribution at 1$\sigma$ confidence when the sources specifically targeted by \textit{NuSTAR} are removed (shown in blue), the \textit{NuSTAR} distribution is slightly higher than the \textit{Chandra} distribution above $>3\times10^{-13}$~erg~cm$^{-2}$~s$^{-1}$.  The fact that this excess is only seen using unabsorbed 3--10~keV fluxes but not the observed 3--10~keV fluxes suggests that, for some sources, we measure $N_{\mathrm{H}}$ values that are too high and thus overcorrect for absorption. \par
The bottom panel of Figure~\ref{fig:lognlogs} shows the log$N$-log$S$ distributions calculated using modeling-derived and non-parametric fluxes in the observed 10--20~keV band.  Since there is very little difference between the observed and unabsorbed 10--20~keV fluxes, the log$N$-log$S$ distribution in the unabsorbed 10--20~keV band is not shown.  Although the 10--20~keV \textit{NuSTAR} distributions deviate from a simple power-law due to the small number of sources (16) detected in this hard X-ray band, overall the slope is still consistent with the NARCS slope.  The normalizations of the different NARCS distributions extrapolated into the 10--20 keV band are distinct depending on the spectral model assumed; for the \textit{NuSTAR} and NARCS normalizations to be consistent, the average spectrum of Norma sources must either have $kT=10-20$~keV or $\Gamma=2$.  This average spectrum is indeed consistent with the individual spectral fits of most of the \textit{NuSTAR} sources and their locations in the \textit{NuSTAR} quantile space.

\subsection{Comparison to the Galactic Center \textit{NuSTAR} population}
Comparing the log$N$-log$S$ distributions of sources in the Norma region and the 1$^{\circ}\times0.6^{\circ}$ Galactic Center (GC) region surveyed by \textit{NuSTAR}, the number density of \textit{NuSTAR} sources is $\approx2$ times higher in the GC \citep{hong16}, which is to be expected since the stellar density in the vicinity of the GC is higher than the stellar density along the line-of-sight of the Norma region.  The power-law slope of the number-count distribution is also steeper in the GC ($\alpha\approx-1.4$; \citealt{hong16}), which is consistent with the trend that is seen for \textit{Chandra} sources in the GC and the field in the 0.5--8~keV band \citep{muno09}.  In order for the normalizations of the GC \textit{NuSTAR} and \textit{Chandra} number-count distributions to be consistent, the typical spectrum of GC sources must either have $kT=20-50$~keV or $\Gamma\approx1.5$, which is harder than the typical spectrum of Norma sources.  \par
\citet{hong16} argue that $40-60$\% of \textit{NuSTAR} GC sources are magnetic CVs, primarily IPs, given their very hard X-ray spectra ($\Gamma\lesssim1.5$) and the presence of strong Fe emission.  All but two of the Norma CV candidates have softer spectra ($\Gamma>1.5$, $kT\lesssim20$ ~eV).  The spectral differences between the \textit{NuSTAR} populations in the Norma and GC regions are mirrored in the differences between the Galactic Ridge X-ray emission (GRXE; \citealt{revnivtsev06a}a; \citealt{revnivtsev06b}b; \citealt{revnivtsev09}) and the central hard X-ray emission (CHXE) discovered by \textit{NuSTAR} in the GC \citep{perez15}. The lower temperatures of the Norma CV candidates are consistent with the thermal spectra of the GRXE, whose hot component has a temperature of $kT\approx15$~keV (\citealt{turler10}; \citealt{yuasa12}), while the high temperatures of GC CVs resemble the $kT>25$~keV emission observed in the inner few parsecs of the Galaxy (\citealt{perez15}; \citealt{hong16}).  \par
However, it is unclear why the X-ray populations in the GC and the disk are different.   Under the assumption that most of the sources contributing to the CHXE and GRXE are IPs, the differences in their typical X-ray temperatures have been attributed to differences in their WD masses, with WDs in the GC CVs having masses $\gtrsim0.8M_{\odot}$ (\citealt{perez15}; \citealt{hong16}) and those in the disk CVs having masses $\approx0.6M_{\odot}$ (\citealt{krivonos07}; \citealt{turler10}; \citealt{yuasa12}).  However, the mean WD mass among all CVs has been measured to be $0.83\pm0.23M_{\odot}$ \citep{zorotovic11}, and the X-ray inferred masses of confirmed field IPs are consistent with this higher value of $\approx0.8M_{\odot}$ \citep{hailey16}.  The discrepancy between the measured WD masses for field CVs and the lower masses inferred from the temperature of the GRXE suggests that it may be incorrect to assume that the GRXE is dominated by IPs \citep{hailey16}.  Thus, it may be similarly incorrect to attribute the temperature differences between the \textit{NuSTAR} CV candidates in the GC and Norma regions to differences in their WD masses. \par
In fact, as discussed in \S~\ref{sec:cv}, a significant fraction of the Norma CV candidates may not be IPs but rather a mixture of polars, non-magnetic CVs, hard-spectrum SBs, and SyXBs.  These types of sources have softer spectra than IPs, and thus the difference in the average temperatures of Norma and GC sources may be explained by variations in the relative fractions of different types of CVs and symbiotic binaries.  It is unclear what physical processes would drive variations in the relative fractions of different types of compact object binaries in these two Galactic regions, but investigating these issues further will first require confirmining the true nature of the CV candidates.  \par
The clearest ways of distinguishing different types of CVs and SBs is by measuring the relative flux ratios of their Fe emission lines \citep{xu16} or measuring both their spin and orbital periods \citep{scaringi10}, but since most of the Norma CV candidates are quite faint, it will be difficult to obtain X-ray spectra or light curves with enough photons to make such measurements with current telescopes.  Monitoring the long-term X-ray and infrared variability of the CV candidates, and determining the spectral types of their counterparts more accurately to estimate distances and luminosities will help to identify the nature of these sources.  



\section{Conclusions}

\begin{list}{}{%
\setlength{\topsep}{0pt}%
\setlength{\leftmargin}{0.0in}%
\setlength{\listparindent}{0.0in}%
\setlength{\itemindent}{0.1in}%
\setlength{\parsep}{\parskip}%
\setlength{\itemsep}{4pt}
}%
\item[]
1. We have detected 28 hard X-ray sources in a square-degree region in the direction of the Norma spiral arm surveyed by \textit{NuSTAR}, which are designated as tier~1 sources.  Twenty-three of these sources were previously detected in observations of the Norma Arm Region \textit{Chandra} survey, one was a well-studied black hole transient (4U~1630-472), and four were newly discovered transients that we followed up and localized with \textit{Chandra}.  Out of 28 sources, 16 of them are detected above 10~keV.  In addition, we found ten NARCS sources with 2--10~keV fluxes $>6\times10^{-6}$~cm$^{-2}$~s$^{-1}$ that did not exceed our formal detection threshold for \textit{NuSTAR} but which displayed significant X-ray emission (S/N$> 3$) in at least one of three energy bands, which are designated as tier~2 sources.  We have provided photometric information for these sources in our catalog but do not include them in our calculation of the log$N$-log$S$ distribution since they do not meet our detection thresholds. 
\item[]
2. The log$N$-log$S$ distribution of \textit{NuSTAR} sources in the 3--10~keV band is consistent with the distribution of 2--1~ keV \textit{Chandra} sources in the Norma region. 
\item[]
3. The \textit{NuSTAR} log$N$-log$S$ distribution in the 10--20~keV band is consistent with the 2--10~keV \textit{Chandra} distribution if the average spectrum of the \textit{NuSTAR} sources can be described by a power-law model with $\Gamma=2$ or a single temperature \texttt{apec} model with a plasma temperature between 10 and 20~keV.  The broadband (3--40~keV) energy quantiles of the \textit{NuSTAR} sources show that the majority of sources have photon indices of $\Gamma=2-3$ for a power-law model or $kT=5-30$~keV for a bremsstrahlung model, which are consistent with the spectral parameters required for good agreement between the 10--20~keV and 2--10~keV log$N$-log$S$ distributions. 
\item[]
4.  We fit the joint \textit{Chandra} and \textit{NuSTAR} spectra of all sources with $>100$ counts in the 3--40 keV band, but find that $>300$ \textit{NuSTAR} counts are required to provide meaningful constraints on spectral model parameters.  We find good agreement between the spectral parameters from our fits and the location of sources in the quantile diagrams.
\item[]
5. Four of the sources detected in the \textit{NuSTAR} Norma Arm Region survey are previously well-studied sources: NNR~1 is the black hole LMXB 4U~1630-472, NNR~2 is the supergiant HMXB IGR~J16393-4643, NNR~3 is the PWN and luminous TeV source HESS~J1640-465, and NNR~24 is the magnetar SGR~J1627-41.  Based on the X-ray variability, spectral fits, and infrared counterpart information for each source, we determined the most likely nature of the fainter sources in our survey, which are summarized in Table~\ref{tab:class}.  Sources NNR~5, 8, and 21 are PWN candidates, NNR~7 and 14 are likely colliding wind binaries, NNR~10 is a possible magnetar, and NNR~15 is a quiescent black hole LMXB candidate.  The other sources are primarily CV candidates, a mixture of IPs, polars, non-magnetic CVs, and symbiotic binaries.  We estimate that five background AGN are present among the tier~1 \textit{NuSTAR} sources.  
\item[]
6.  Compared to the \textit{NuSTAR} sources that are detected in the Galactic Center region, the sources in the Norma region have softer spectra on average.  Even restricting the comparison to the CV candidates in these two regions, the Norma CVs exhibit lower plasma temperatures than those in the GC.  The $kT\approx15$~keV temperatures of Norma CV candidates resemble the hot component of the GRXE spectrum.  
\item[]
7.  If most of the Norma CV candidates are IPs, then their plasma temperatures indicate the white dwarfs in these systems have masses of $\approx0.6M_{\odot}$, which are lower than the WD masses of $\gtrsim0.8M_{\odot}$ estimated for the GC IPs.  However, we argue that it is more likely that the fraction of IPs relative to polars, non-magnetic CVs, and symbiotic binaries is lower among Norma CV candidates than in the GC region.  Since IPs have the hardest X-ray spectra of all these types of sources, a lower fraction of IPs in the Norma region would result in lower plasma temperatures for the average source. 
\item[]
8.  In order to understand the nature of the hard X-ray sources in the Norma region and why they differ from the hard X-ray sources in the GC region, it is necessary to continue monitoring the X-ray variability of the Norma CV candidates, better characterize the variability and spectral types of their infrared counterparts, and obtain higher quality spectra, especially at Fe line energies, for the brighter sources.  Follow-up multiwavelength observations of the candidate PWN, CWBs, and quiescent black hole binary would be useful in furthering our understanding of compact stellar remnants and the evolution of massive stars.
\end{list}

\acknowledgments

We thank the referee for feedback which helped improve the clarity of the work presented in this paper.  This work made use of data from the \textit{NuSTAR} mission, a project led by the California Institute of Technology, managed by the Jet Propulsion Laboratory, and funded by the National Aeronautics and Space Administration.  We thank the \textit{NuSTAR} Operations, Software and Calibration teams for support with the execution and analysis of these observations. This research has made use of the \textit{NuSTAR} Data Analysis Software (NuSTARDAS) jointly developed by the ASI Science Data Center (ASDC, Italy) and the California Institute of Technology (USA).  We also made use of observations taken by the \textit{Chandra X-ray Observatory} and of software provided by the Chandra X-ray Center (CXC) in the application packages CIAO and Sherpa.  This work also made use of data products from observations made with ESO Telescopes at the La Silla or Paranal Observatories under ESO programme ID 179.B-2002.  In addition, FMF acknowledges support from the National Science Foundation Graduate Research Fellowship and thanks G. K. Keating for helpful conversations on some of the statistical measures and figures in the paper.  JAT acknowledges support from \textit{Chandra} grants GO4-15138X and GO5-16152X.  FEB acknowledges support from CONICYT-Chile grants Basal-CATA PFB-06/2007, FONDECYT Regular 1141218, ``EMBIGGEN'' Anillo ACT1101, and the Ministry of Economy, Development, and Tourism's Millenium Science Initiative through grant IC120009, awarded to The Millenium Institute of Astrophysics, MAS.  JCS ackowledges support from FONDECYT 3140310.  RK acknowledges support from Russian Science Foundation (grant 14-12-01315).  DB thanks the French Space Agency (CNES) for financial support.  

\bibliographystyle{jwapjbib}
\bibliography{refs}

\appendix
In this appendix, we present the \textit{Chandra} and \textit{NuSTAR} spectra of sources NNR~10--27 and residuals for the best-fitting spectral models listed in Table~\ref{tab:spectra}.  Our spectral analysis is described in \S~\ref{sec:spectral}.  The spectra of NNR~1, 2, and 3 are shown in \citealt{king14}, B16, and G14, respectively, while the spectra of sources NNR~4--9 are shown in Figure~\ref{fig:examplespec}.  In the figures below, \textit{Chandra} data is shown in black, \textit{NuSTAR} FPMA data is shown in red, and FPMB data is shown in blue.  For NNR~19, black points show the \textit{Chandra} spectrum from ObsID 7591 while orange points shows the \textit{Chandra} spectrum from ObsID 16170.  For NNR~21, black points denote the \textit{Chandra} spectrum for the point source and extended emission combined while orange points display the point source contribution alone.

\begin{figure*}[h]
\makebox[\textwidth]{ %
	\centering
	\subfigure{
		\includegraphics[angle=270,width=0.47\textwidth]{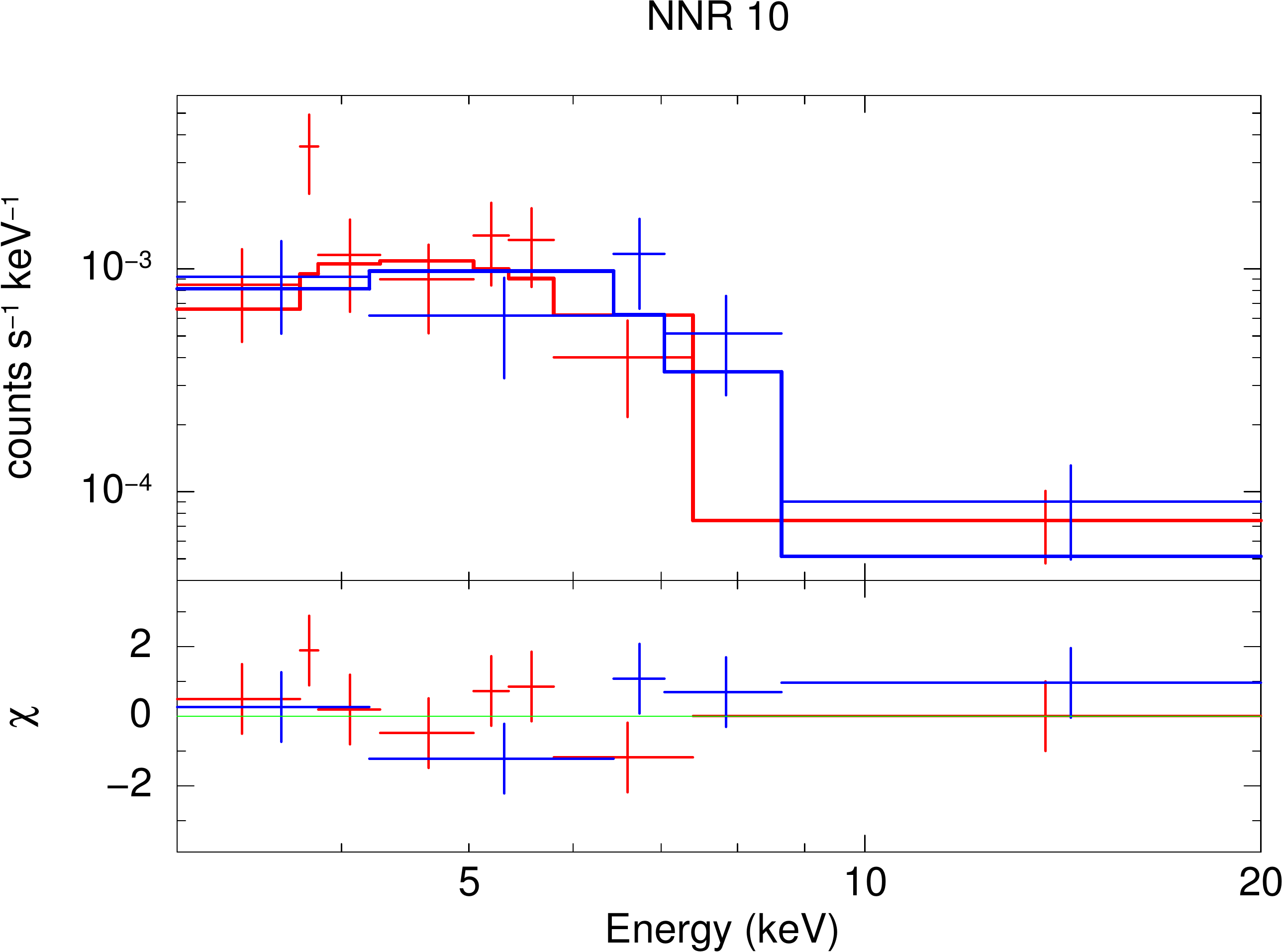}}
	\subfigure{
		\includegraphics[angle=270,width=0.47\textwidth]{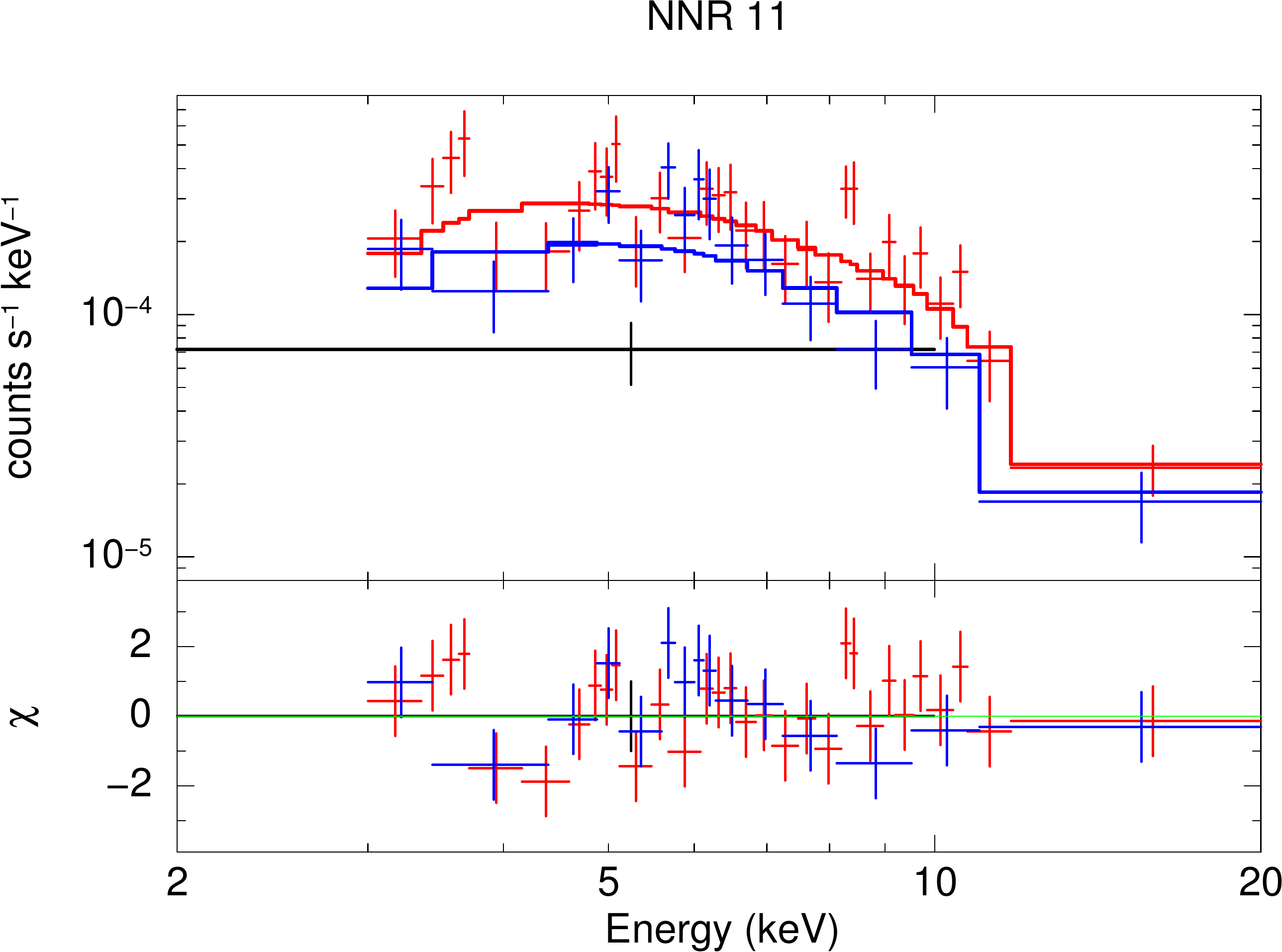}}}
\end{figure*}
\begin{figure*}
\makebox[\textwidth]{ %
	\centering
	\subfigure{
		\includegraphics[angle=270,width=0.47\textwidth]{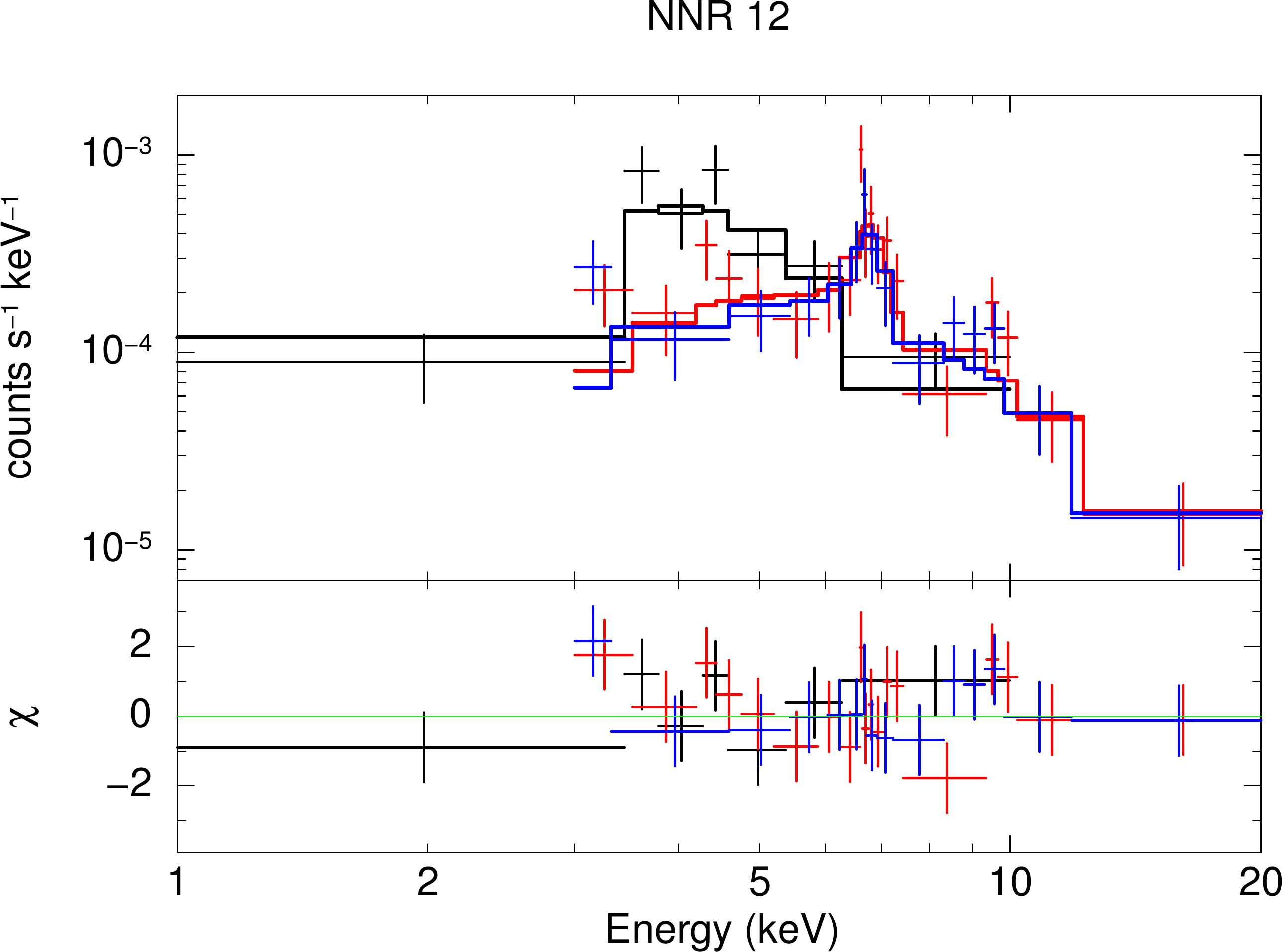}}
	\subfigure{
		\includegraphics[angle=270,width=0.47\textwidth]{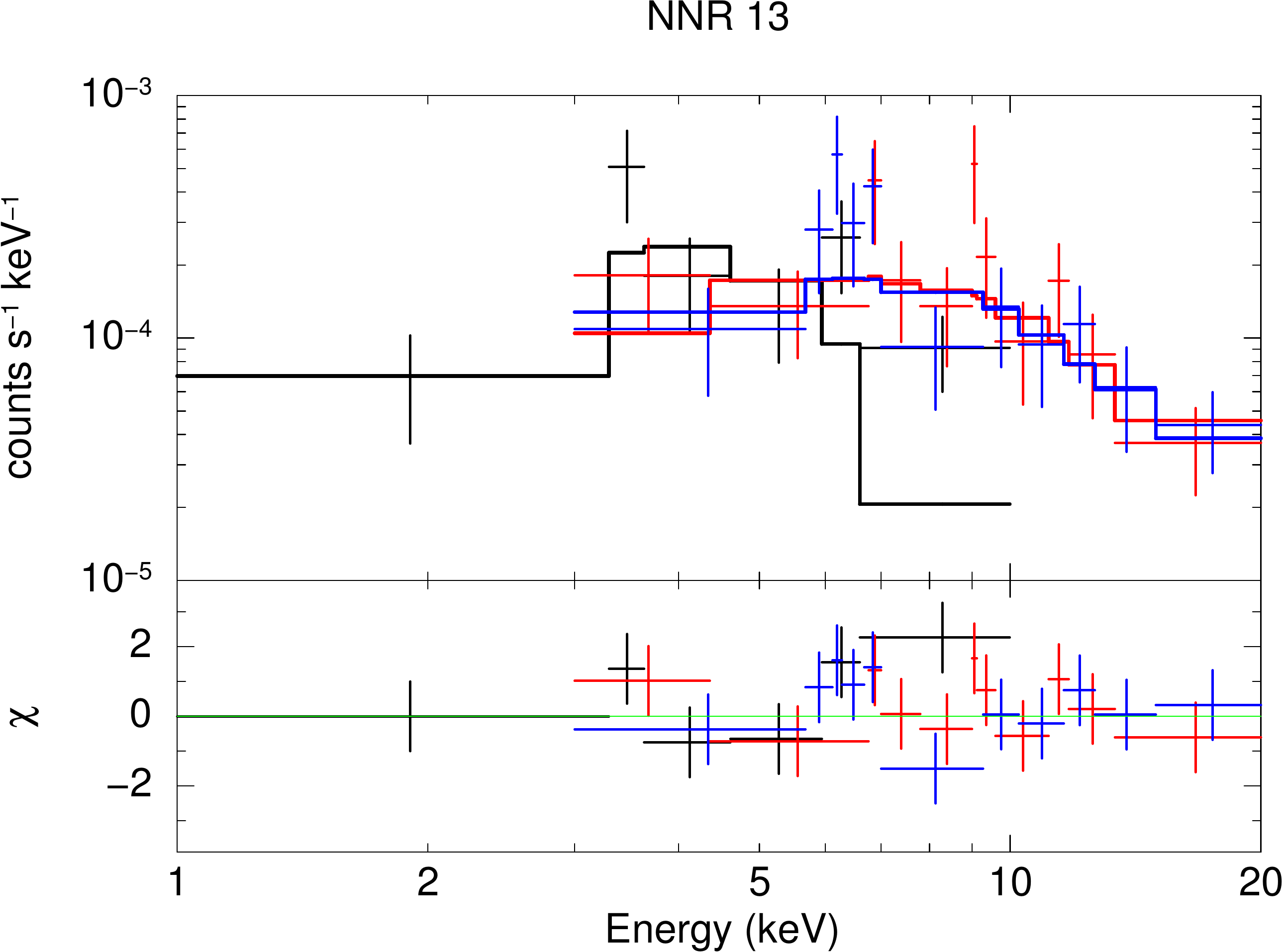}}}
\end{figure*}
\begin{figure*}
\makebox[\textwidth]{ %
	\centering
	\subfigure{
		\includegraphics[angle=270,width=0.47\textwidth]{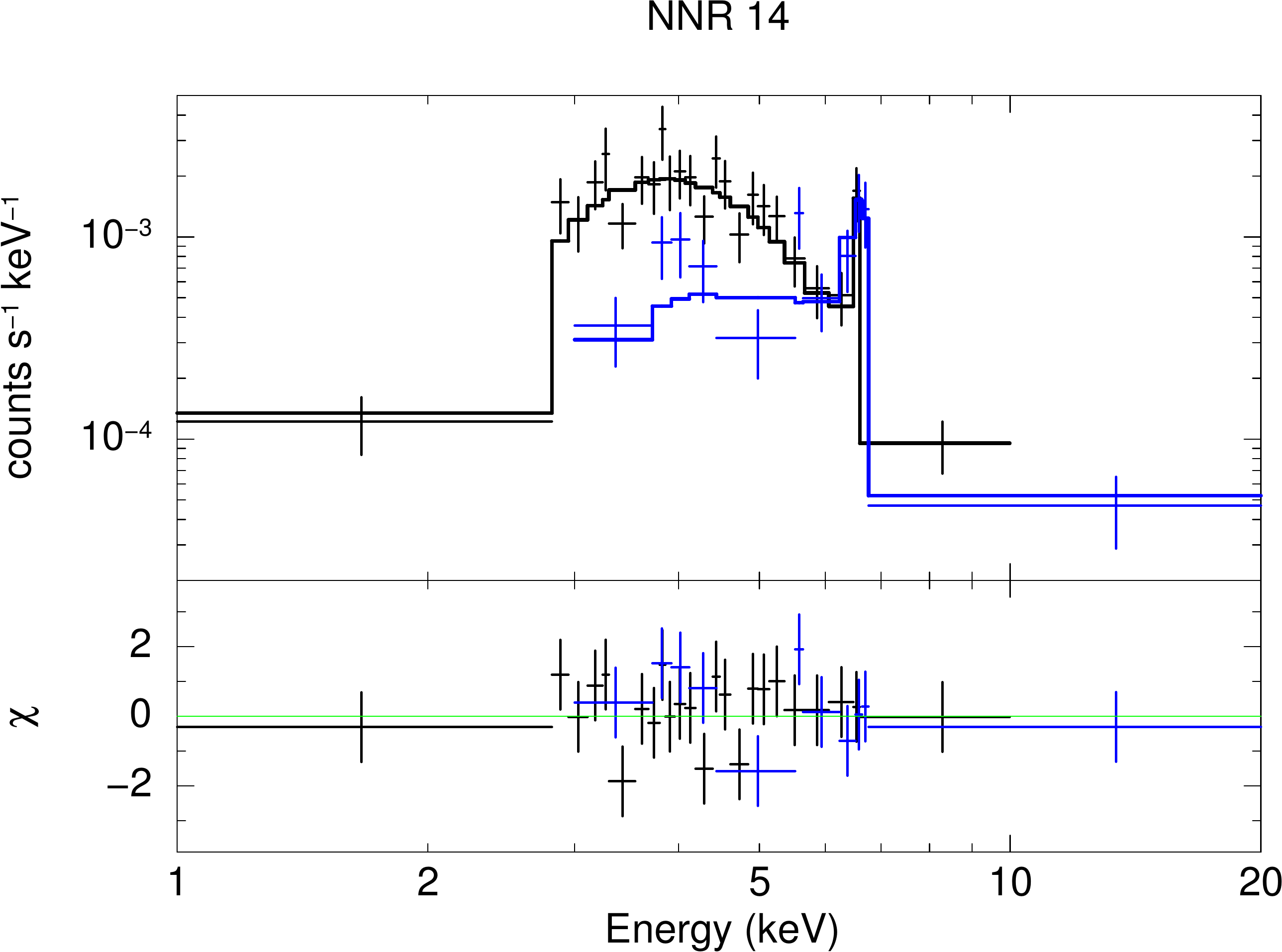}}
	\subfigure{
		\includegraphics[angle=270,width=0.47\textwidth]{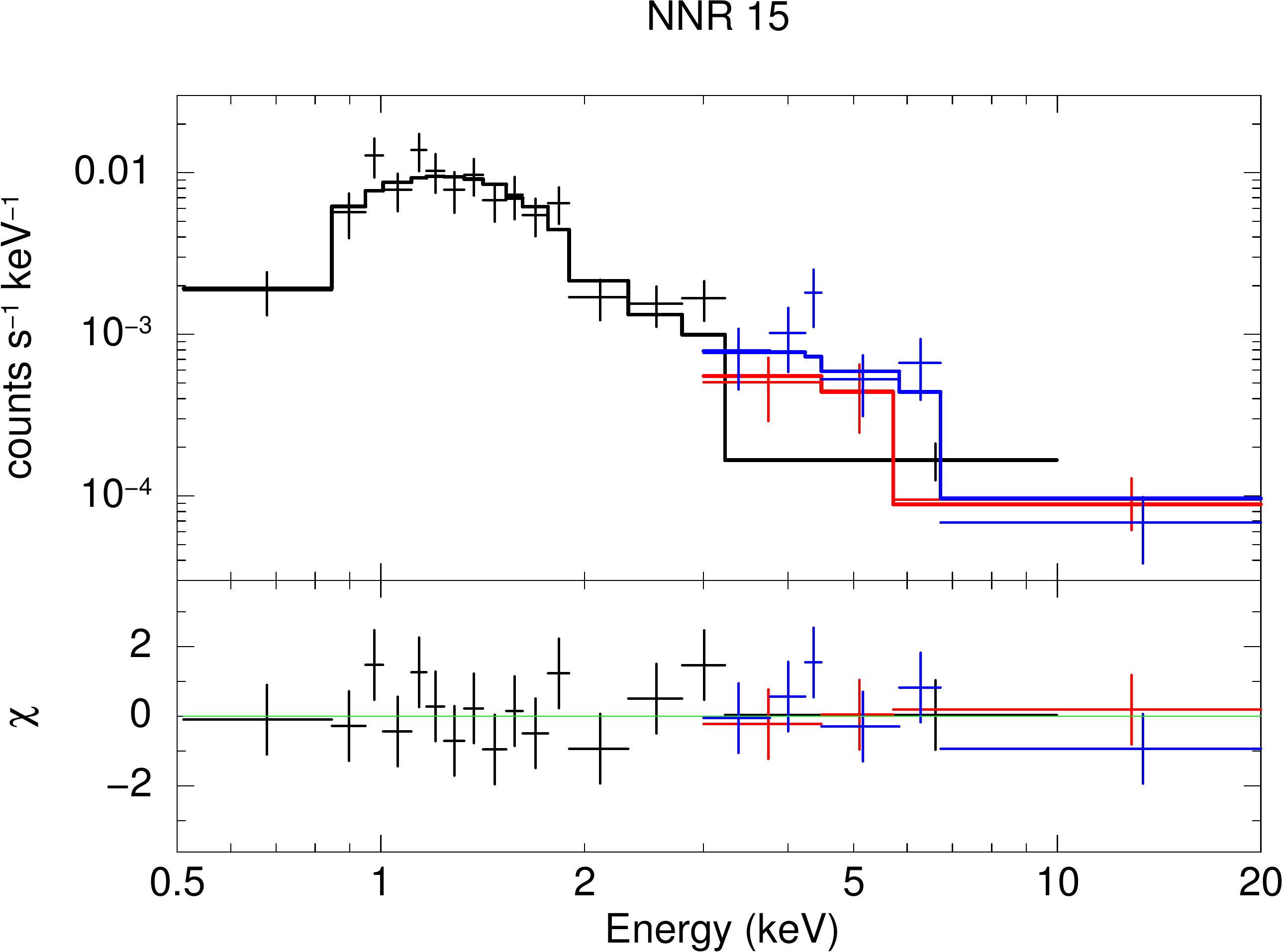}}}
\end{figure*}

\begin{figure*}
\makebox[\textwidth]{ %
	\centering
	\subfigure{
		\includegraphics[angle=270,width=0.47\textwidth]{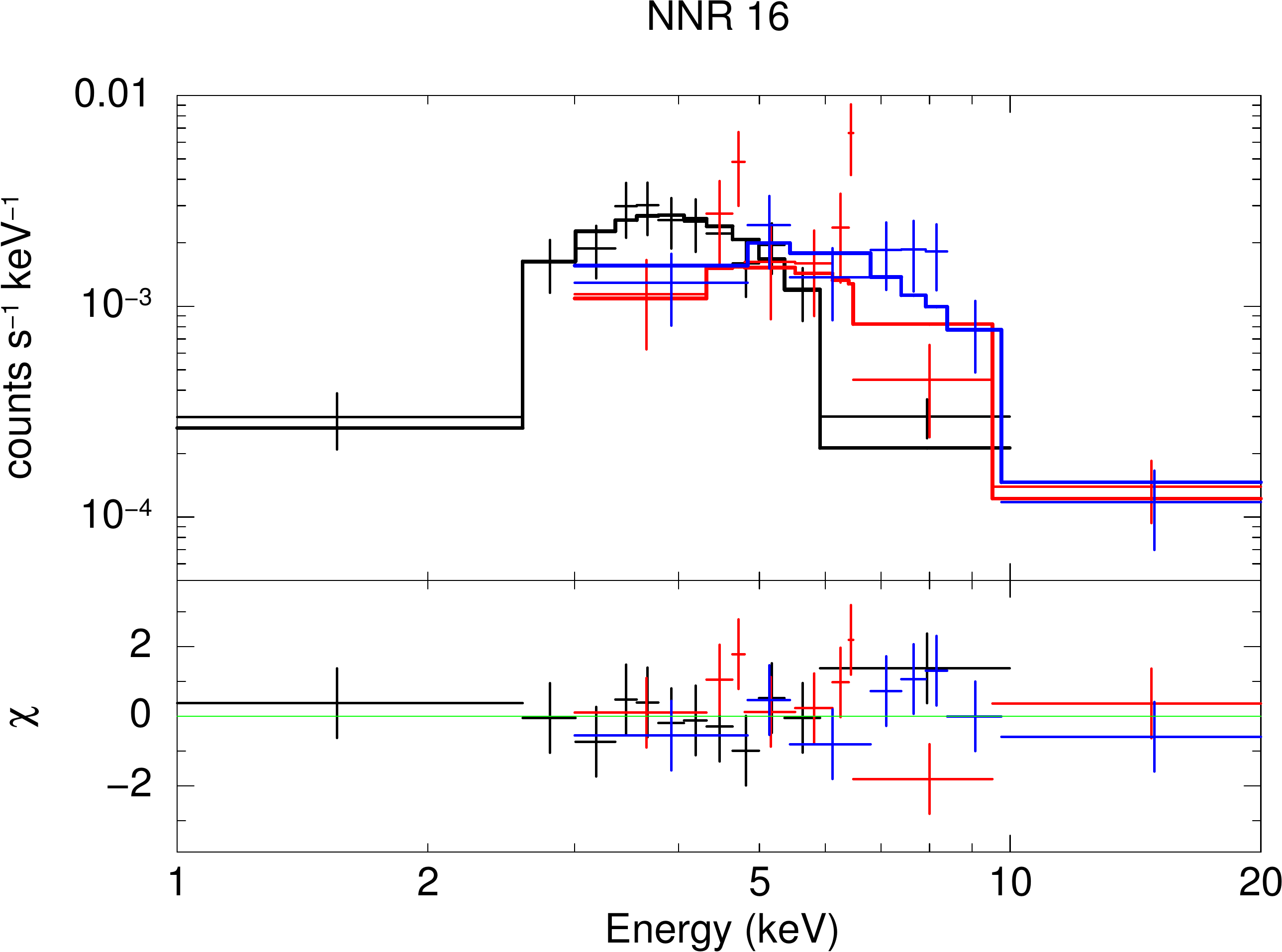}}
	\subfigure{
		\includegraphics[angle=270,width=0.47\textwidth]{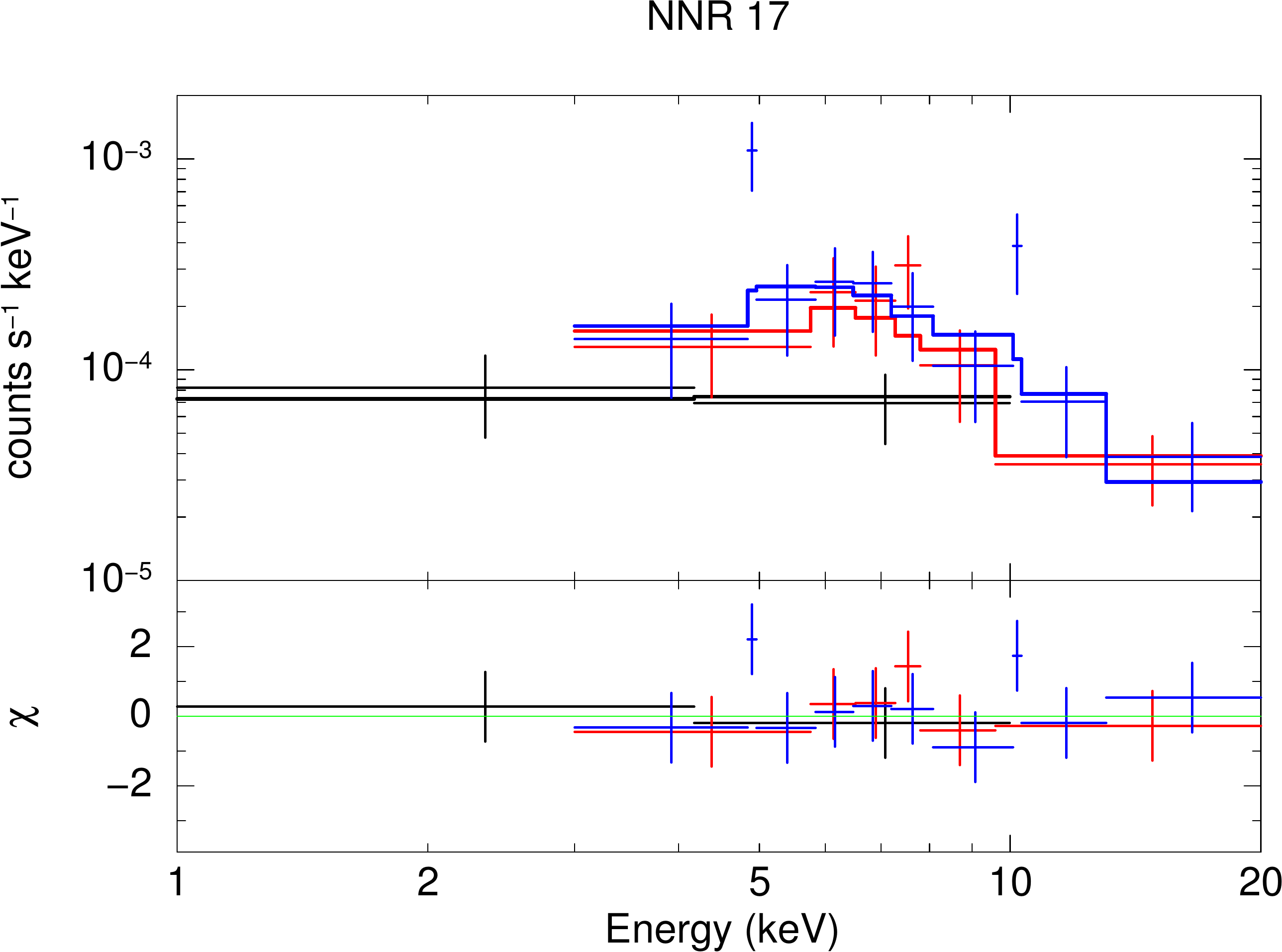}}}
\end{figure*}
\begin{figure*}
\makebox[\textwidth]{ %
	\centering
	\subfigure{
		\includegraphics[angle=270,width=0.47\textwidth]{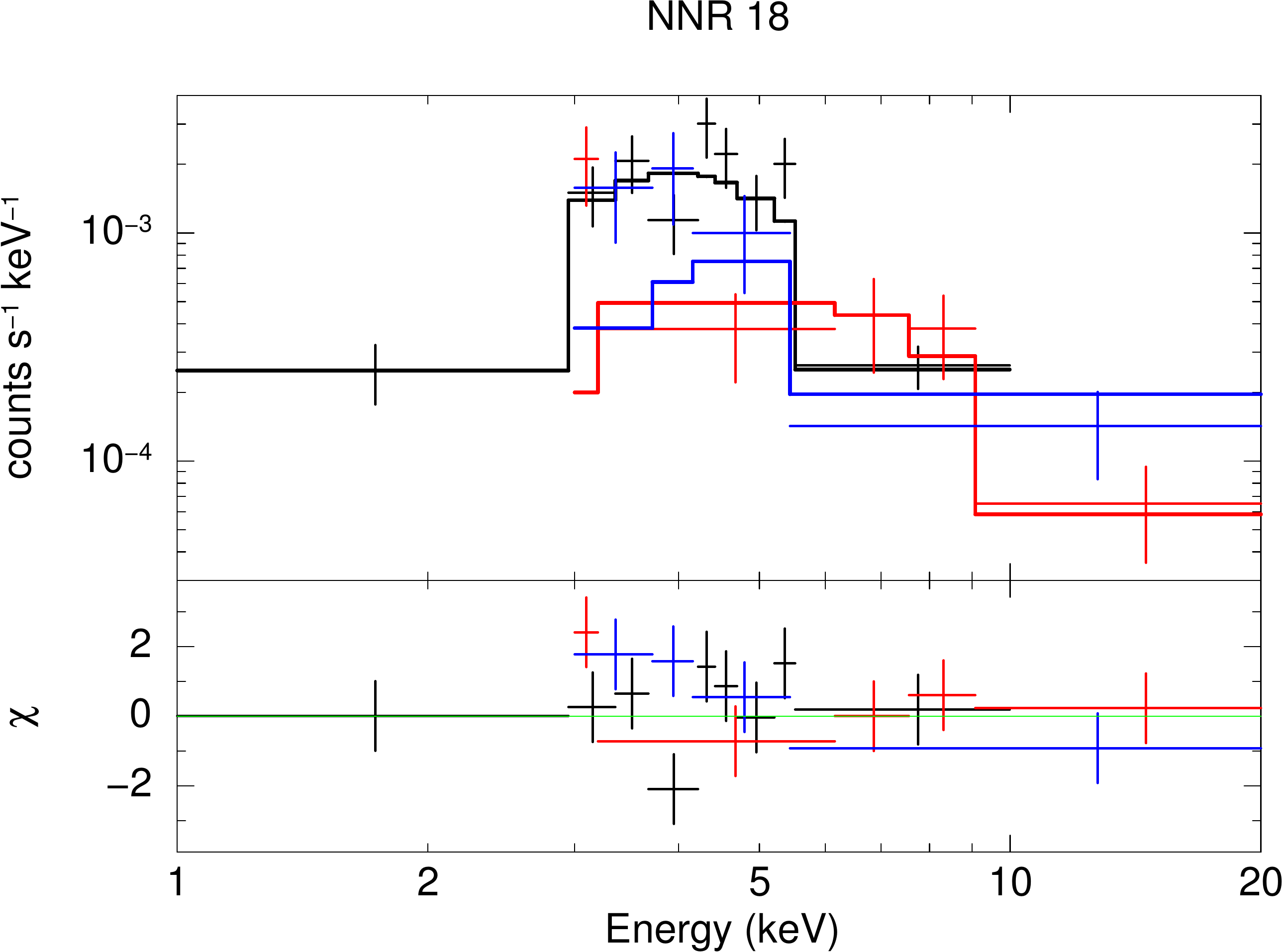}}
	\subfigure{
		\includegraphics[angle=270,width=0.47\textwidth]{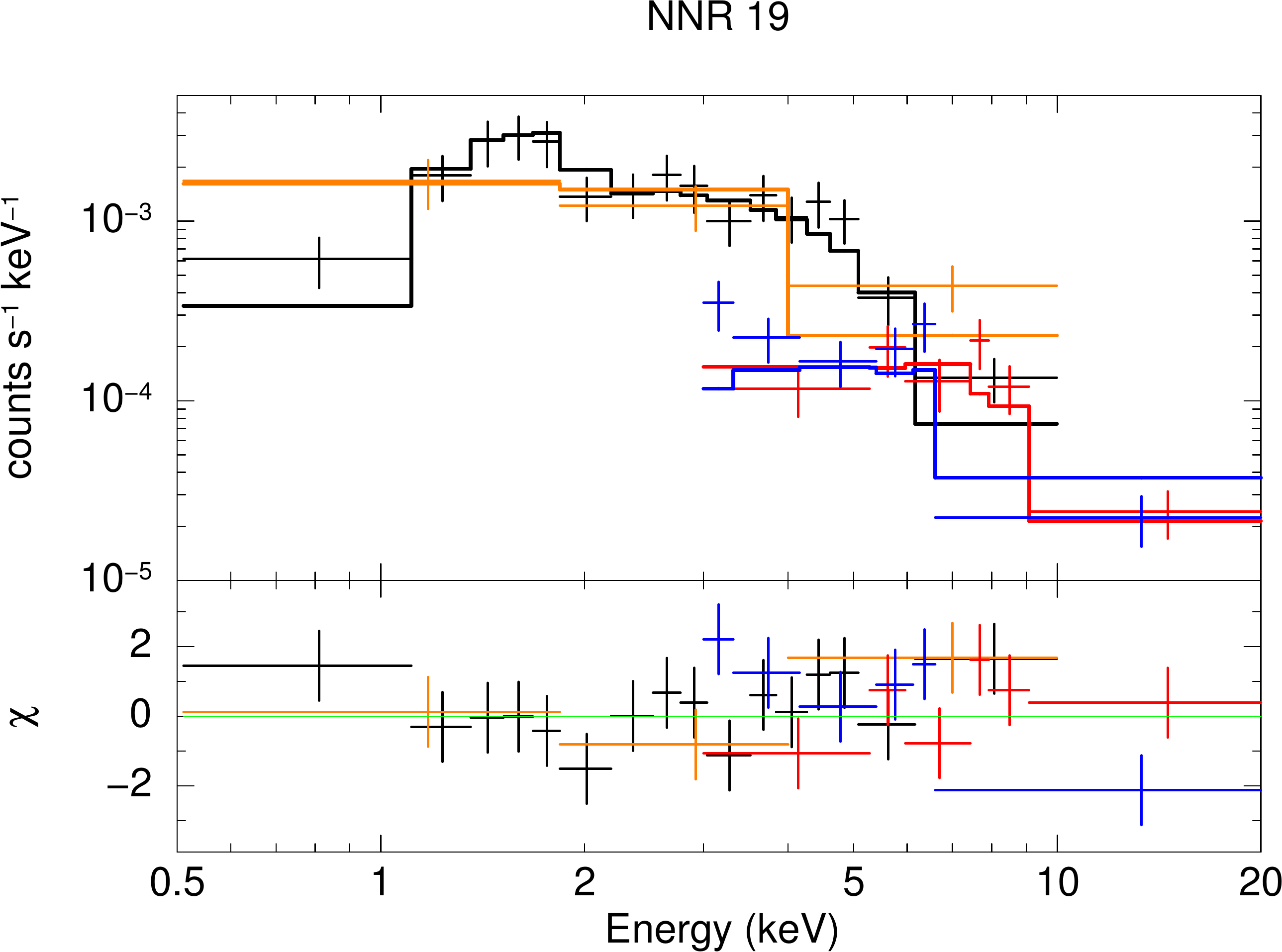}}}
\end{figure*}
\begin{figure*}
\makebox[\textwidth]{ %
	\centering
	\subfigure{
		\includegraphics[angle=270,width=0.47\textwidth]{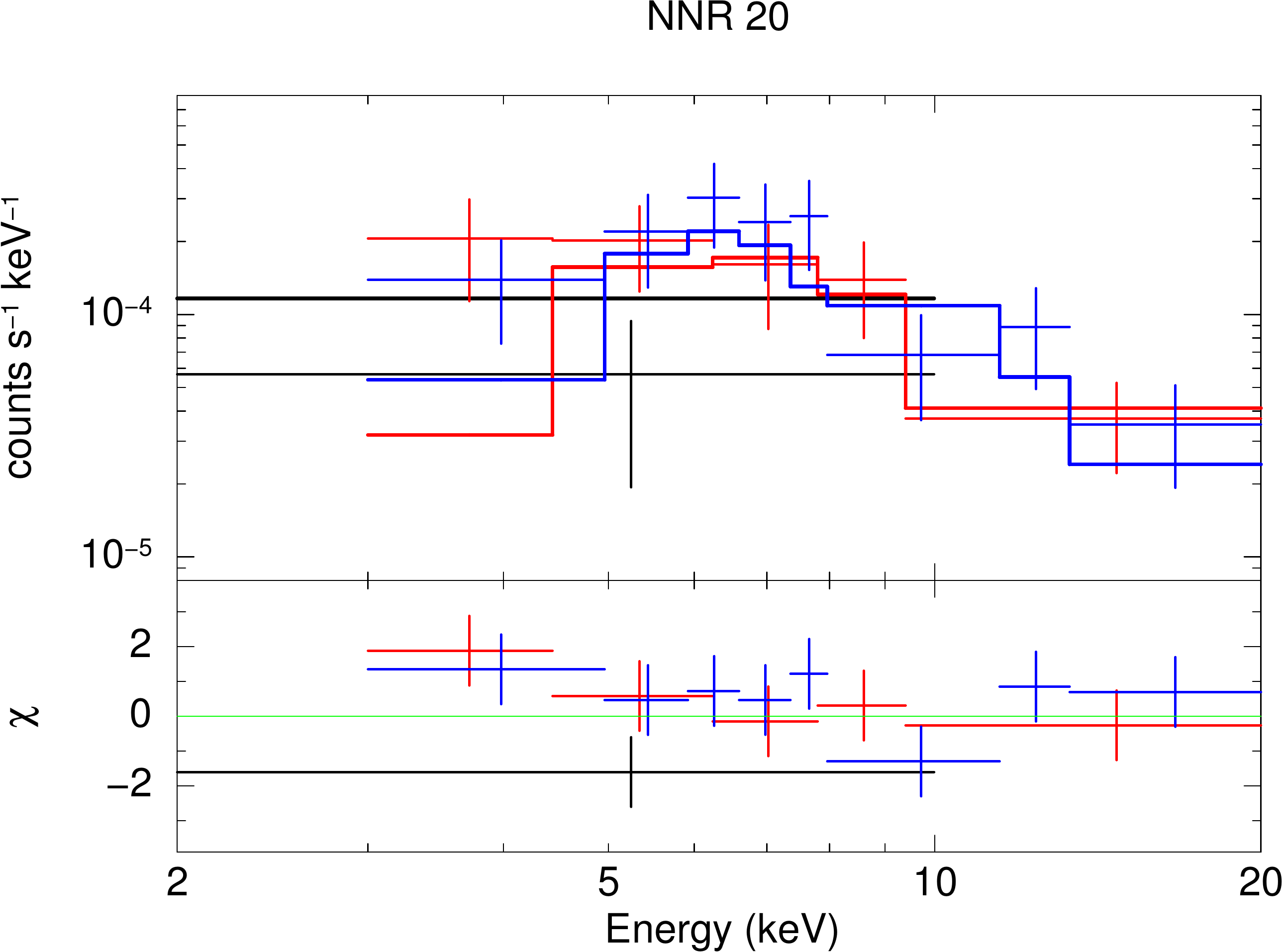}}
	\subfigure{
		\includegraphics[angle=270,width=0.47\textwidth]{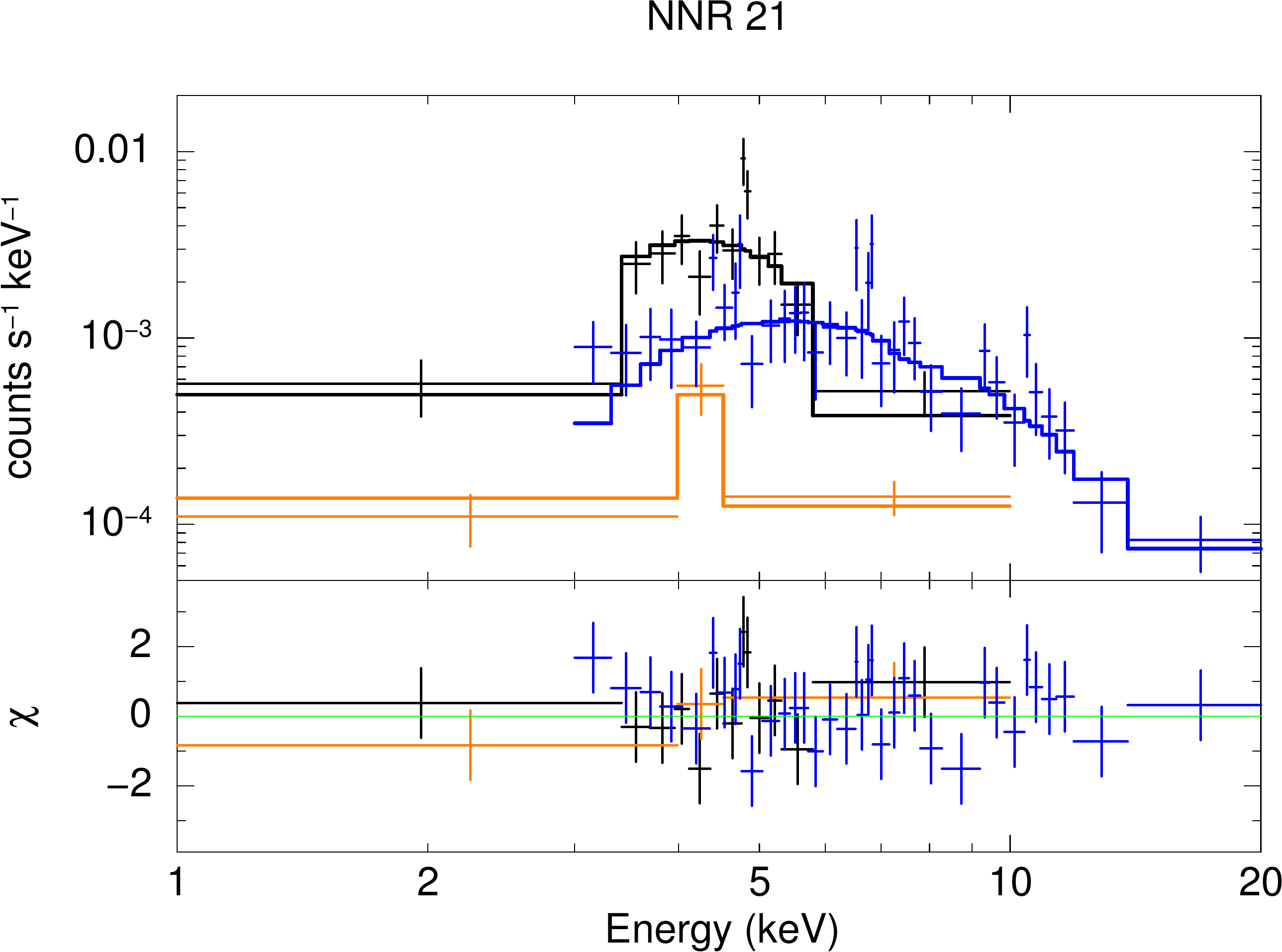}}}
\end{figure*}

\begin{figure*}
\makebox[\textwidth]{ %
	\centering
	\subfigure{
		\includegraphics[angle=270,width=0.47\textwidth]{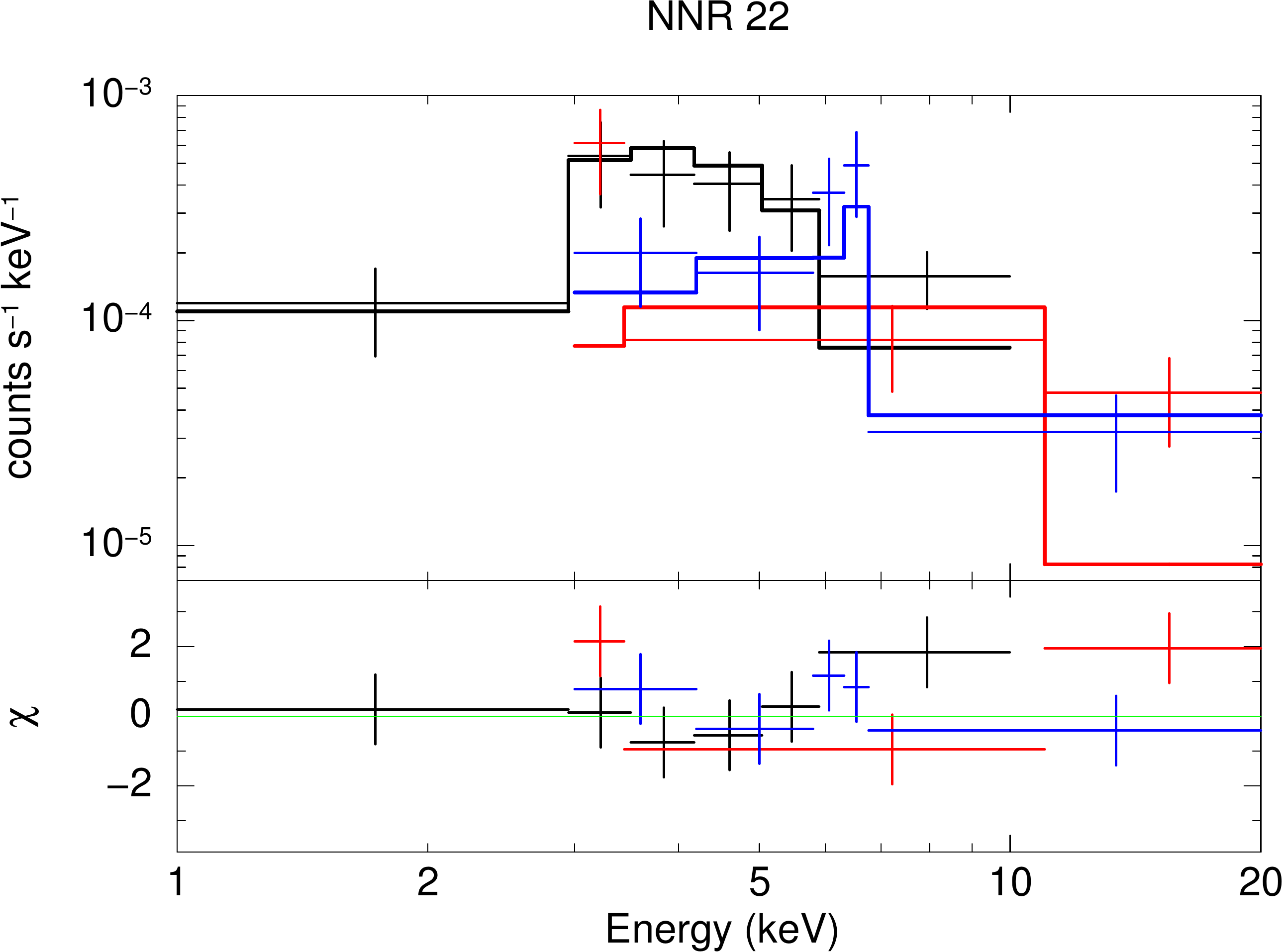}}
	\subfigure{
		\includegraphics[angle=270,width=0.47\textwidth]{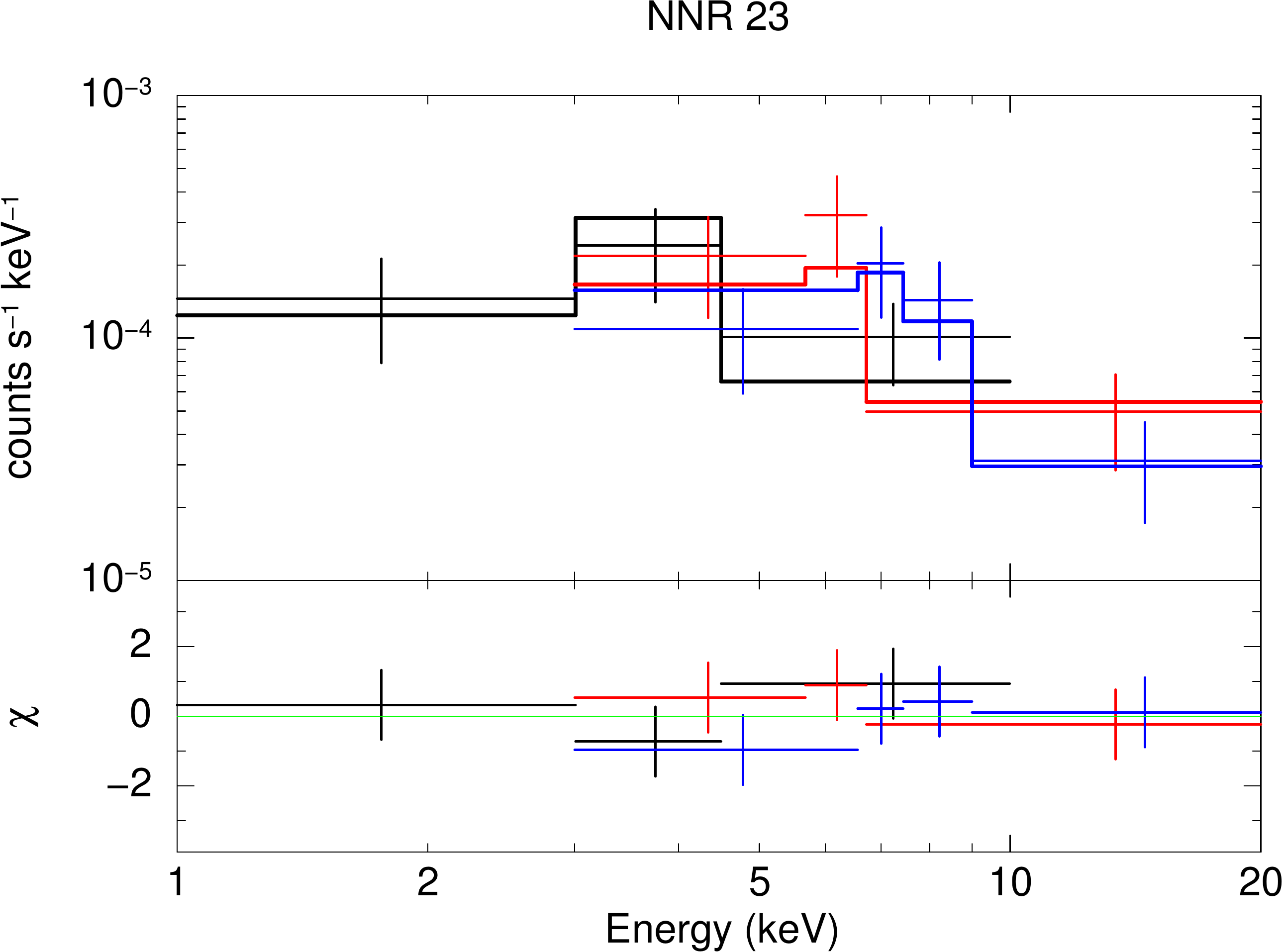}}}
\end{figure*}
\begin{figure*}
\makebox[\textwidth]{ %
	\centering
	\subfigure{
		\includegraphics[angle=270,width=0.47\textwidth]{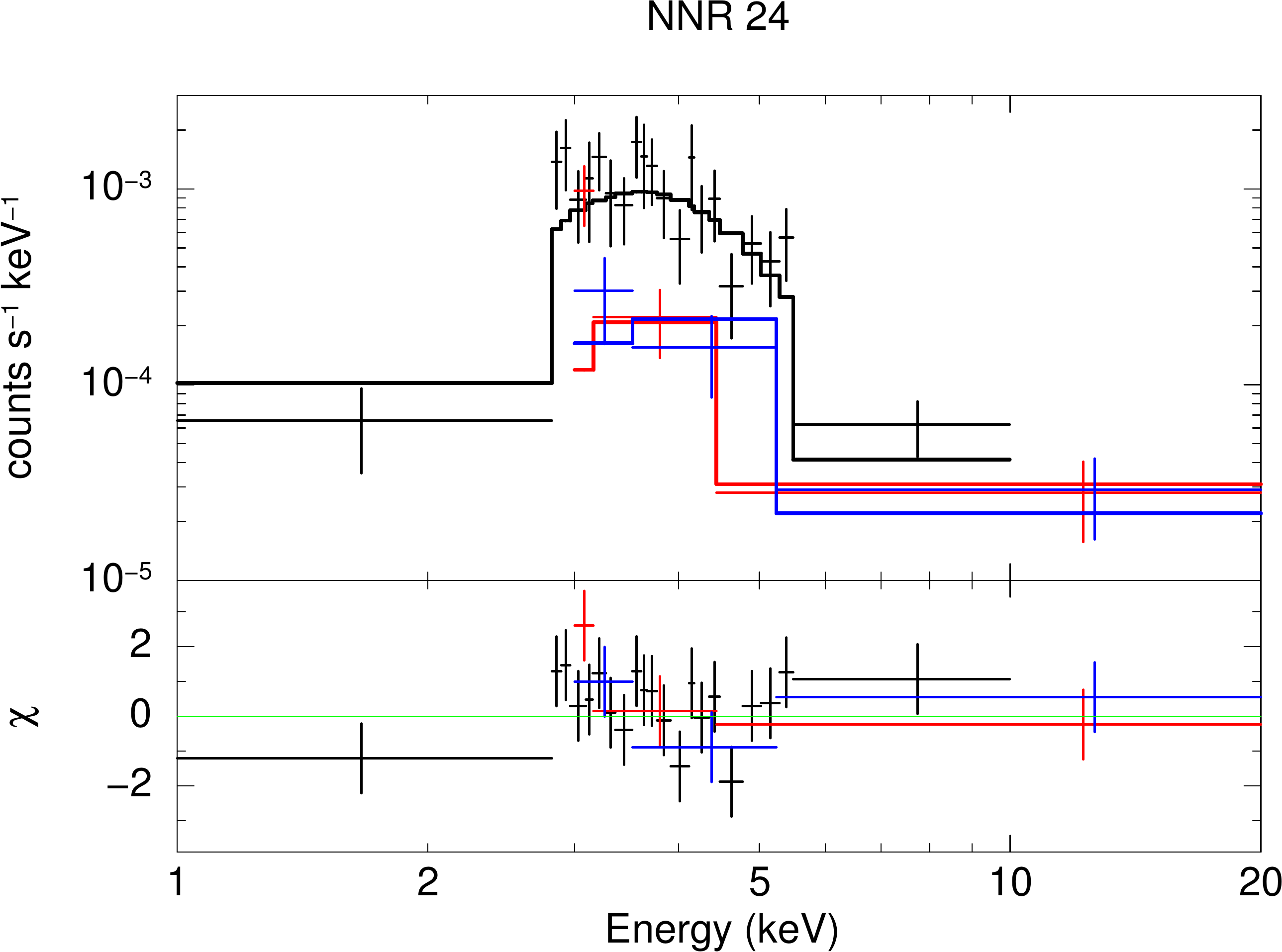}}
	\subfigure{
		\includegraphics[angle=270,width=0.47\textwidth]{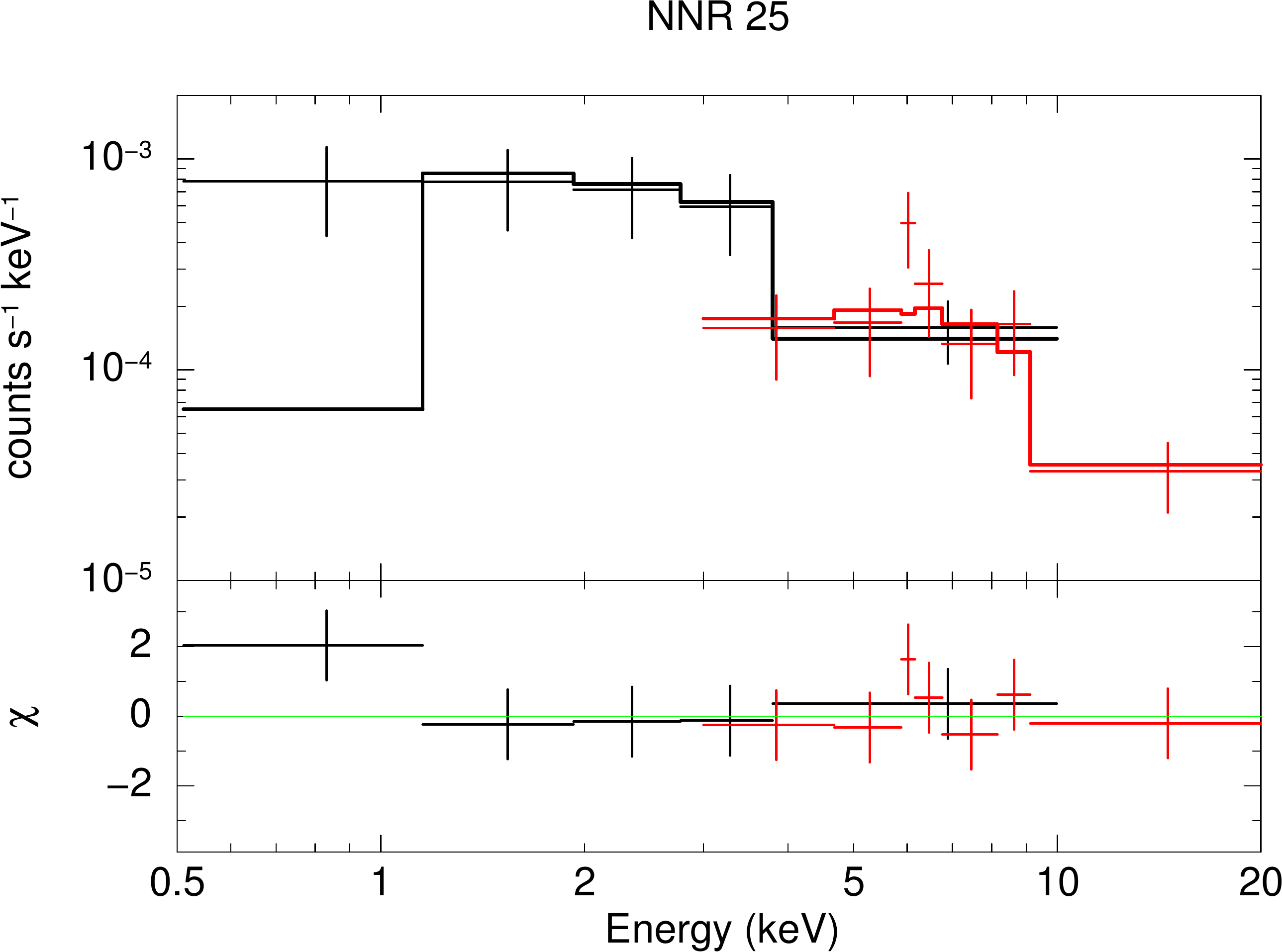}}}
\end{figure*}
\begin{figure*}
\makebox[\textwidth]{ %
	\centering
	\subfigure{
		\includegraphics[angle=270,width=0.47\textwidth]{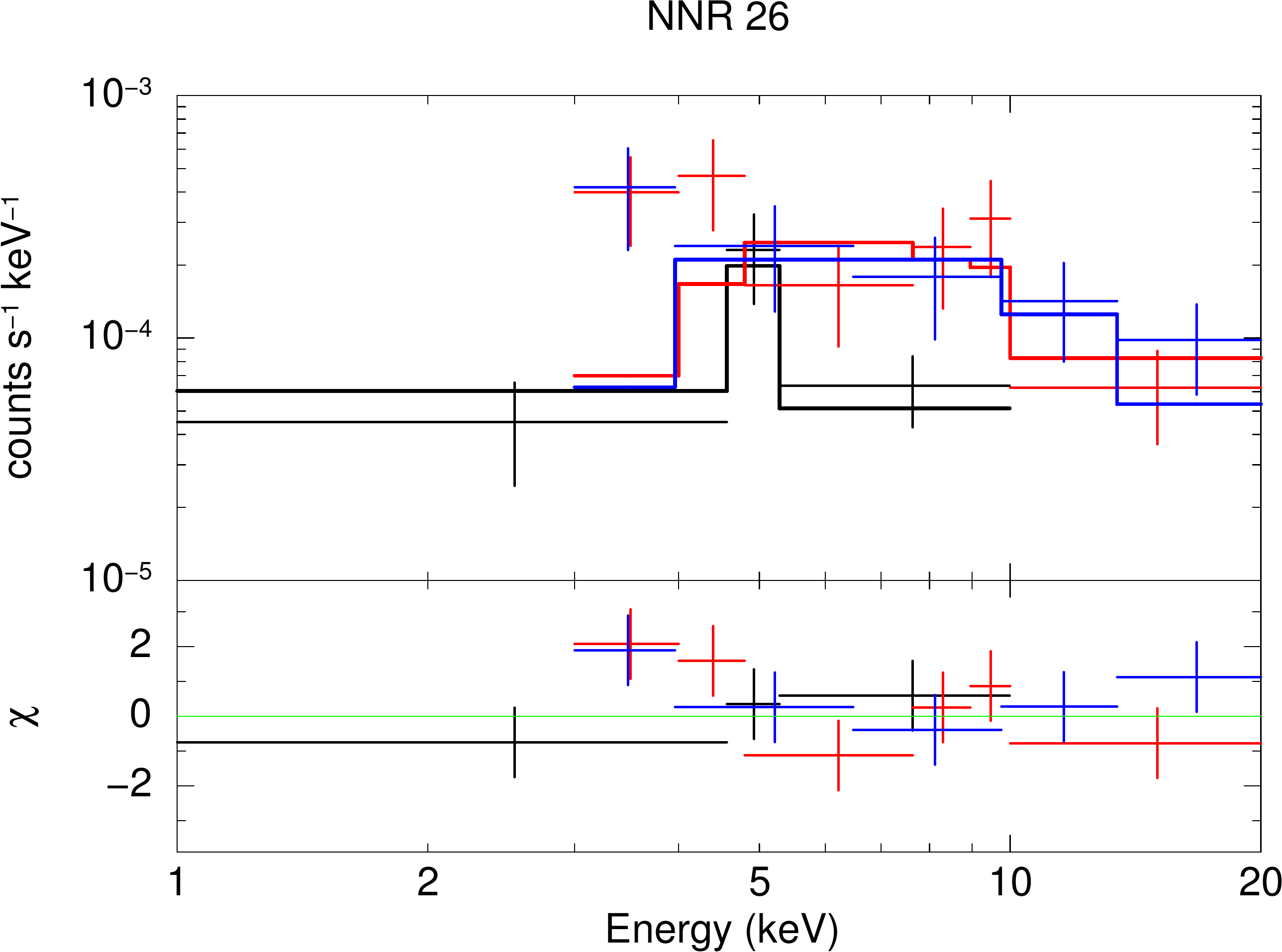}}
	\subfigure{
		\includegraphics[angle=270,width=0.47\textwidth]{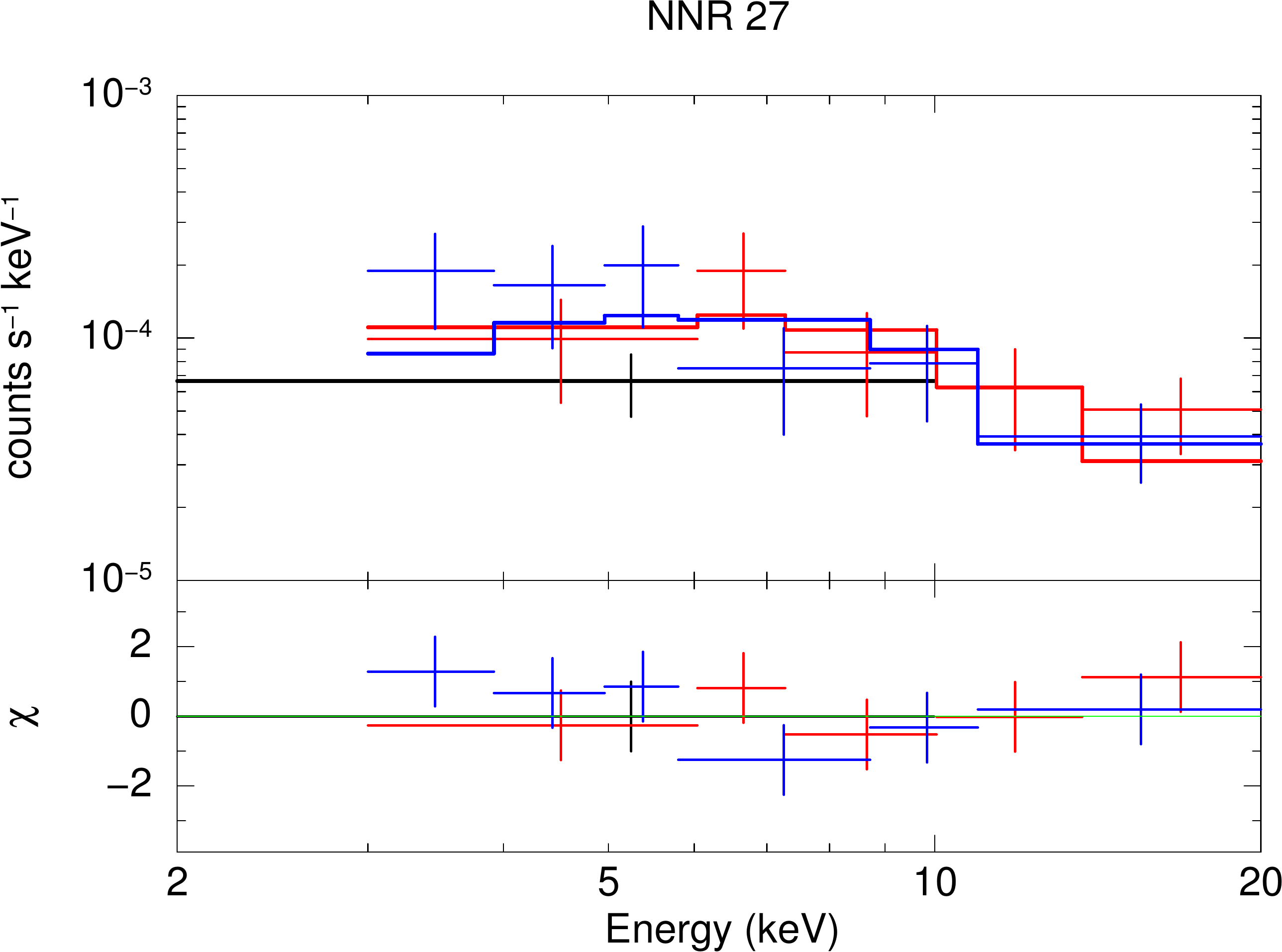}}}
\label{fig:morespec}
\end{figure*}

\clearpage
\newpage
\centering
{\normalsize
\begin{longtable}{cccccccp{0.6in}p{1.2in}} 
\kill
\caption{NuSTAR Observations of the Norma Arm Region} \\
\hline \hline
\T ObsID & \multicolumn{3}{c}{Pointing (J2000)} & Start Time & Exposure & SL Removal & \hspace{0.2in}SL & \hspace{0.4in}Other \\ \cline{2-4}
\T& R.A. ($^{\circ}$) & Dec. ($^{\circ}$)  & PA ($^{\circ}$) & (UT) & (ks) & (FPM) & \hspace{0.1in}Source & \hspace{0.2in} Contamination\\
\B(1) & (2) & (3) & (4) & (5) &(6) & (7) & \hspace{0.2in}(8) & \hspace{0.5in}(9) \\
\hline
\endfirsthead
\endhead
\endfoot
\endlastfoot
\multicolumn{2}{l}{\textbf{\textit{Wide Shallow Survey}}}&&&&&&&\T \\ 
\multicolumn{2}{l}{\hspace{0.2in}\textit{First mini-survey}} \\ 
40014001001&248.4829&-47.7204&160.1494&2013-02-24 01:46&18.4&&&Ghost rays from 4Ub in AB\\
40014002001&248.3623&-47.6444&160.1471&2013-02-24 11:31&19.5&&&Ghost rays from 4Ub in AB\\
40014003001&248.2407&-47.5669&160.1266&2013-02-21 20:31&20.8&&&Ghost rays from 4Ub in AB\\
40014004001&248.5977&-47.6374&160.1231&2013-02-22 07:46&19.5&&&Ghost rays from 4Ub in AB\\
40014005001&248.4775&-47.5622&160.1304&2013-02-22 17:31&21.3&&&Ghost rays from 4Ub in AB\\
40014006001&248.3529&-47.4868&160.1393&2013-02-23 04:46&18.9&&&Ghost rays from 4Ub in AB\\
40014007001&248.7099&-47.5554&160.1350&2013-02-23 14:31&22.7&&&Ghost rays from 4Ub in AB\\
40014008002&248.5845&-47.4826&160.1196&2013-02-20 23:32&16.6&&&Ghost rays from 4Ub in AB\\
40014009001&248.4670&-47.4038&160.1198&2013-02-21 10:46&14.7&&&Ghost rays from 4Ub in AB\\ 
\multicolumn{2}{l}{\hspace{0.2in}\textit{Later observations}} \\ 
40014011002&250.0712&-46.4909&280.7063&2013-06-20 00:06&21.5&AB&4Ua&\\
40014012001&250.0006&-46.4546&280.7266&2013-06-20 14:21&19.7&AB&4Ua&\\
40014013001&249.9200&-46.4004&281.4251&2013-06-21 03:16&20.5&A&4Ua&\\
40014014001&250.2358&-46.3989&285.7049&2013-06-21 17:46&16.4&&&1.5$'$ streak in AB\\
40014015001&250.0770&-46.3706&285.7091&2013-06-22 08:21&19.2&&&\\
40014016001&250.2620&-46.5441&285.6937&2013-06-23 21:21&19.6&AB&4Ua&\\
40014017001&250.1762&-46.5238&285.6774&2013-06-23 11:51&24.2&AB&4Ua&\\
40014018001&249.9326&-46.3469&286.8740&2013-06-24 00:51&23.8&&&\\
40014019001&250.1520&-46.3873&286.8743&2013-06-24 15:21&25.6&&&\\
40014021002&249.1106&-47.1553&168.0928&2014-03-09 21:56&29.1&AB&GX, 4Ua&\\
40014022001&249.0348&-46.9577&168.0985&2014-03-10 15:31&28.4&AB&GX, 4Ua&\\
40014023001&249.2029&-47.1072&168.1000&2014-03-11 07:41&28.8&AB&GX, 4Ua&\\
40014024001&248.8388&-46.9903&168.1169&2014-03-11 23:46&28.1&AB&GX, 4Ua&\\
40014025001&248.8796&-47.0734&168.1144&2014-03-12 17:36&29.1&AB&GX, 4Ua&\\
40014026001&249.1206&-46.9073&168.0941&2014-03-13 11:26&30.2&AB&GX, 4Ua&\\
40014027001&249.1610&-47.0090&168.1171&2014-03-14 03:31&30.2&AB&GX, 4Ua&\\
40014028002&249.0277&-47.1640&168.1544&2014-03-18 12:36&29.6&AB&GX, 4Ua&\\
40014029001&249.9367&-46.8984&168.2590&2014-03-19 04:41&29.2&AB&GX, 4Ua&\\
40014030001&250.2174&-46.7179&168.3050&2014-03-19 20:46&27.3&B&4Ua&\\
40014031001&250.5222&-46.7773&168.3521&2014-03-20 13:01&30.0&B&4Ua&\\
40014032001&250.4317&-46.7896&168.4181&2014-03-21 05:01&30.9&B&4Ua&\\
40014033002&250.4849&-46.6649&168.2038&2014-03-24 10:41&31.5&B&4Ua&\\
40014034001&250.2701&-46.8265&168.1849&2014-03-25 02:46&31.2&AB&GX, 4Ua&\\
40014035001&250.0454&-46.8714&168.1523&2014-03-25 18:56&39.2&AB&GX, 4Ua&\\
30001008002&249.8301&-46.6567&295.0558&2014-06-26 02:21&50.4&&&\\
30001012002&248.6712&-47.6364&171.9830&2013-03-23 08:31&16.3&A&GX&Ghost rays from 4Ub in AB\\
30001016002&248.5333&-47.3795&164.6452&2014-03-06 22:56&21.3&AB&GX, 4Ua&\\
30001017002&248.8967&-47.3836&210.3881&2014-05-12 21:31&49.0&AB&GX, 4Ua&\\
30001033002&249.4897&-46.9015&145.8254&2015-01-28 05:16&51.8&AB&GX, 4Ua&\\
30160001002&249.3137&-47.5723&267.7851&2015-06-11 14:46&49.4&AB&GX, 4Ua&\\
30160002002&248.9436&-47.5918&261.9558&2015-06-07 23:46&97.1&AB&GX, 4Ua&\\
30160003002&249.0412&-47.8404&244.4634&2015-05-31 11:11&76.7&AB&GX, 4Ua&\\
40001022002&249.5341&-47.2183&164.5577&2014-03-07 11:51&100.6&AB&GX, 4Ua&\\ \hline
\multicolumn{2}{l}{\textbf{\textit{Deep HESS Field}}} &&&&&&&\T\\
30002021002&250.1049&-46.5763&353.7407&2013-09-29 6:56&62.8&AB&GX, 4Ua&SL of unknown origin in A\\
30002021003&250.1324&-46.5412&353.7551&2013-09-30 16:31&20.8&A&4Ua&SL of unknown origin in A\\
30002021005&250.2036&-46.5095&161.2653&2014-02-38 23:16&99.5&AB&4Ua&\\
30002021007&250.2027&-46.5145&161.2702&2014-03-06 01:51&35.9&AB&4Ua&\\
30002021009&250.2175&-46.5088&166.7254&2014-03-14 21:21&32.5&AB&4Ua&\\
30002021011&250.2296&-46.5012&179.7925&2014-04-11 13:11&22.5&AB&4Ua&\\
30002021013&250.1923&-46.5268&227.3736&2014-05-25 01:56&21.6&&&\\
30002021015&250.1802&-46.5601&289.9801&2014-06-23 12:51&29.2&A&4Ua&\\
30002021017&250.1814&-46.5644&295.1336&2014-06-25 13:31&22.0&&&\\
30002021019&250.1913&-46.5447&295.1661&2014-06-28 01:20&19.5&&&\\
30002021021&250.1762&-46.5687&295.1179&2014-06-30 01:41&19.8&&&\\
30002021023&250.1892&-46.5586&311.5738&2014-07-11 02:21&22.1&&&\\
30002021025&250.1569&-46.5398&330.9082&2014-08-10 05:36&21.9&A&IGR, 4Ua&6$'$ streak in B\\
30002021027&250.1477&-46.5524&344.3607&2014-09-11 10:56&20.0&AB&GX, 4Ua&\\
30002021029&250.1247&-46.5392&356.5397&2014-10-11 01:01&22.1&AB&4Ua&\\
30002021031&250.1400&-46.5260&15.7729&2014-11-05 07:56&4.3&AB&4Ua&\\
30002021033&250.2058&-46.4950&129.7677&2015-01-08 04:46&4.2&&&\\
30002021034&250.2115&-46.4858&129.7237&2015-01-12 18:16&16.7&&&\\
30002031036&250.2188&-46.5018&154.4146&2015-02-16 02:41&31.8&AB&4Ua, 4Ub&\\
\hline \hline

\multicolumn{9}{p{6.0in}}{\T Notes: (4) Position angle (east of North). 

(7) Focal plane module(s) from which stray light background photons from sources in column~8 were removed. 

(8) Stray light background sources: GX = GX~340+0, 4Ua = 4U~1624-49, 4Ub =~1630-472, IGR = IGR~J16318-4848.  Although additional stray light from IGR~J16320-4751 was present in some of the first mini-survey observations, and stray light from 4U~1624-49 and GX~340+0 was present in observation 30001008002, this stray light background was not removed since real sources could be seen in the raw data residing in the stray light-contaminated regions.  The contamination in observations 30002021002A, 30002021003A, 30002021036B, and 30001012002 was so extensive that these observations were not included in our analysis. }
\label{tab:obs}
\end{longtable}}

\begin{table}
\begin{minipage}{\linewidth}
\centering
{\footnotesize
\begin{threeparttable}
\caption{Archival \textit{Chandra} observations used in this study}
\begin{tabular}{cccccc} \hline \hline
\T \textit{Chandra} & \multicolumn{2}{c}{Pointing (J2000)} &Start Time & Exposure & References \\
ObsID & R.A. ($^{\circ}$) & Dec. ($^{\circ}$) & (UT) & (ks) & \\ 
\B (1) & (2) & (3) & (4) & (5) & (6) \\
\hline
\T 7591&250.187126&-46.520108&2007-05-11 11:01&28.8 & Lemiere et al. 2009\\
11008&250.134287&-46.393394&2010-06-19 22:10&39.6 & Rahoui et al. 2014 \\
\hline
\multicolumn{3}{l}{\textbf{\textit{Norma Arm Region Chandra Survey (NARCS)}}}&&& Fornasini et al. 2014 \T \\ 
12507&250.373201&-46.662951&2011-06-06 10:15&18.8\\
12508&250.155011&-46.530604&2011-06-06 15:57&18.5\\
12509&249.937805&-46.397816&2011-06-06 21:22&19.4\\
12510&250.180190&-46.812896&2011-06-09 12:29&19.9\\
12511&249.961646&-46.681456&2011-06-17 11:15&19.3\\
12512&249.743370&-46.550407&2011-06-27 04:52&20.5\\
12513&249.984947&-46.965904&2011-06-27 11:00&20.2\\
12514&249.767582&-46.829470&2011-06-10 16:07&19.8\\
12515&249.550110&-46.695978&2011-06-10 22:04&19.5\\
12516&249.790838&-47.111874&2011-06-11 03:46&19.5\\
12517&249.572205&-46.978413&2011-06-11 09:28&19.5\\
12518&249.354673&-46.844540&2011-06-11 15:10&19.5\\
12519&249.594334&-47.262081&2011-06-13 04:25&19.3\\
12520&249.375577&-47.128273&2011-06-13 10:13&19.0\\
12521&249.157932&-46.994022&2011-06-13 15:46&19.0\\
12522&249.396933&-47.410725&2011-06-13 21:20&19.0\\
12523&249.178061&-47.276529&2011-06-14 02:53&19.0\\
12524&248.960334&-47.141940&2011-06-14 08:27&19.5\\
12525&249.198427&-47.559064&2011-06-14 14:08&19.5\\
12526&248.979417&-47.424468&2011-06-14 19:50&19.0\\
12527&248.761625&-47.289491&2011-06-15 19:36&19.3\\
12528&248.998831&-47.707016&2011-06-16 01:24&19.0\\
12529&248.779750&-47.572056&2011-06-16 06:58&19.0\\
12530&248.561776&-47.436667&2011-06-16 12:31&19.3\\
12531&248.798050&-47.854617&2011-06-16 18:09&19.5\\
12532&248.578823&-47.719259&2011-06-16 23:51&19.5\\
12533&248.360823&-47.583518&2011-06-17 05:32&19.5\\
 \hline \hline
\end{tabular}
\begin{tablenotes}[flushleft]
\item (6) References in which archival observations were previously presented and analyzed.
\end{tablenotes}
\label{tab:archivalchandra}
\end{threeparttable}}
\end{minipage}
\end{table}

\begin{table}
\begin{minipage}{\linewidth}
\centering
{\footnotesize
\begin{threeparttable}
\caption{\textit{Chandra} follow-up observations of \textit{NuSTAR} transients}
\begin{tabular}{ccccccc} \hline \hline
\T \textit{Chandra} & Source &\multicolumn{2}{c}{Pointing (J2000)} & Start Time & Exposure & Delay between \textit{NuSTAR} \& \textit{Chandra} \\
ObsID & No. & R.A. ($^{\circ}$) & Dec. ($^{\circ}$) & (UT) & (ks) & observations (days)\\
\B (1) & (2) & (3) & (4) & (5) & (6) & (7)  \\
\hline
\T 16170 & 19 & 250.315079 & -46.540562 & 2014-03-17 05:44 & 4.9 & 3 \\
16171 & 20 & 250.591644 & -46.716049 & 2014-10-20 06:31 & 4.9 & 210 \\
17242 & 25 & 248.999542 & -47.807671 & 2015-07-04 10:26 & 9.8 & 34 \\
\hline \hline
\end{tabular}
\begin{tablenotes}[flushleft]
\item Notes: (2) NNR source that triggered the \textit{Chandra} observation.  

\item (7) Time elapsed between \textit{NuSTAR} observation where source is detected and \textit{Chandra} follow-up observation.  These times vary significantly because some of these sources were obvious in the raw images while others required mosaicking and careful photometric analysis to determine that they were significant detections.  
\end{tablenotes}
\label{tab:chandra}
\end{threeparttable}}
\end{minipage}
\end{table}

\begin{table}
\centering
{\footnotesize
\begin{threeparttable}
\caption{Boresight Corrections}
\begin{tabular}{ccccc} \hline \hline
\T ObsID & Total shift & R.A. Shift & Dec. Shift & Reference Source\\
& ($''$) & ($''$) & ($''$) & (NARCS ID)\\
\B(1) & (2) & (3) & (4) & (5)\\
\hline
\T 30001008002A&5.5&-6.5&3.25&999\\
30001008002B&6.9&-0.1&6.7&999\\
30001033002A&5.6&-1.9&-5.4&750\\
30001033002B&3.2&-2.1&-2.8&750\\
30002021002B&4.2&6.0&0.0&1321\\
30002021003B&10.8&13.7&-5.2&1321\\
30002021005A&4.5&5.0&-2.9&1321\\
30002021005B&3.7&-5.3&0.5&1321\\
30002021007A&4.2&3.1&-3.6&1321\\
30002021007B&3.7&-5.4&-0.8&1321\\
30002021009A&1.7&-0.3&-1.7&1321\\
30002021009B&4.4&-6.1&1.2&1321\\
30002021011A&4.3&-4.4&3.1&1321\\
30002021011B&4.7&-6.9&0.3&1321\\
30002021013A&7.9&8.2&5.5&1321\\
30002021013B&6.1&1.8&6.0&1321\\
30002021015A&4.5&3.7&3.7&1321\\
30002021015B&6.0&0.6&5.9&1321\\
30002021017A&2.2&2.1&1.6&1321\\
30002021017B&2.9&4.1&0.6&1321\\
30002021019A&7.1&10.0&-1.6&1321\\
30002021019B&10.0&11.2&6.4&1321\\
30002021021A&1.8&-0.6&-1.8&1321\\
30002021021B&7.2&9.4&3.2&1321\\
30002021023A&1.2&1.8&0.1&1321\\
30002021023B&7.9&-6.2&-6.6&1321\\
30002021025A&7.7&11.2&0.0&1321\\
30002021025B&9.3&13.3&1.0&1321\\
30002021027A&0.6&0.9&0.1&1321\\
30002021027B&8.7&11.7&-3.2&1321\\
30002021029A&10.2&12.1&-5.9&1321\\
30002021029B&5.9&5.9&-4.3&1321\\
30002021031A&9.3&10.9&5.5&1321\\
30002021031B&7.2&0.8&7.2&1321\\
30002021033A&7.2&1.3&-7.1&1321\\
30002021033B&14.4&-20.6&-2.6&1321\\
30002021034A&10.5&-8.7&-8.7&1321\\
30002021034B&9.8&10.8&6.3&1321\\
30002021036A&5.7&-8.4&-0.1&1321\\
40001022002A&4.9&-5.9&-2.9&786\\
40001022002B&6.4&-9.5&0.3&786\\
40014017001A&9.0&6.9&-7.7&1321\\
40014017001B&7.7&6.6&6.2&1321\\
\hline \hline
\end{tabular}
\begin{tablenotes}[flushleft]
\item Notes: The 90\% confidence statistical uncertainties of the astrometric corrections are estimated to be $<2^{\prime\prime}$ for NARCS 999 and $5-6^{\prime\prime}$ for all other NARCS sources.

\item (2) Angular distance between original pointing and boresight corrected pointing.  

\item (5) NARCS ID of source used to determine astrometric correction.

\end{tablenotes}
\label{tab:shift}
\end{threeparttable}}
\end{table}
\newpage

\clearpage
\begin{landscape}
\setlength\LTleft{-0.5in}
\centering
{\normalsize
\begin{longtable}{ccccccccccccccl}
\kill
\caption{Source List} \\
\hline \hline
\T Src & R.A. & Dec. & $\ell$ & \textit{b} & Unc. & Source & NARCS & Offset & Exp. & No. Trials & Band & EEF & No. \\
No. & \multicolumn{2}{c}{(J2000 $^{\circ}$)} & \multicolumn{2}{c}{(J2000 $^{\circ}$)} & ($''$) & Name & ID & ($''$) & (ks) & (10$^X$) & (keV) & (\%) & Det. \\
\B(1) & (2) & (3) & (4) & (5) & (6) & (7) & (8) & (9) & (10) & (11) & (12) & (13) & (14) \\
\hline
\endfirsthead
\caption{Source List (continued)} \\
\hline
\T Src & R.A. & Dec. & $\ell$ & \textit{b} &Unc. & Source & NARCS & Offset & Exp. & No. Trials & Band & EEF & No.  \\
\B No. & \multicolumn{2}{c}{(J2000 $^{\circ}$)} & \multicolumn{2}{c}{(J2000 $^{\circ}$)} & ($''$) & Name & ID & ($''$) & (ks) & (10$^X$) & (keV) & (\%) & Det. \\
\hline
\endhead
\endfoot
\endlastfoot
\multicolumn{2}{l}{\textbf{\textit{Tier 1}}} &&&&&&&&&&&&\T\\
\T 1&248.5070&-47.3923&336.9119&0.2506&8&4U 1630-472&$-$&$-$&63&1596934.6&3-78&30&18\\
\M 2&249.7733&-46.7041&338.0014&0.0746&3$^{\star}$&IGR  J16393-4643&999&2&101&15406.9&3-78&30&18\\
\M 3&250.1813&-46.5272&338.3198&-0.0173&6$^{\star}$&CXOU J164043.5-463135&1321$^a$&2&1039&1180.9&3-78&30&17\\
\M 4&249.4627&-46.9299&337.6914&0.0821&6$^{\star}$&CXOU J163750.8-465545&750&3&96&141.2&3-10&30&10\\
\M 5&249.5112&-47.2327&337.4885&-0.1451&6$^{\star}$&CXOU J163802.6-471358&786&1&200&132.6&3-10&30&14\\
\M 6&248.4812&-47.6342&336.7224&0.0993&8&CXOU J163355.1-473804&78&4&43&92.3&3-78&30&13\\
\M 7&250.1214&-46.3929&338.3930&0.1026&8&CXOU J164029.5-462329&1278/9$^b$&7&215&77.6&3-10&30&6\\
\M 8&248.9483&-47.6217&336.9445&-0.1241&9&CXOU J163547.0-473739&365$^c$&22$^*$&94&64.9&3-78&30&12\\
\M 9&249.8060&-46.4027&338.2412&0.2586&8&CXOU J163912.9-462357&1024&13&87&45.4&3-10&30&12\\
\M 10&248.6407&-47.6439&336.7881&0.0138&8&NuSTAR J163433-4738.7&$-$&$-$&45$^{\dag}$&40.6&3-10&30&6\\
\M 11&250.1467&-46.4991&338.3251&0.0191&9&CXOU J164035.5-462951&1301&7&1123&34.9&3-10&30&10\\
\M 12&250.1143&-46.4226&338.3676&0.0865&9&CXOU J164027.8-462513&1276&9&654&31.1&3-10&30&10\\
\M 13&249.9911&-46.4329&338.3035&0.1432&9&CXOU J163957.8-462549&1181&8&208&28.4&3-78&30&12\\
\M 14&249.9943&-46.8584&337.9869&-0.1410&9&CXOU J163957.2-465126&1180&14&69&28.0&3-10&30&6\\
\M 15&250.3823&-46.5145&338.4208&-0.1127&9&CXOU J164130.8-463048&1379&10&39&27.7&3-10&30&6\\
\M 16&248.4639&-47.7762&336.6102&0.0115&10&CXOU J163350.9-474638&72&6&37&21.8&3-78&30&11\\
\M 17&249.9421&-46.4023&338.3039&0.1888&9&CXOU J163946.1-462359&1137&8&161&19.4&3-10&30&6\\
\M 18&248.3743&-47.5569&336.7301&0.2048&9&CXOU J163329.5-473332&38&9&37&18.4&3-78&30&6\\
\M 19&250.3176&-46.5373&338.3743&-0.0943&9&NuSTAR J164116-4632.2&$-$&13&424$^{\dag}$&15.7&3-10&30&5\\
\M 20&250.5927&-46.7153&338.3652&-0.3538&9&NuSTAR J164222-4642.9&$-$&4&123&14.9&3-10&30&8\\
\M 21&248.9882&-47.3188&337.1864&0.0601&12&CXOU J163555.4-471907&402/4$^d$&18$^*$&47&14.8&3-10&30&3\\
\M 22&250.1156&-46.8060&338.0812&-0.1684&10&CXOU J164027.6-464814&1273&7&66&13.4&3-10&20&4\\
\M 23&249.0619&-46.8736&337.5493&0.3228&11&CXOU J163614.2-465222&454&7&86&12.8&3-10&30&6\\
\M 24&248.9650&-47.5894&336.9760&-0.1106&11&CXOU J163551.8-473523&391&3&187&12.8&3-78&30&4\\
\M 25&249.0020&-47.8078&336.8313&-0.2763&10&NuSTAR J163600-4748.4&$-$&6&77&11.8&3-78&30&5\\
\M 26&249.8911&-46.9254&337.8899&-0.1330&9&CXOU J163933.2-465530&1090&7&58&10.8&3-78&30&9\\
\M 27&250.1304&-46.8142&338.0817&-0.1814&13&CXOU J164031.0-464845&1291&6&121&9.9&10-20&30&3\\
\M 28&249.2382&-46.8161&337.6730&0.2722&11&CXOU J163657.1-464903&585&6&26&8.5&3-10&20&2\\
\multicolumn{2}{l}{\textbf{\textit{Tier 2}}} &&&&&&&&&&&&\T\\
\M 29&250.0101&-46.5335&338.23700&0.0666&$-$&CXOU J164002.4-463200&1203&$-$&212&8.7&3-10&30&0\\
\M 30&250.5191&-46.7281&338.32231&-0.3243&$-$&CXOU J164204.5-464341&1408&$-$&177&7.2&3-10&30&0\\
\M 31&248.3784&-47.4266&336.82764&0.2912&$-$&CXOU J163330.8-472535&40&$-$&11&6.0&3-78&30&0\\
\M 32&248.6447&-47.2967&337.04544&0.2468&$-$&CXOU J163434.7-471748&139&$-$&20&5.6&10-20&30&0\\
\M 33&250.0287&-46.4872&338.28012&0.0878&$-$&CXOU J164006.8-462913&1216&$-$&434&5.3&3-10&30&0\\
\M 34&249.8351&-46.8352&337.93184&-0.0443&$-$&CXOU J163920.4-465006&1039&$-$&29&5.1&3-10&30&0\\
\M 35&248.9010&-47.0967&337.31056&0.2536&$-$&CXOU J163536.2-470548&325&$-$&115&4.6&10-20&20&0\\
\M 36&250.3453&-46.7582&338.22112&-0.2546&$-$&CXOU J164122.8-464529&1374&$-$&178&4.4&3-10&30&0\\
\M 37&248.9518&-47.3590&337.14012&0.0512&$-$&CXOU J163548.4-472132&373&$-$&89&3.6&40-78&15&0\\
\M \B 38&248.4062&-47.4119&336.85116&0.2874&$-$&CXOU J163337.4-472442&52&$-$&21&2.3&3-10&30&0\\
\hline\hline
\multicolumn{15}{p{7.0in}}{\T Notes: (1) \textit{NuSTAR} Norma Region (NNR) source ID. 

(2)-(5) Right ascension, declination, Galactic longitude, and Galactic latitude of source determined from centroid algorithm for tier 1 sources and adopting \textit{Chandra} positions from \citet{fornasini14} for tier 2 sources. 

(6) 90\% confidence positional uncertainty, including statistical and systematic uncertainties summed in quadrature.  In most cases, the 90\% confidence systematic uncertainty is 8$^{\prime\prime}$; however for sources that were used to derive astrometric corrections ($^{\star}$), the 90\% systematic uncertainty is estimated based on simulations (2$^{\prime\prime}$ for NARCS 999 and 6$^{\prime\prime}$ for all sources marked with a $^{\star}$).  Uncertainties for tier 2 sources are not provided since the positions of these sources are simply set to the \textit{Chandra} positions.  

(7) NARCS source name or other commonly used name for source.  For \textit{NuSTAR} discoveries, a \textit{NuSTAR} name is provided.  

(8) NARCS catalog ID number. 

(9) Angular distance between the source positions in \textit{NuSTAR} and \textit{Chandra} observations.  For tier 2 sources, no offset is shown since the \textit{Chandra}-determined position in adopted for the \textit{NuSTAR} analysis. 

(10) Total \textit{NuSTAR} exposure, including both modules (FPMA and FPMB) and all observations used in measuring photometric properties of the source (see \S~\ref{sec:phot} for details). 

(11) The maximum value from the trial maps at the location of the source; this value is the number of random trials required to produce the observed counts from a random background fluctuation.  For extended sources, this is the maximum trial map value within 30$^{\prime\prime}$ of the listed source location.  

(12) The energy band of the trial map in which the maximum trial value for the source is measured. 

(13) The PSF enclosed energy fraction of the trial map in which the maximum trial value for the source is measured.

(14) The total number of trial maps in which the source exceeds the detection threshold.  There are 18 trial maps in total, using six different energy bands and three different PSF enclosure fractions.  

(15) Tier 1 sources are those detected in at least two trial maps.  Tier 2 sources are NARCS sources with 2--10 keV fluxes $>6\times10^{-6}$ ph cm$^{-2}$ s$^{-1}$ which do not meet the NuSTAR detection threshold requirements but have S/N$>3$ in the 3--10, 10--20, or 3--40 keV bands (S/N values can be found in Table~\ref{tab:phot}).

(a) Point source embedded in extended emission.  We treat it as a point source and leave the detailed analysis of the extended emission to Gotthelf et al. (2014).  

(b) Blend of two \textit{Chandra} sources, which are also blended in NARCS but resolved in \textit{Chandra} ObsID 11008 \citep{rahoui14}. 

(c) Extended source. 

(d) In \textit{Chandra}, point source 402 is resolved within extended emission (404), but in \textit{NuSTAR} the two are not distinguishable so we treat it as extended source.  

($^*$) These large offsets are due to the fact that the positions for these extended sources were determined by eye in NARCS.  

($^{\dag}$) For these transients sources, the exposure times listed only include observations in which the source was detected at $>2\sigma$ level.
}
\label{tab:srclist}
\end{longtable}}
\newpage

\clearpage
\centering
{\normalsize
\begin{longtable}{ccccccccccccp{0.25in}p{0.25in}c}
\kill
\caption{Photometry} \\
\hline \hline
\T Source & S/N & S/N & S/N & Net Counts & \multicolumn{2}{c}{Ph. Flux (10$^{-6}$ cm$^{-2}$ s$^{-1}$)} & 
\multicolumn{2}{c}{En. Flux (10$^{-14}$ erg cm$^{-2}$ s$^{-1}$)} & Hardness & $E_{50}$ & $QR$ & \multicolumn{2}{c}{Var. Flag} & Aperture \\ \cline{6-7} \cline{13-14} \cline{8-9}
\T No. & 3--40 keV & 3--10 keV & 10--20 keV & 3--40 keV & 3--10 keV & 10--20 keV & 3--10 keV & 10--20 keV & Ratio & (keV) &  &\textit{NuST.}&\textit{Chan.}& Flag \\
\B (1) & (2) & (3) & (4) & (5) & (6) & (7) & (8) & (9) & (10) & (11) & (12) & (13) & (14) & (15) \\
\hline
\endfirsthead
\caption{Photometry (continued)} \\
\hline
\T Source & S/N & S/N & S/N & Net Counts & \multicolumn{2}{c}{Ph. Flux (10$^{-6}$ cm$^{-2}$ s$^{-1}$)} & \multicolumn{2}{c}{En. Flux (10$^{-14}$ erg cm$^{-2}$ s$^{-1}$)} & Hardness & $E_{50}$ & $QR$ & \multicolumn{2}{c}{Var. Flag} & Aperture \\ \cline{6-7} \cline{13-14} \cline{8-9}
\T\B No. & 3--40 keV & 3--10 keV & 10--20 keV & 3--40 keV & 3--10 keV & 10--20 keV & 3--10 keV & 10--20 keV & Ratio & (keV) &  &\textit{NuST.}&\textit{Chan.}& Flag \\
\hline
\endhead
\endfoot
\endlastfoot
\multicolumn{2}{l}{\textbf{\textit{Tier 1}}} &&&&&&&&&&&&&\T\\
\U\multirow{2}{*}{1}&134534.5&142889.4&15742.0&3214900$\pm$18000&598180$\pm$340&19112$\pm$55&473890$\pm$270&38210$\pm$110&-0.9246&5.3245&1.0334&\multirow{2}{*}{l}&\multirow{2}{*}{$-$}&\multirow{2}{*}{pcm} \\
\M&130019.1&138317.1&15422.4&4079200$\pm$2000&603320$\pm$300&19060$\pm$49&477130$\pm$240&38119$\pm$100&$\pm$0.0008&$\pm$0.0006&$\pm$0.0004&&&\\
\U\multirow{2}{*}{2}&616.4&350.0&648.6&37360$\pm$200&1634$\pm$14&1710$\pm$14&1748$\pm$14&3954$\pm$32&-0.112&9.83&1.077&\multirow{2}{*}{sp$^*$}&\multirow{2}{*}{slp}&\multirow{2}{*}{pcm}\\
\M&581.4&331.0&614.8&46720$\pm$240&1623$\pm$13&1694$\pm$13&1737$\pm$13&3914$\pm$29&$\pm$0.006&$\pm$0.03&$\pm$+0.006&&&\\
\U\multirow{2}{*}{3}&144.2&128.5&84.4&9590$\pm$120&50.1$\pm$0.8&21.5$\pm$0.5&46.4$\pm$0.7&48.7$\pm$1.2&\multirow{2}{*}{-0.41$\pm$0.01}&\multirow{2}{*}{8.0$\pm$0.1}&\multirow{2}{*}{0.90$\pm$+0.02}&\multirow{2}{*}{sp$^*$}&\multirow{2}{*}{}&\multirow{2}{*}{pcm}\\
\M&153.3&140.4&87.4&13550$\pm$150&57.9$\pm$0.8&23.3$\pm$0.5&53.3$\pm$0.7&52.7$\pm$1.2&&&&&&\\
\U\multirow{2}{*}{4}&21.3&24.6&6.5&556$\pm$32&49.1$\pm$2.7&7.6$\pm$1.3&40.3$\pm$2.2&18.8$\pm$3.2&\multirow{2}{*}{-0.72$\pm$0.06}&\multirow{2}{*}{6.4$\pm$0.1}&\multirow{2}{*}{0.92$\pm$+0.10}&\multirow{2}{*}{l}&\multirow{2}{*}{slp}&\multirow{2}{*}{pcm}\\
\M&20.3&23.3&5.9&723$\pm$40&50.8$\pm$2.7&7.8$\pm$1.4&41.8$\pm$2.2&19.7$\pm$3.4&&&&&&\\
\U\multirow{2}{*}{5}&23.0&22.9&9.6&842$\pm$42&23.6$\pm$1.3&6.4$\pm$0.8&21.6$\pm$1.2&13.7$\pm$2.0&\multirow{2}{*}{-0.55$\pm$0.06}&\multirow{2}{*}{7.8$\pm$0.3}&\multirow{2}{*}{0.93$\pm$+0.06}&\multirow{2}{*}{}&\multirow{2}{*}{}&\multirow{2}{*}{p}\\
\M&21.4&21.3&9.1&1087$\pm$55&24.0$\pm$1.4&6.9$\pm$0.9&22.0$\pm$1.2&14.7$\pm$2.2&&&&&&\\
\U\multirow{2}{*}{6}&14.3&12.8&6.9&359$\pm$29&77.8$\pm$7.7&18.0$\pm$3.2&67.7$\pm$6.2&40.0$\pm$7.1&\multirow{2}{*}{-0.63$\pm$0.09}&\multirow{2}{*}{6.5$\pm$0.2}&\multirow{2}{*}{0.91$\pm$+0.14}&\multirow{2}{*}{}&\multirow{2}{*}{}&\multirow{2}{*}{pc}\\
\M&13.4&11.7&7.3&464$\pm$38&76.0$\pm$8.0&21.2$\pm$3.3&67.6$\pm$6.5&46.4$\pm$7.5&&&&&&\\
\U\multirow{2}{*}{7}&17.5&20.6&1.7&621$\pm$40&37.4$\pm$2.2&1.1$\pm$0.8&29.6$\pm$1.7&1.7$^{+2.0}_{-1.7}$&\multirow{2}{*}{-0.92$\pm$0.08}&\multirow{2}{*}{5.5$\pm$0.2}&\multirow{2}{*}{0.90$\pm$+0.06}&\multirow{2}{*}{}&\multirow{2}{*}{}&\multirow{2}{*}{pc}\\
\M&17.0&19.6&3.0&835$\pm$53&38.1$\pm$2.2&2.3$\pm$1.0&30.6$\pm$1.8&4.4$\pm$2.2&&&&&&\\
\U\multirow{2}{*}{8}&24.9&22.3&14.4&884$\pm$41&40.5$\pm$2.3&17.0$\pm$1.4&37.6$\pm$2.0&37.4$\pm$3.3&\multirow{2}{*}{-0.41$\pm$0.05}&\multirow{2}{*}{8.0$\pm$0.2}&\multirow{2}{*}{0.90$\pm$+0.06}&\multirow{2}{*}{}&\multirow{2}{*}{}&\multirow{2}{*}{e}\\
\M&21.9&20.0&12.5&1083$\pm$52&44.9$\pm$2.5&17.6$\pm$1.6&40.7$\pm$2.2&38.3$\pm$3.7&&&&&&\\
\U\multirow{2}{*}{9}&13.4&13.3&7.1&303$\pm$26&33.9$\pm$3.2&14.1$\pm$2.5&32.1$\pm$3.0&30.8$\pm$5.6&\multirow{2}{*}{-0.47$\pm$0.09}&\multirow{2}{*}{7.5$\pm$0.4}&\multirow{2}{*}{1.02$\pm$+0.11}&\multirow{2}{*}{}&\multirow{2}{*}{}&\multirow{2}{*}{p}\\
\M&11.9&12.5&5.8&371$\pm$34&37.1$\pm$3.4&12.9$\pm$2.5&33.4$\pm$3.0&28.5$\pm$5.9&&&&&&\\
\U\multirow{2}{*}{10}&9.7&10.6&1.7&240$\pm$27&84.0$\pm$9.1&4.0$^{+3.1}_{-2.8}$&67.3$\pm$7.1&8.0$^{+7.0}_{-6.3}$&\multirow{2}{*}{-0.89$^{+0.14}_{-0.11}$}&\multirow{2}{*}{5.6$\pm$0.3}&\multirow{2}{*}{0.83$\pm$+0.11}&\multirow{2}{*}{l}&\multirow{2}{*}{$-$}&\multirow{2}{*}{p}\\
\M&6.5&6.9&1.6&220$\pm$35&56.2$\pm$9.1&3.8$\pm$3.1&46.5$\pm$7.1&6.9$\pm$6.9&&&&&&\\
\U\multirow{2}{*}{11}&17.1&18.8&6.8&1310$\pm$81&9.8$\pm$0.6&1.9$\pm$0.3&8.3$\pm$0.5&4.1$\pm$0.8&\multirow{2}{*}{-0.64$\pm$0.07}&\multirow{2}{*}{6.4$\pm$0.1}&\multirow{2}{*}{0.92$\pm$+0.08}&\multirow{2}{*}{l}&\multirow{2}{*}{}&\multirow{2}{*}{pcm}\\
\M&17.1&19.0&6.6&1830$\pm$110&10.9$\pm$0.6&2.1$\pm$0.4&9.4$\pm$0.5&4.6$\pm$0.8&&&&&&\\
\U\multirow{2}{*}{12}&12.6&13.9&5.0&687$\pm$58&11.1$\pm$0.9&2.2$\pm$0.5&9.7$\pm$0.8&4.5$\pm$1.2&\multirow{2}{*}{-0.65$\pm$0.09}&\multirow{2}{*}{6.6$\pm$0.2}&\multirow{2}{*}{1.06$\pm$+0.15}&\multirow{2}{*}{}&\multirow{2}{*}{}&\multirow{2}{*}{pcm}\\
\M&12.2&13.7&5.0&929$\pm$79&12.1$\pm$1.0&2.5$\pm$0.6&10.5$\pm$0.8&5.3$\pm$1.4&&&&&&\\
\U\multirow{2}{*}{13}&10.5&8.6&6.6&339$\pm$35&10.1$\pm$1.5&5.8$\pm$1.0&9.7$\pm$1.3&13.4$\pm$2.4&\multirow{2}{*}{-0.34$\pm$0.11}&\multirow{2}{*}{8.9$\pm$0.7}&\multirow{2}{*}{0.95$\pm$+0.11}&\multirow{2}{*}{}&\multirow{2}{*}{}&\multirow{2}{*}{p}\\
\M&9.3&7.4&6.1&418$\pm$47&9.5$\pm$1.6&5.9$\pm$1.0&9.2$\pm$1.3&13.5$\pm$2.5&&&&&&\\
\U\multirow{2}{*}{14}&7.7&9.9&0.9&159$\pm$23&20.9$\pm$2.5&$<$3.6&17.2$\pm$2.1&$<$8.8&\multirow{2}{*}{$>$-1}&\multirow{2}{*}{5.7$\pm$0.4}&\multirow{2}{*}{1.11$\pm$+0.19}&\multirow{2}{*}{}&\multirow{2}{*}{}&\multirow{2}{*}{p}\\
\M&6.5&9.3&0.3&187$\pm$30&21.6$\pm$2.5&$<$3.2&17.7$\pm$2.1&$<$7.8&&&&&&\\
\U\multirow{2}{*}{15}&6.0&7.8&0.6&89$\pm$16&28.6$\pm$4.4&$<$3.9&23.5$\pm$3.6&$<$8.5&\multirow{2}{*}{$>$-1}&\multirow{2}{*}{5.6$\pm$0.6}&\multirow{2}{*}{0.87$\pm$+0.13}&\multirow{2}{*}{s}&\multirow{2}{*}{}&\multirow{2}{*}{p}\\
\M&6.0&7.5&0.3&125$\pm$22&31.2$\pm$4.6&$<$3.9&24.7$\pm$3.7&$<$9.0&&&&&&\\
\U\multirow{2}{*}{16}&9.6&8.5&4.2&287$\pm$32&60.8$\pm$8.9&10.2$^{+3.0}_{-2.8}$&54.9$\pm$7.1&22.5$^{+6.7}_{-6.1}$&\multirow{2}{*}{-0.71$\pm$0.14}&\multirow{2}{*}{6.4$\pm$0.3}&\multirow{2}{*}{0.85$\pm$+0.10}&\multirow{2}{*}{}&\multirow{2}{*}{}&\multirow{2}{*}{p}\\
\M&9.4&8.5&3.4&393$\pm$44&65.7$\pm$9.6&8.6$\pm$2.8&60.2$\pm$7.7&18.7$\pm$6.3&&&&&&\\
\U\multirow{2}{*}{17}&7.8&8.1&3.0&215$\pm$30&9.0$\pm$1.5&2.5$\pm$1.0&9.2$\pm$1.3&5.0$\pm$2.3&\multirow{2}{*}{-0.62$\pm$0.15}&\multirow{2}{*}{7.3$\pm$0.5}&\multirow{2}{*}{1.13$\pm$+0.22}&\multirow{2}{*}{}&\multirow{2}{*}{}&\multirow{2}{*}{p}\\
\M&7.5&7.3&3.9&292$\pm$40&9.2$\pm$1.7&3.8$\pm$1.1&9.3$\pm$1.4&8.3$\pm$2.6&&&&&&\\
\U\multirow{2}{*}{18}&6.3&6.6&1.9&134$\pm$23&47.7$\pm$8.0&3.7$^{+3.1}_{-2.7}$&38.1$\pm$6.2&5.8$^{+6.6}_{-5.8}$&\multirow{2}{*}{-0.78$^{+0.21}_{-0.20}$}&\multirow{2}{*}{5.9$\pm$0.7}&\multirow{2}{*}{0.54$\pm$+0.09}&\multirow{2}{*}{}&\multirow{2}{*}{s}&\multirow{2}{*}{pc}\\
\M&5.4&5.8&1.5&159$\pm$30&47.8$\pm$8.5&3.3$^{+3.3}_{-3.1}$&37.5$\pm$6.6&5.6$^{+7.4}_{-5.6}$&&&&&&\\
&&&&&&&&&&&&&&\\
\U\multirow{2}{*}{19}&10.3&11.7&2.7&399$\pm$41&11.0$\pm$1.1&1.6$\pm$0.6&9.5$\pm$0.9&3.7$\pm$1.5&\multirow{2}{*}{-0.77$\pm$0.12}&\multirow{2}{*}{6.6$\pm$0.2}&\multirow{2}{*}{1.09$\pm$+0.18}&\multirow{2}{*}{l}&\multirow{2}{*}{$-$}&\multirow{2}{*}{p}\\
\M&9.0&10.5&2.0&487$\pm$56&11.1$\pm$1.2&1.3$\pm$0.7&9.4$\pm$1.0&3.1$\pm$1.6&&&&&&\\
\U\multirow{2}{*}{20}&5.9&6.7&3.3&126$\pm$23&10.0$\pm$1.9&4.5$^{+1.5}_{-1.3}$&9.4$\pm$1.6&11.7$^{+3.6}_{-3.3}$&\multirow{2}{*}{-0.53$\pm$0.17}&\multirow{2}{*}{6.8$\pm$0.6}&\multirow{2}{*}{1.27$\pm$+0.39}&\multirow{2}{*}{l}&\multirow{2}{*}{$-$}&\multirow{2}{*}{p}\\
\M&6.4&7.1&3.5&191$\pm$31&11.8$\pm$2.0&5.0$\pm$1.4&11.1$\pm$1.7&12.4$\pm$3.3&&&&&&\\
\U\multirow{2}{*}{21}&13.4&13.5&6.1&312$\pm$26&46.4$\pm$4.1&12.1$\pm$2.4&39.9$\pm$3.5&26.0$\pm$5.4&\multirow{2}{*}{-0.58$\pm$0.09}&\multirow{2}{*}{6.7$\pm$0.3}&\multirow{2}{*}{0.79$\pm$+0.11}&\multirow{2}{*}{}&\multirow{2}{*}{}&\multirow{2}{*}{e}\\
\M&12.3&12.7&5.5&408$\pm$35&52.3$\pm$4.5&13.4$\pm$2.7&45.4$\pm$3.8&29.9$\pm$6.3&&&&&&\\


\U\multirow{2}{*}{22}&6.0&7.4&1.4&96$\pm$18&17.5$\pm$2.7&2.1$^{+2.2}_{-2.0}$&15.0$\pm$2.3&4.3$^{+5.2}_{-4.3}$&\multirow{2}{*}{-0.75$^{+0.22}_{-0.21}$}&\multirow{2}{*}{6.9$\pm$0.5}&\multirow{2}{*}{1.30$\pm$+0.43}&\multirow{2}{*}{}&\multirow{2}{*}{}&\multirow{2}{*}{p}\\
\M&5.9&7.6&1.4&132$\pm$23&20.0$\pm$2.9&2.3$\pm$2.2&17.3$\pm$2.4&4.3$^{+5.1}_{-4.3}$&&&&&&\\
\U\multirow{2}{*}
{23}&6.0&5.9&2.7&108$\pm$20&8.4$\pm$1.8&2.6$^{+1.2}_{-1.1}$&7.6$\pm$1.5&6.0$^{+2.8}_{-2.5}$&\multirow{2}{*}{-0.55$\pm$0.20}&\multirow{2}{*}{7.4$\pm$0.9}&\multirow{2}{*}{0.91$\pm$+0.30}&\multirow{2}{*}{}&\multirow{2}{*}{}&\multirow{2}{*}{p}\\
\M&4.8&4.9&2.2&122$\pm$26&7.5$\pm$1.9&2.1$\pm$1.1&7.1$\pm$1.6&4.4$\pm$2.6&&&&&&\\
\U\multirow{2}{*}{24}&6.7&5.0&4.0&198$\pm$31&6.8$\pm$1.2&2.3$\pm$0.6&5.3$\pm$1.0&4.7$\pm$1.5&\multirow{2}{*}{-0.37$\pm$0.18}&\multirow{2}{*}{9.0$\pm$2.1}&\multirow{2}{*}{0.35$\pm$+0.10}&\multirow{2}{*}{}&\multirow{2}{*}{}&\multirow{2}{*}{pcm}\\
\M&5.4&3.5&3.0&222$\pm$42&5.9$\pm$1.3&1.9$\pm$0.7&4.6$\pm$1.0&3.7$\pm$1.6&&&&&&\\
\U\multirow{2}{*}{25}&6.0&5.8&1.4&98$\pm$18&6.4$\pm$1.5&1.4$^{+0.9}_{-0.8}$&6.4$\pm$1.3&3.8$^{+2.4}_{-2.1}$&\multirow{2}{*}{-0.74$^{+0.25}_{-0.24}$}&\multirow{2}{*}{8.1$\pm$0.9}&\multirow{2}{*}{0.60$\pm$+0.25}&\multirow{2}{*}{l}&\multirow{2}{*}{$-$}&\multirow{2}{*}{p}\\
\M&6.2&5.5&2.0&144$\pm$24&6.8$\pm$1.6&1.8$^{+1.0}_{-0.9}$&6.7$\pm$1.3&4.6$^{+2.4}_{-2.2}$&&&&&&\\
\U\multirow{2}{*}{26}&6.0&5.7&4.3&107$\pm$19&11.5$\pm$2.2&6.6$\pm$1.7&9.5$\pm$1.8&16.6$\pm$4.1&\multirow{2}{*}{-0.25$\pm$0.16}&\multirow{2}{*}{8.3$\pm$0.9}&\multirow{2}{*}{0.55$\pm$+0.16}&\multirow{2}{*}{}&\multirow{2}{*}{}&\multirow{2}{*}{pcm}\\
\M&6.1&6.1&4.1&152$\pm$26&13.2$\pm$2.3&6.4$\pm$1.7&11.5$\pm$2.0&14.8$\pm$4.1&&&&&&\\
\U\multirow{2}{*}{27}&7.3&6.0&5.3&179$\pm$26&11.0$\pm$2.1&8.5$\pm$1.8&10.0$\pm$1.8&20.3$\pm$4.3&\multirow{2}{*}{-0.20$\pm$0.14}&\multirow{2}{*}{8.3$\pm$1.1}&\multirow{2}{*}{0.75$\pm$+0.18}&\multirow{2}{*}{l}&\multirow{2}{*}{}&\multirow{2}{*}{p}\\
\M&7.4&6.5&5.0&252$\pm$35&13.9$\pm$2.3&9.1$\pm$1.9&12.4$\pm$2.0&22.5$\pm$4.6&&&&&&\\
\U\multirow{2}{*}{28}&3.5&2.7&2.7&32$^{+11}_{-10}$&15.3$^{+7.0}_{-5.9}$&11.6$^{+6.1}_{-4.9}$&11.9$^{+5.7}_{-4.9}$&28.7$^{+\
15.3}_{-12.2}$&\multirow{2}{*}{-0.17$^{+0.36}_{-0.30}$}&\multirow{2}{*}{10.3$\pm$4.1}&\multirow{2}{*}{0.42$\pm$+0.26}&\multirow{2}{*}{l}&\multirow{2}{*}{}&\multirow{2}{*}{p}\\
\M&3.0&2.1&2.2&37$\pm$13&11.5$^{+6.6}_{-5.8}$&10.4$^{+5.9}_{-4.9}$&9.4$^{+5.6}_{-4.9}$&25.7$^{+14.8}_{-12.2}$&&&&&&\\
&&&&&&&&&&&&&&\\
\hline
\multicolumn{2}{l}{\textbf{\textit{Tier 2}}} &&&&&&&&&&&&&\T\\
\U\multirow{2}{*}{29}&5.5&6.0&2.3&169$\pm$32&9.1$\pm$1.7&2.3$\pm$1.1&8.1$\pm$1.4&5.0$\pm$2.7&\multirow{2}{*}{-0.60$\pm$0.20}&\multirow{2}{*}{7.5$\pm$0.8}&\multirow{2}{*}{0.94$\pm$+0.30}&\multirow{2}{*}{l}&\multirow{2}{*}{}&\multirow{2}{*}{pcm}\\
\M&4.7&5.5&3.0&201$\pm$43&9.0$\pm$1.8&3.5$\pm$1.3&8.2$\pm$1.5&7.5$\pm$3.0&&&&&&\\
\U\multirow{2}{*}{30}&4.7&6.1&0.9&128$\pm$28&6.6$\pm$1.3&$<$2.4&6.1$\pm$1.1&$<$15.9&\multirow{2}{*}{$>$-1}&\multirow{2}{*}{6.3$\pm$0.6}&\multirow{2}{*}{0.93$\pm$+0.41}&\multirow{2}{*}{}&\multirow{2}{*}{}&\multirow{2}{*}{pcm}\\
\M&3.9&5.8&0.8&147$\pm$39&7.9$\pm$1.5&$<$2.5&6.6$\pm$1.2&$<$6.2&&&&&&\\
\U\multirow{2}{*}{31}&4.6&3.9&1.5&29$^{+9}_{-8}$&37.7$^{+15.8}_{-13.3}$&10.4$^{+10.2}_{-7.5}$&35.3$^{+13.7}_{-11.6}$&23.7$\
^{+23.5}_{-17.2}$&\multirow{2}{*}{-0.61$^{+0.42}_{-0.34}$}&\multirow{2}{*}{7.0$\pm$2.3}&\multirow{2}{*}{0.67$\pm$+0.44}&\multirow{2}{*}{}&\multirow{2}{*}{}&\multirow{2}{*}{pc}\\
\M&4.5&3.7&1.7&38$^{+11}_{-10}$&29.9$^{+13.9}_{-12.2}$&12.3$^{+9.7}_{-7.7}$&33.1$^{+12.6}_{-11.0}$&28.1$^{+22.4}_{-17.6}$&&&&&&\\
\U\multirow{2}{*}{32}&2.5&1.9&1.9&18$^{+9}_{-8}$&9.0$^{+6.4}_{-5.3}$&8.5$^{+6.2}_{-4.9}$&8.4$^{+5.6}_{-4.6}$&19.1$^{+14.6}
_{-11.5}$&\multirow{2}{*}{-0.11$^{+0.49}_{-0.39}$}&\multirow{2}{*}{10.1$\pm$3.8}&\multirow{2}{*}{0.68$^{+0.56}_{-0.68}$}&\multirow{2}{*}{}&\multirow{2}{*}{}&\multirow{2}{*}{p}\\
\M&3.0&1.6&2.3&32$^{+12}_{-11}$&10.0$^{+6.8}_{-5.8}$&11.7$^{+6.5}_{-5.4}$&8.1$^{+5.7}_{-4.8}$&26.4$^{+15.3}_{-12.6}$&&&&&&\\
\U\multirow{2}{*}{33}&2.5&3.4&0.5&113$\pm$45&3.2$\pm$1.3&$<$1.6&3.2$\pm$1.0&$<$17.4&\multirow{2}{*}{$>$-1}&\multirow{2}{*}{6.6$\pm$0.4}&\multirow{2}{*}{1.62$\pm$+0.57}&\multirow{2}{*}{}&\multirow{2}{*}{l}&\multirow{2}{*}{p}\\
\M&2.0&2.8&0.0&125$\pm$63&2.9$\pm$1.4&$<$1.3&3.0$\pm$1.1&$<$3.1&&&&&&\\
\U\multirow{2}{*}{34}&3.1&3.3&1.5&46$\pm$15&14.6$\pm$4.7&5.1$^{+4.2}_{-3.7}$&12.6$\pm$4.1&11.1$^{+10.0}_{-8.9}$&\multirow{2}{*}{-0.49$^{+0.34}_{-0.33}$}&\multirow{2}{*}{6.2$\pm$1.4}&\multirow{2}{*}{0.83$\pm$+0.38}&\multirow{2}{*}{}&\multirow{2}{*}{}&\multirow{2}{*}{p}\\
\M&2.4&3.2&0.5&49$\pm$21&17.4$\pm$5.2&$<$7.9&13.9$\pm$4.5&$<$18.7&&&&&&\\
\U\multirow{2}{*}{35}&4.1&3.6&3.3&85$\pm$22&5.4$\pm$1.6&3.4$^{+1.2}_{-1.1}$&4.4$\pm$1.3&8.3$^{+2.8}_{-2.5}$&\multirow{2}{*}{-0.25$\pm$0.23}&\multirow{2}{*}{7.4$\pm$1.0}&\multirow{2}{*}{0.82$\pm$+0.37}&\multirow{2}{*}{l}&\multirow{2}{*}{}&\multirow{2}{*}{p}\\
\M&3.6&4.2&1.5&104$\pm$30&6.6$\pm$1.8&1.6$\pm$1.1&5.7$\pm$1.5&3.8$\pm$2.5&&&&&&\\
&&&&&&&&&&&&&&\\
\U\multirow{2}{*}{36}&3.5&1.7&3.2&101$\pm$29&3.3$\pm$1.5&3.5$\pm$1.2&2.1$\pm$1.2&8.4$\pm$2.9&\multirow{2}{*}{0.17$\pm$0.31}&\multirow{2}{*}{11.8$\pm$3.3}&\multirow{2}{*}{0.48$^{+0.42}_{-0.48}$}&\multirow{2}{*}{l}&\multirow{2}{*}{sl}&\multirow{2}{*}{p}\\
\M&4.2&2.4&3.5&168$\pm$41&4.7$\pm$1.7&4.4$\pm$1.3&3.1$\pm$1.4&10.6$\pm$3.2&&&&&&\\
\U\multirow{2}{*}{37}&2.2&2.2&1.0&40$\pm$18&3.4$\pm$1.6&1.2$^{+1.1}_{-1.0}$&2.7$\pm$1.3&3.2$^{+2.7}_{-2.4}$&\multirow{2}{*}{-0.56$^{+0.51}_{-0.44}$}&\multirow{2}{*}{6.5$\pm$2.0}&\multirow{2}{*}{0.32$^{+0.50}_{-0.32}$}&\multirow{2}{*}{}&\multirow{2}{*}{}&\multirow{2}{*}{p}\\
\M&3.4&3.2&1.4&84$\pm$25&6.7$\pm$1.9&1.8$^{+1.2}_{-1.1}$&5.0$\pm$1.6&4.8$^{+2.9}_{-2.7}$&&&&&&\\
\U\multirow{2}{*}{38}&2.8&3.7&0.5&23$^{+10}_{-9}$&11.4$^{+4.3}_{-3.8}$&$<$4.4&10.7$^{+3.7}_{-3.2}$&$<$11.5&\multirow{2}{*}{$>$-1}&\multirow{2}{*}{7.1$\pm$1.6}&\multirow{2}{*}{0.61$^{+0.60}_{-0.61}$}&\multirow{2}{*}{}&\multirow{2}{*}{}&\multirow{2}{*}{pc}\\
\M\B&3.2&3.3&1.6&37$\pm$12&11.4$^{+4.3}_{-3.9}$&3.6$^{+2.6}_{-2.2}$&10.0$^{+3.6}_{-3.2}$&8.9$^{+6.2}_{-5.1}$&&&&&&\\
\hline\hline
\multicolumn{15}{p{8.9in}}{\T Notes: (2)-(9) The signal-to-noise ratios, net counts, photon flux, and energy flux of the source in the specified energy bands. Values in the top (bottom) row for each entry are based on using source aperture regions with small (large) radius.  All other table column values are based on using small aperture regions. 

(10) The hardness ratio is defined as (H-S)/(H+S), where H represents the net counts in the 10--20 keV band and S represents the net counts in the 3--10 keV band. 

(11)-(12) Median energy in the 3--40 keV band, and the y-value of the quantile plot, defined as 3(E$_{25}$-3 keV)/(E$_{75}$-3 keV). 

(13) Flags indicating source variability: `s' - short timescale ($<$ few hours) variability, `l' - long timescale (weeks-years) variability, `p' - periodic modulations detected.  See~\ref{sec:var} for details.

(14) Variability flags from Fornasini et al. (2014): `s' - short timescale ($<$ few hours) variability; within a single observation, KS test probability that the source is constant is $<0.3$\%, `l' - long timescale (days-weeks) variability; the 0.5--10, 0.5--2, or 2--10 keV photon flux varies by $>3\sigma$ between NARCS observations, `p' - periodic modulations detected by $Z_n^2$ test, `$-$' - source not detected in NARCS. 

(15) `p' - point source region aperture; circle with 30$''$/40$''$ radius, `e' - extended source aperture; circle with 45$''$/60$''$ radius, `c' - background region is a circle with 70$''$ radius offset from the source rather than an annulus centered on the source, `m' - stray light and background spatial variations require background regions to be modified for each observation.

($^*$) Periodic variability for NNR 2 detected by Bodaghee et al. (2016), and for NNR 3 by Gotthelf et al. (2014). 
}
\label{tab:phot}
\end{longtable}}
\newpage


\begin{table}
\begin{minipage}{0.8\linewidth}
\centering
\footnotesize
\begin{threeparttable}
\caption{X-ray Variability}
\begin{tabular}{cccccccc} \hline \hline
\T Source & \textit{NuSTAR} & Maximum 3--10 keV Flux & Variability Amplitude & Criteria for Long-term \\
No. & Var. Flag & ($10^{-6}$ ph cm$^{-2}$ s$^{-1}$) &  3--10 keV & Var. Detection \\
\B (1) & (2) & (3) & (4) & (5) \\
\hline
\multicolumn{2}{l}{\textbf{\textit{Tier 1}}} &&&&&&\T\\
\T 1&l&641200$\pm$700&$>427500$ & $T$, $N$\\
\M 2&sp&10100$\pm$700&$>34$&$-$\\
\M 4&l&71$^{+2}_{-5}$&1.5$^{+0.1}_{-0.2}$&$CS$\\
\M 10&l&84$\pm$9&$>56$&$T$, $N$\\
\M 11&l&26$\pm$4&$>18$&$N$, $CS$\\
\M 15&s&220$\pm$40&$>6$&$-$\\
\M 19&l&11$\pm$1&$>7$&$T$, $N$\\
\M 20&l&10$\pm$2&$>2$&$CS$\\
\M 25&l&6$\pm$1&$>4$&$T$\\
\M 27&l&11$\pm$2&2.2$^{+1.6}_{-0.9}$&$CS$\\
\M\B 28&l&15$^{+7}_{-6}$&11$^{+8}_{-6}$&$CQ$\\
\hline
\multicolumn{2}{l}{\textbf{\textit{Tier 2}}} &&&&&&\T\\
\M 29&l&40$\pm$7&6.5$\pm$1.3&$N$\\
\M 35&l&13$\pm$2&2.5$\pm$0.6&$CQ$\\
\M\B 36&l&9$^{+2}_{-1}$&2.8$\pm$1.4&$CQ$\\
\hline \hline
\end{tabular}
\begin{tablenotes}[flushleft]
\item (2) \textit{NuSTAR} variability flag: `s' - short timescale ($<$ few hours) variability, long timescale (weeks-years) variability, `p' - periodic modulations detected.  See~\ref{sec:var} for details.  
\item (3) Maximum 3--10 keV photon flux either from \textit{Chandra} photometry or \textit{NuSTAR} photometry (based on 30$^{\prime\prime}$-radius aperture regions).
\item (4) Ratio of maximum to minimum 3--10 keV photon fluxes.  
\item (5) Criteria by which long-term variability was determined for sources flagged with ``l'': $T$ - transient source detected by \textit{NuSTAR} but falling below the survey sensitivity of NARCS, $N$ - photon flux varies by $>3\sigma$ between different \textit{NuSTAR} observations, $CS$ - cross-normalization between \textit{Chandra} and \textit{NuSTAR} spectra is inconsistent at $>90$\% confidence, $CQ$ - \textit{Chandra} 2--10 keV and \textit{NuSTAR} 3--10 keV photon fluxes are inconsistent at $>90$\% confidence when adopting a range of spectral models consistent with the quantile values of the source
\end{tablenotes}
\label{tab:var}
\end{threeparttable}
\end{minipage}
\end{table}

\begin{table}
\begin{minipage}{0.8\linewidth}
\centering
\footnotesize
\begin{threeparttable}
\caption{Properties of \textit{Chandra} counterparts to \textit{NuSTAR} discoveries}
\begin{tabular}{cccccccc} \hline \hline
\T Source & R.A. & Dec. & Position & Significance & Net counts & $E_{50}$ & $QR$ \\
No. & \multicolumn{2}{c}{J2000 ($^{\circ}$)} & uncertainty ($^{\prime\prime}$) & 0.5--10 keV & 0.5--10 keV & (keV) & \\
\B(1) & (2) & (3) & (4) & (5) & (6) & (7) & (8) \\
\hline
\T 19&250.315033&-46.540543&0.68&15&245$^{+17}_{-16}$&2.9$\pm$0.2&0.92$\pm$0.06\\
\M20&250.591644&-46.716049&0.87&2.9&3$^{+3}_{-2}$&$-^*$&$-$\\
\B\M 25&248.999542&-47.807671&0.71&6&33$^{+7}_{-6}$&2.3$\pm$0.4&0.9$\pm$0.3\\
\hline \hline
\end{tabular}
\begin{tablenotes}[flushleft]
\item (4) 90\% statistical and systematic positional uncertainties summed in quadrature.
\item ($^*$) The \textit{Chandra} counterpart of NNR~20 has too few counts to perform quantile analysis.  The energies of the three photons attributed to this source are 4.2, 5.7, and~7.0 keV; since the \textit{Chandra} effective area is higher at softer energies, the fact that no photons are detected with energies $<4$~keV suggests that this source is subject to high levels of absorption.
\end{tablenotes}
\label{tab:chandrasrc}
\end{threeparttable}
\end{minipage}
\end{table}
\end{landscape}

\newpage
\begin{landscape}
\setlength\LTleft{-0.5in}
\centering
\footnotesize
\begin{longtable}{cp{0.6in}ccccccccccp{1.7in}}
\kill
\caption{Spectral Fits} \\
\hline \hline
\T Src & \hspace{0.1in}Model & \textit{N/C} & FPMA/B & $N_{\mathrm{H}}$ & $\Gamma$ & E$_\mathrm{{cut}}$& Power-law & $kT_{BB}$ & Bbody & $\chi^2_{\nu}$/dof & Bin & \hspace{0.4in}Comments \\
 No. & \hspace{0.05in}\texttt{tbabs*X} & norm & norm & (10$^{22}$cm$^{-2}$) & & (keV) & norm & (keV) & norm & & ($\sigma$) & \\
\B(1) & \hspace{0.18in}(2) & (3) & (4) & (5) & (6) & (7) & (8) & (9) & (10) & (11) & (12) & \hspace{0.6in}(13) \\
\hline
\endfirsthead
\caption{Spectral Fits (continued)} \\
\hline
\T Src & \hspace{0.1in}Model & \textit{N/C} & FPMA/B & $N_{\mathrm{H}}$ & $\Gamma$ or & Norm & Line En. & Line Eq. & Line norm & $\chi^2_{\nu}$/dof & Bin& \hspace{0.4in}Comments\\ 
\B No. & \hspace{0.05in}\texttt{tbabs*X} & norm & norm & (10$^{22}$cm$^{-2}$) & $kT$ (keV) & (10$^{-5}$) & (keV) & (keV) & (10$^{-6}$) & & ($\sigma$) & \\ 
\hline
\endhead
\\
\endfoot
\endlastfoot
\U\D 1&\textbf{\texttt{PL +diskbb}}&$-$&0.978$^{+0.01}_{-0.02}$&12.47$\pm$0.08&2.15$\pm$0.03&$-$&0.22$\pm$0.02&1.425$^{+0.002}_{-0.003}$&192$\pm$2&2.68/806&10&See King et al. (2014) for fit including disk reflection and wind absorption.\\ 
\cline{13-13}

\U\D 2&\textbf{\texttt{cutoffpl +bbodyrad}}&0.67$^{+0.02}_{-0.01}$&1.02$^{+0.03}_{-0.02}$&46.0$\pm$1.5&-2.5$^{+0.4}_{-0.5}$&4.05$^{+0.33}_{-0.06}$&1.3$^{+1.5}_{-0.1}\times10^{-5}$&1.56$^{+0.06}_{-0.08}$&0.75$^{+0.12}_{-0.08}$&1.14/1096&5,5&See Bodaghee et al. (2016) for fit including cyclotron absorption line.\\
\hline \hline
\T Src & \hspace{0.1in}Model & \textit{N/C} & FPMA/B & $N_{\mathrm{H}}$ & $\Gamma$ or & Norm & Line En. & Line Eq. & Line norm & $\chi^2_{\nu}$/dof & Bin& \hspace{0.4in}Comments\\ 
No. & \hspace{0.05in}\texttt{tbabs*X} & norm & norm & (10$^{22}$cm$^{-2}$) & $kT$ (keV) & (10$^{-5}$) & (keV) & (keV) & (10$^{-6}$) & & ($\sigma$) & \\ 
\B&&&&&(14)&(15)&(16)&(17)&(18)&&&\\
\hline
\T\B\multirow{4}{*}{3}&\multirow{4}{*}{\textbf{\texttt{PL}}}&\multirow{4}{*}{3.4$^{+1.0}_{-0.7}$}&\multirow{4}{*}{1.04$^{+0.04}_{-0.03}$}&\multirow{4}{*}{12$\pm$2}&\multirow{4}{*}{1.71$\pm$0.06}&\multirow{4}{*}{6.7$^{+1.0}_{-0.6}$}&&&&\multirow{4}{*}{1.02/263}&\multirow{4}{*}{3,5}&\textit{Chandra} only includes point source while \textit{NuSTAR} includes extended emission.  See Gotthelf et al. (2014) for detailed analysis.\\ 
\cline{13-13}

\T\multirow{3}{*}{4}&\textbf{\texttt{PCA*(PL+G)}}&0.59$^{+0.10}_{-0.09}$&0.92$^{+0.15}_{-0.13}$&0.35$^{+0.11}_{-0.10}$& 2.34$\pm$0.22&56$^{+26}_{-17}$&6.65$^{+0.10}_{-0.06}$&0.9$^{+0.2}_{-0.1}$&6.2$^{+2.4}_{-1.9}$& 1.19/154&3,3&\multirow{4}{1.7in}{\texttt{pcfabs} reduces $\chi^2_{\nu}$ by $\approx$0.2.  For \texttt{PL}, $N_{\mathrm{H,cvr}}=$ $6^{+2}_{-1}\times10^{22}$ cm$^{-2}$, cvrf= $0.77^{+0.06}_{-0.08}$.  For \texttt{BR} and \texttt{AP}, $N_{\mathrm{H,cvr}}=$ $5\pm2\times10^{22}$ cm$^{-2}$, cvrf= $0.5\pm0.1$.}\\
\M&\texttt{PCA*(BR+G)}&0.57$^{+0.09}_{-0.08}$&0.90$^{+0.15}_{-0.12}$&0.14$\pm$0.08&7.9$^{+2.4}_{-1.7}$ &24$^{+4}_{-3}$ &6.65$^{+0.09}_{-0.06}$&0.8$\pm$0.2&5.6$^{+2.1}_{-1.8}$&1.20/154&3,3&\\
\M&\texttt{PCA*(AP+G)}&0.56$^{+0.04}_{-0.08}$&0.90$^{+0.16}_{-0.12}$&0.13$^{+0.09}_{-0.08}$&7.4$^{+2.1}_{-1.5}$ &68$^{+11}_{-9}$&6.56$^{+0.12}_{-0.17}$&0.2$\pm$0.1&2.1$^{+2.2}_{-1.6}$&1.19/154&3,3&\\
&&&&&&&&&&&&\B\\
\cline{13-13}

\U \multirow{3}{*}{5} &\textbf{\texttt{PL}}&1.3$^{+0.5}_{-0.3}$&0.9$^{+0.3}_{-0.2}$&27$^{+10}_{-8}$&2.3$\pm$0.3&28$^{+40}_{-16}$&6.3-7.1 &$<0.36$&$<1.3$&1.07/47&3,3&\\
\M&\texttt{BR}&1.3$^{+0.5}_{-0.3}$&0.9$^{+0.3}_{-0.2}$&21$^{+8}_{-6}$&10$^{+5}_{-3}$&9$^{+5}_{-3}$&&& &1.07/47&3,3&\\
\M\D&\texttt{AP}&1.2$^{+0.5}_{-0.3}$&0.9$^{+0.3}_{-0.2}$&17$^{+6}_{-5}$&13$^{+5}_{-3}$&21$^{+8}_{-7}$&&& &1.15/47&3,3&\\
\cline{13-13}

\U \multirow{3}{*}{6}&\texttt{PL+G}&1.0$\pm$0.2&0.9$^{+0.3}_{-0.2}$&5$\pm$1&1.5$\pm$0.3&13$^{+7}_{-4}$&6.5$^{+0.3}_{-1.7}$ &1.5$\pm$0.5&11$^{+62}_{-5}$&1.79/27&5,3&\\
\M&\texttt{BR+G}&1.0$^{+0.3}_{-0.2}$&0.98$\pm$0.25&4.3$^{+0.9}_{-1.5}$&$>15$&18$^{+3}_{-2}$&6.5$^{+0.4}_{-0.3}$ &1.3$\pm$0.4&10$^{+7}_{-5}$&1.72/27&5,3&\\
\M\D&\textbf{\texttt{AP+G}}&1.0$\pm$0.2&1.0$^{+0.3}_{-0.2}$&4.3$^{+0.9}_{-0.7}$&$>15$&51$^{+9}_{-6}$&6.4$\pm$0.4& 1.2$\pm$0.5&9$^{+6}_{-5}$&1.69/27&5,3&\\
\cline{13-13}

\U \multirow{3}{*}{7}&\texttt{PL+G}&1.0$\pm$0.2&0.8$^{+0.2}_{-0.1}$&15$^{+3}_{-2}$&3.4$^{+0.4}_{-0.3}$&220$^{+80}_{-90}$ &6.76$\pm$0.12&0.65$\pm$0.20&2.1$^{+1.1}_{-0.9}$&0.92/75&2.5,2.5&\multirow{3}{1.7in}{\texttt{apec} abundance = 0.5$\pm$0.3. NARCS~1278 flux is 30\% of total \citep{rahoui14}.}\\
\M&\texttt{BR+G}&1.0$\pm$0.2&0.9$^{+0.2}_{-0.1}$&11$^{+2}_{-1}$&3.4$^{+0.7}_{-0.6}$&32$^{+6}_{-7}$&6.76$\pm$0.12 &0.5$\pm$0.2&1.8$^{+1.1}_{-0.9}$&0.93/75&2.5,2.5&\\
\M\D&\textbf{\texttt{AP}}&1.0$\pm$0.2&0.9$^{+0.2}_{-0.1}$&11$\pm$2&3.2$^{+0.8}_{-0.5}$&100$^{+30}_{-25}$&&& &0.89/77&2.5,2.5&\\
 \cline{13-13}

\U\multirow{3}{*}{8}&\textbf{\texttt{PL}}&1.0$\pm$0.2&$-$&14$^{+7}_{-5}$&1.8$\pm$0.2&18$^{+15}_{-8}$&6.3-7.1&$<0.26$&$<$1.9 &1.01/27&3,5&\multirow{3}{1.7in}{Only FPMA used.}\\ 
\M&\texttt{BR}&1.0$\pm$0.2&$-$&12$^{+5}_{-4}$&25$^{+22}_{-9}$&15$^{+5}_{-4}$&&&&1.03/27&3,5&\\
\M\D&\texttt{AP}&1.0$\pm$0.2&$-$&10$^{+4}_{-3}$&$>21$&44$^{+11}_{-10}$&&&&1.14/27&3,5&\\
\cline{13-13}

\U\multirow{3}{*}{9}&\texttt{PL+G}&0.9$^{+0.3}_{-0.2}$&0.9$^{+0.3}_{-0.2}$&7$^{+3}_{-2}$&1.5$\pm$0.3&9$^{+7}_{-4}$ &6.5$\pm$1.2&0.6$\pm$0.4&3$^{+25}_{-2}$&0.84/29&3,2.5&\\
\M&\texttt{BR+G}&0.9$^{+0.3}_{-0.2}$&0.9$^{+0.3}_{-0.2}$&7$^{+2}_{-1}$&$>15$&11$^{+6}_{-2}$ &6.4$^{+0.7}_{-0.3}$&0.5$\pm$0.3&3$^{+8}_{-2}$&0.83/29&3,2.5&\\
\M\D&\textbf{\texttt{AP+G}}&0.8$^{+0.3}_{-0.2}$&0.9$\pm$0.2&7$^{+2}_{-1}$&$>15$&33$^{+8}_{-6}$&6.4$^{+0.5}_{-0.4}$ &0.4$\pm$0.2&2.2$^{+2.7}_{-1.7}$&0.82/29&3,2.5&\\
\cline{13-13}

\U\multirow{3}{*}{10}&\textbf{\texttt{PL}}&$-$&0.9$^{+0.6}_{-0.3}$&28&4.1$^{+0.9}_{-0.8}$&1800$^{+6400}_{-1400}$ &6.3-7.1&$<0.61$&$<4.2$&1.09/10&2&\multirow{3}{1.7in}{$N_{\mathrm{H}}$ set to values from Tomsick et al. (2014)}\\
\M&\texttt{BR}&$-$&0.9$^{+0.6}_{-0.3}$&17&3$^{+2}_{-1}$&70$^{+100}_{-40}$&&&&1.14/10&2&\\
\M\D&\texttt{AP}&$-$&0.9$^{+0.6}_{-0.4}$&17&1.9$^{+5.0}_{-0.7}$&330$^{+800}_{-290}$&&&&1.77/10&2&\\
\cline{13-13}

\U\multirow{3}{*}{11}&\texttt{PL}&5$^{+7}_{-2}$&1.3$^{+0.9}_{-0.8}$&11$^{+11}_{-9}$&2.3$\pm$0.4&2.4$^{+6.1}_{-2.0}$&&&&1.32/42&3,3&\\
\M&\textbf{\texttt{BR}}&6$^{+21}_{-3}$&1.3$^{+1.1}_{-1.0}$&$<14$&10$^{+6}_{-3}$&0.7$^{+0.9}_{-0.6}$&6.3-7.1&$<0.35$&$<0.1$&1.27/42&3,3&\\
\M\D&\texttt{AP}&17$^{+15}_{-12}$&1.4$^{+3.4}_{-0.6}$&$<5$&14$^{+5}_{-3}$&0.6$^{+1.6}_{-0.3}$&&&&1.33/42&3,3&\\
\cline{13-13}


\U\multirow{3}{*}{12}&\textbf{\texttt{PL+G}}&1.1$^{+0.5}_{-0.3}$&1.0$^{+0.4}_{-0.3}$&20$^{+9}_{-6}$&2.4$\pm$0.5&14$^{+26}_{-9}$ &6.78$^{+0.14}_{-0.12}$&1.2$\pm$0.4&1.7$^{+1.2}_{-0.8}$&1.14/33&2.5,2.5&\\
\M&\texttt{BR+G}&1.0$^{+0.5}_{-0.3}$&1.0$^{+0.4}_{-0.3}$&16$^{+7}_{-5}$&9$^{+8}_{-3}$&5$\pm$2 &6.77$^{+0.13}_{-0.12}$&1.2$^{+0.5}_{-0.3}$&1.6$^{+1.2}_{-0.8}$&1.17/33&2.5,2.5&\\
\M\D&\texttt{AP}&1.1$^{+0.5}_{-0.3}$&1.0$^{+0.4}_{-0.3}$&19$^{+7}_{-5}$&6$^{+3}_{-1}$&17$^{+8}_{-6}$&&&&1.21/36&2.5,2.5&\\
\cline{13-13}

\U\multirow{3}{*}{13}&\textbf{\texttt{PL}}&1.5$^{+1.1}_{-0.6}$&1.0$^{+0.7}_{-0.4}$&9$^{+17}_{-6}$&1.0$\pm$0.5&0.7$^{+1.8}_{-0.5}$ &6.3$-$7.1&$<2.7$&$<3.4$&1.22/23&2,2&\\
\M&\texttt{BR}&1.6$^{+1.1}_{-0.6}$&1.0$^{+0.7}_{-0.4}$&11$^{+11}_{-6}$&$>31$&2.9$^{+2.2}_{-0.4}$&&&&1.25/23&2,2&\\
\M\D&\texttt{AP}&1.6$\pm$0.7&1.0$^{+0.7}_{-0.4}$&13$^{+24}_{-7}$&$>21$&7$^{+1}_{-3}$&&&&1.28/23&2,2&\\
\cline{13-13}

\U\multirow{3}{*}{14}&\textbf{\texttt{PL+G}}&0.7$^{+0.2}_{-0.1}$&$-$&29$^{+9}_{-7}$&4.1$^{+1.2}_{-0.9}$&760$^{+5700}_{-610}$ &6.59$^{+0.08}_{-0.06}$&1.8$\pm$0.5&6$^{+3}_{-2}$&1.07/28&3,2.5&\multirow{3}{1.7in}{Only FPMB used.}\\ 
\M&\texttt{BR+G}&0.7$^{+0.2}_{-0.1}$&$-$&22$^{+7}_{-5}$&2.4$^{+1.4}_{-0.9}$&60$^{+130}_{-30}$& 6.59$^{+0.10}_{-0.06}$&1.7$^{+0.6}_{-0.4}$&5$^{+3}_{-2}$&1.08/28&3,2.5&\\
\M\D&\texttt{AP}&0.8$\pm$0.2&$-$&25$^{+7}_{-5}$&2.1$^{+0.9}_{-0.5}$&190$^{+240}_{-90}$&&&&1.11/31&3,2.5&\\
\cline{13-13}

\U\multirow{3}{*}{15}&\textbf{\texttt{PL}}&1.9$^{+1.4}_{-0.8}$&0.7$^{+0.4}_{-0.3}$&$<0.4$&2.6$\pm$0.4&13$^{+6}_{-4}$&6.3-7.1& $<1.7$&$<2.1$&0.75/20&3,2&\\
\M&\texttt{BR}&1.8$^{+1.1}_{-0.8}$&0.6$^{+0.3}_{-0.2}$&$<0.08$&2.9$^{+1.0}_{-0.7}$&11$\pm$2&&&&0.90/20&3,2&\\
\M\D&\texttt{AP}&1.8$^{+1.1}_{-0.8}$&0.6$^{+0.3}_{-0.2}$&$<0.10$&2.9$^{+0.8}_{-0.7}$&28$^{+9}_{-7}$&&&&0.82/19&3,2&\\
\cline{13-13}\\

\U\multirow{3}{*}{16}&\texttt{PL}&1.6$^{+0.7}_{-0.5}$&0.7$^{+0.3}_{-0.2}$&19$^{+6}_{-5}$&2.9$^{+0.6}_{-0.5}$&130$^{+240}_{-80}$&&&&1.05/24&3,2.5&\multirow{3}{1.7in}{Harder spectrum than found by Bodaghee et al. (2014) due to different background regions.}\\
\M&\textbf{\texttt{BR}}&1.6$^{+0.6}_{-0.5}$&0.7$^{+0.3}_{-0.2}$&14$^{+4}_{-3}$&5$^{+3}_{-1}$&25$^{+14}_{-8}$&6.3-7.1 &$<0.8$&$<4.8$&0.95/24&3,2.5&\\
\M\D&\texttt{AP}&1.2$^{+0.5}_{-0.4}$&0.7$\pm$0.2&13$\pm$3&6$^{+4}_{-2}$&58$^{+27}_{-15}$&&&&1.06/24&3,2.5&\\
\cline{13-13}

\U\multirow{3}{*}{17}&\textbf{\texttt{PL}}&1.1$^{+1.2}_{-0.5}$&0.8$^{+0.6}_{-0.4}$&21$^{+32}_{-16}$&2.0$^{+1.0}_{-0.8}$& 6$^{+65}_{-5}$&6.3-7.1&$<1.0$&$<1.3$&0.94/13&2,2&\\
\M&\texttt{BR}&1.1$^{+1.3}_{-0.5}$&0.8$^{+0.6}_{-0.4}$&16$^{+24}_{-12}$&$>6$&3$^{+6}_{-2}$&&&&0.95/13&2,2&\\
\M\D&\texttt{AP}&1.1$^{+1.3}_{-0.5}$&0.8$^{+0.6}_{-0.4}$&14$^{+17}_{-10}$&$>8$&9$^{+9}_{-5}$&&&&0.94/13&2,2&\\
\cline{13-13}

\U\multirow{3}{*}{18}&\texttt{PL}&1.1$^{+0.6}_{-0.5}$&0.8$^{+0.7}_{-0.4}$&19$^{+9}_{-6}$&2.6$^{+1.0}_{-0.8}$&50$^{+260}_{-40}$&&&&1.87/13&3,2&\\
\M&\textbf{\texttt{BR}}&1.1$^{+0.6}_{-0.4}$&0.8$^{+0.7}_{-0.4}$&16$^{+7}_{-4}$&6$^{+11}_{-3}$&16$^{+19}_{-6}$& 6.3-7.1&$<0.9$&$<3$&1.81/13&3,2&\\
\M\D&\texttt{AP}&1.0$^{+0.5}_{-0.4}$&0.8$^{+0.8}_{-0.4}$&13$^{+7}_{-3}$&9$^{+24}_{-6}$&33$^{+51}_{-8}$&&&&1.97/13&3,2&\\
\cline{13-13}

\U\multirow{3}{*}{19}&\texttt{PL}&1.0$\pm$0.3&1.2$^{+0.6}_{-0.4}$&1.7$^{+0.8}_{-0.6}$&1.7$^{+0.3}_{-0.4}$&4$\pm$2&&&& 1.66/25&3,3&\multirow{3}{1.7in}{\textit{N}/\textit{C}=0.8$^{+0.3}_{-0.2}$ for \textit{Chandra} Obs 7591.}\\ 
\M&\texttt{BR}&1.0$^{+0.4}_{-0.2}$&1.1$^{+0.6}_{-0.3}$&1.4$^{+0.5}_{-0.4}$&13$^{+18}_{-5}$&4.0$^{+0.8}_{-0.6}$& &&&1.44/25&3,3&\\
\M\D&\textbf{\texttt{AP}}&1.0$\pm$0.3&1.1$^{+0.6}_{-0.3}$&1.4$^{+0.5}_{-0.4}$&11$^{+18}_{-4}$&12$\pm$2&6.3-7.1&$<1.3$&$<2.4$& 1.44/25&3,3&\\
\cline{13-13}

\U\multirow{3}{*}{20}&\textbf{\texttt{PL}}&1&1&70$^{+130}_{-50}$&2.6$^{+2.1}_{-1.4}$&52$^{+15000}_{-50}$&6.3-7.1&$<0.6$&$<6.4$&1.26/11&2,2&\multirow{3}{1.7in}{If the cross-normalization constant between \textit{Chandra} and \textit{NuSTAR} is left as a free parameter, $N/C>2$ at 90\% confidence.}\\
\M&\texttt{BR}&1&1&60$^{+90}_{-40}$&$>3$&8$^{+80}_{-5}$&&&&1.31/11&2,2&\\
\M\D&\texttt{AP}&1&1&50$^{+50}_{-30}$&$>6$&18$^{+27}_{-8}$&&&&1.30/11&2,2&\\
\cline{13-13}

\U\multirow{3}{*}{21}&\textbf{\texttt{PL}}&0.9$\pm$0.2&$-$&26$^{+9}_{-7}$&2.6$^{+0.5}_{-0.4}$&120$^{+200}_{-70}$&6.3-7.1&$<0.5$&$<4.6$ &1.01/47&3,3&\multirow{3}{1.7in}{Only FPMB used. Point source (NARCS~402) flux is 20$\pm$5\% of total.}\\ 
\M&\texttt{BR}&0.9$\pm$0.2&$-$&20$^{+7}_{-5}$&8$^{+5}_{-2}$&31$^{+15}_{-9}$&&&&1.04/47&3,3&\\
\M\D&\texttt{AP}&0.8$\pm$0.2&$-$&17$^{+5}_{-4}$&10$^{+7}_{-3}$&71$^{+22}_{-16}$&&&&1.13/47&3,3&\\
\cline{13-13}

\U\multirow{3}{*}{22}&\texttt{PL}&1.7$^{+1.5}_{-0.8}$&1&13$^{+12}_{-7}$&2.0$^{+1.3}_{-1.2}$&5.1$^{+41}_{-4.4}$&&&&1.92/10&2,2&\\
\M&\texttt{BR}&1.6$^{+1.4}_{-0.8}$&1&11$^{+10}_{-5}$&$>4$&3$^{+6}_{-1}$&&&&1.91/10&2,2&\\
\M\D&\textbf{\texttt{AP}}&1.8$^{+1.3}_{-0.7}$&1&13$^{+13}_{-5}$&5$^{+23}_{-3}$&12$^{+24}_{-6}$&6.3-7.1&$<3.8$&$<4.4$&1.67/10&2,2&\\
\cline{13-13}


\U\multirow{3}{*}{23}&\texttt{PL}&1.7$^{+2.1}_{-0.9}$&1&7$^{+65}_{-5}$&1.8$^{+2.0}_{-0.8}$&1.6$^{+340}_{-1.3}$&&&&0.83/6&2,2&\\
\M&\texttt{BR}&1.7$^{+2.0}_{-0.9}$&1&6$^{+49}_{-4}$&$>4$&1.5$^{+14.5}_{-0.9}$&&&&0.75/6&2,2&\\
\M\D&\textbf{\texttt{AP}}&1.7$^{+2.0}_{-0.9}$&1&7$^{+29}_{-5}$&$>5$&4.6$^{+13}_{-2.6}$&6.3-7.1&$<3.9$&$<2.9$&0.64/6&2,2&\\
\cline{13-13}

\U\multirow{3}{*}{24}&\textbf{\texttt{PL}}&0.7$\pm$0.3&1&28$^{+12}_{-8}$&5.0$^{+2.2}_{-1.4}$&1180$^{+41220}_{-1070}$&6.3-7.1&$<16$&$<6.2$&1.25/23&2,2&\\
\M&\texttt{BR}&0.7$\pm$0.3&1&20$^{+8}_{-6}$&1.7$^{+1.3}_{-0.7}$&50$^{+290}_{-30}$&&&&1.33/23&2,2&\\
\M\D&\texttt{AP}&0.7$\pm$0.3&1&24$^{+7}_{-6}$&1.4$^{+0.8}_{-0.4}$&160$^{+500}_{-110}$&&&&1.32/23&2,2&\\
\cline{13-13}

\U\multirow{3}{*}{25}&\texttt{PL}&1.3$^{+1.3}_{-0.6}$&$-$&3.1$^{+3.8}_{-2.8}$&1.8$\pm$0.7&1.9$^{+4.2}_{-1.5}$& &&&1.14/8&2,2&\multirow{3}{1.7in}{Only FPMA used.}\\
\M&\texttt{BR}&1.3$^{+1.7}_{-0.6}$&$-$&2.3$^{+3.0}_{-2.2}$&$>6$&1.7$^{+1.1}_{-1.0}$&&&&1.05/8&2,2&\\
\M\D&\textbf{\texttt{AP}}&1.3$^{+1.7}_{-0.6}$&$-$&2.3$^{+2.9}_{-2.1}$&$>6$&5$\pm$3&6.3-7.1&$<2.1$&$<1.8$&1.02/8&2,2&\\
\cline{13-13}

\U\multirow{3}{*}{26}&\textbf{\texttt{PL}}&1.3$^{+2.8}_{-0.7}$&1&30$^{+35}_{-23}$&1.5$^{+1.0}_{-0.9}$&2.8$^{+30}_{-2.5}$ &6.3-7.1&$<1.2$&$<1.9$&1.57/10&2,2&\\
\M&\texttt{BR}&1.3$^{+2.6}_{-0.6}$&1&28$^{+31}_{-21}$&$>9$&3.5$^{+5.5}_{-2.5}$&&&&1.58/10&2,2&\\
\M\D&\texttt{AP}&1.3$^{+2.5}_{-0.6}$&1&28$^{+28}_{-19}$&$>13$&11$^{+12}_{-8}$&&&&1.59/10&2,2&\\
\cline{13-13}

\U\multirow{3}{*}{27}&\textbf{\texttt{PL}}&4.4$^{+5.9}_{-3.5}$&1&$<23$&0.9$^{+0.8}_{-0.4}$&0.18$^{+2.68}_{-0.09}$ &6.3-7.1&$<1.1$&$<0.5$&0.85/8&2,2&\\
\M&\texttt{BR}&6.3$^{+6.2}_{-5.2}$&1&$<27$&$>23$&0.7$^{+6.0}_{-0.3}$&&&&0.95/8&2,2&\\
\M\D&\texttt{AP}&7.4$^{+10.2}_{-2.7}$&1&$<34$&$>20$&1.4$^{+13.8}_{-0.7}$&&&&1.08/8&2,2&\\

\hline \hline
\multicolumn{13}{p{8.9in}}{\T Notes: Errors provided are 90\% confidence intervals, except for errors on the line equivalent widths which are 1$\sigma$ confidence intervals.

(2) For sources NNR 1-3, which have been analyzed in more detail in other papers, we present the results of simplified models, used to derive the fluxes and conversion factors in Table~\ref{tab:convfactors}.  For all other sources, we present fits using power-law (\texttt{PL}), bremsstrahlung (\texttt{BR}), and collisionally-ionized apec models (\texttt{AP}).  Some models include a Gaussian line (\texttt{G}) or partial covering absorption (\texttt{PCA}). The best-fitting model for each source is written in bold text.

(3) Multiplicative \texttt{constant} included in all spectral models.  The constant is set to 1.0 for \textit{Chandra} and, if enough spectral bins are available, it is allowed to vary independently for \textit{NuSTAR} FPMA and FPMB.  \textit{N/C} provides the ratio of the \textit{NuSTAR} FPMA constant relative to \textit{Chandra}. 
 
(4) Ratio of the FPMA to FPMB fitting constants, providing the cross-calibration of the two \textit{NuSTAR} modules.  For sources with insufficient photon statistics, this ratio is set to 1.0.

(11) Reduced $\chi^2$ statistic and degrees of freedom for the best-fitting model.

(12) Minimum significance of bins for \textit{Chandra}, \textit{NuSTAR} spectra.

(16-18)  The central energy, equivalent width, and normalization of a Gaussian model accounting for iron line emission.  In cases where an iron line is clearly visible in the spectrum, a Gaussian line (\texttt{G}) is included in the model; otherwise, we provide the results of the best-fit models without a Gaussian line and an upper limit to the Fe line equivalent width derived by adding a Gaussian component as described in \S~\ref{sec:spectral}.
}
\label{tab:spectra}
\end{longtable}
\newpage
\end{landscape}
\newpage

\begin{table}
\begin{minipage}{\linewidth}
\begin{adjustwidth}{-0.9in}{-0.9in}
\centering
\footnotesize
\begin{threeparttable}
\caption{Spectrally derived fluxes}
\begin{tabular}{cccccccccc} \hline \hline
\T &\multicolumn{3}{c}{Ph. Flux (10$^{-6}$cm$^{-2}$s$^{-1}$)}&\multicolumn{3}{c}{\underline{Abs. Flux (10$^{-14}$erg cm$^{-2}$s$^{-1}$)}}&\multicolumn{3}{c}{\underline{Ph. flux to unabs. flux (10$^{-9}$erg/ph)}}\\ \cline{2-4} 
\T Src&\textit{Chandra}&\textit{NuSTAR}&\textit{NuSTAR}&\textit{Chandra}&\textit{NuSTAR}&\textit{NuSTAR}&\textit{Chandra}&\textit{NuSTAR}&\textit{NuSTAR}\\
No.&2-10 keV&3-10 keV&10-20 keV&2-10 keV&3-10 keV&10-20 keV&2-10 keV&3-10 keV&10-20 keV\\
\B(1)&(2)&(3)&(4)&(5)&(6)&(7)&(8)&(9)&(10)\\
\hline
\T 1&$-$&614000$\pm$300&19950$^{+30}_{-50}$&$-$&490300$^{+200}_{-300}$&39590$^{+60}_{-100}$&$-$&12.2&20.9\\
\M 2&2200$^{+200}_{-500}$&1500$^{+100}_{-300}$&1700$^{+500}_{-700}$&2500$^{+300}_{-600}$&1700$^{+200}_{-300}$&4000$^{+1000}_{-2000}$&27.0&24.7&26.3\\
\M 3&19$^{+0.2}_{-0.3}$&56.0$^{+0.3}_{-1.9}$&22.8$^{+0.1}_{-0.8}$&16$\pm$2&52.1$^{+0.2}_{-1.8}$&51.3$^{+0.3}_{-1.9}$&14.4&12.6&23.4\\
\M 4&101$^{+3}_{-7}$&43$^{+1}_{-4}$&7.0$^{+0.4}_{-0.7}$&71$^{+2}_{-6}$&36$^{+1}_{-3}$&15.3$^{+0.8}_{-1.6}$&9.3&9.7&22.3\\
\M 5&16$^{+2}_{-4}$&21.8$^{+0.4}_{-3.6}$&7.1$^{+0.2}_{-1.4}$&15$^{+1}_{-4}$&20.6$^{+0.3}_{-3.4}$&15.7$^{+0.3}_{-3.0}$&26.5&19.0&24.1\\
\M 6&82$^{+3}_{-7}$&68$^{+4}_{-9}$&21$^{+2}_{-5}$&68$^{+3}_{-7}$&63$^{+3}_{-8}$&47$^{+4}_{-11}$&9.9&10.2&22.7\\
\M 7&42$^{+1}_{-3}$&36$^{+1}_{-3}$&1.7$^{+0.2}_{-0.4}$&32$^{+1}_{-3}$&29.1$^{+0.9}_{-2.5}$&3.3$^{+0.4}_{-0.8}$&14.3&11.7&20.7\\
\M 8&43$^{+3}_{-7}$&40.3$^{+0.6}_{-6.4}$&16.6$^{+0.3}_{-2.6}$&39$^{+3}_{-6}$&37.8$^{+0.5}_{-5.8}$&37.5$^{+0.8}_{-6.0}$&15.7&13.4&23.5\\
\M 9&43$\pm$3&33$^{+2}_{-3}$&12.5$^{+0.9}_{-1.1}$&37$\pm$3&31$\pm$2&28$\pm$2&11.2&10.9&22.9\\
\M 10&$-$&55$^{+3}_{-24}$&3.6$^{+0.3}_{-2.1}$&$-$&44$^{+2}_{-19}$&7.5$^{+0.7}_{-4.4}$&$-$&23.2&23.0\\
\M 11&2.6$^{+0.6}_{-1.1}$&10.3$^{+0.2}_{-4.1}$&2.15$^{+0.03}_{-0.98}$&2.0$^{+0.6}_{-0.9}$&9.0$^{+0.2}_{-3.5}$&4.65$^{+0.06}_{-2.13}$&9.8&10.0&22.0\\
\M 12&10.4$^{+0.8}_{-3.9}$&10.4$^{+0.2}_{-3.1}$&2.35$^{+0.03}_{-0.8}$&9$^{+2}_{-3}$&9.7$^{+0.1}_{-2.8}$&5.18$^{+0.06}_{-1.70}$&20.7&15.6&23.5\\
\M 13&6.4$^{+0.9}_{-2.8}$&8.4$^{+0.2}_{-3.4}$&5.8$^{+0.1}_{-2.4}$&6$^{+1}_{-3}$&8.27$^{+0.09}_{-3.38}$&13.5$^{+0.2}_{-5.5}$&12.5&11.9&23.7\\
\M 14&28.3$^{+0.9}_{-4.1}$&20$^{+1}_{-4}$&1.1$^{+0.2}_{-0.4}$&23.5$^{+0.9}_{-3.7}$&16.9$^{+0.9}_{-3.4}$&2.4$^{+0.4}_{-0.8}$&48.5&22.5&23.1\\
\M 15&25$\pm$4&30$^{+3}_{-5}$&3.6$^{+0.5}_{-0.9}$&14$^{+3}_{-2}$&23$^{+3}_{-4}$&8$^{+1}_{-2}$&5.9&7.8&21.7\\
\M 16&41$^{+2}_{-7}$&69$^{+4}_{-13}$&9.1$^{+0.9}_{-2.2}$&34$^{+2}_{-6}$&60$^{+3}_{-11}$&19$^{+2}_{-5}$&16.0&13.1&21.8\\
\M 17&8$^{+2}_{-3}$&9.5$^{+0.5}_{-2.8}$&3.6$^{+0.2}_{-1.1}$&7$^{+2}_{-3}$&9.0$^{+0.4}_{-2.6}$&8.2$^{+0.5}_{-2.6}$&20.3&15.9&23.8\\
\M 18&26$^{+1}_{-11}$&31$^{+3}_{-12}$&5.1$^{+0.6}_{-2.5}$&22$^{+1}_{-10}$&28$^{+2}_{-11}$&11$^{+1}_{-5}$&17.0&13.9&22.2\\
\M 19&15$^{+3}_{-4}$&10.1$^{+0.4}_{-1.8}$&1.9$^{+0.3}_{-0.9}$&11$^{+2}_{-3}$&8.8$^{+0.4}_{-1.7}$&4.2$^{+0.7}_{-2.0}$&7.8&9.0&21.8\\
\M 20&7.3$^{+0.2}_{-5.5}$&7.3$^{+0.1}_{-5.2}$&4.4$^{+0.2}_{-2.5}$&7.8$^{+0.2}_{-5.8}$&7.79$^{+0.09}_{-5.52}$&9.7$^{+0.6}_{-5.6}$&80.5&52.4&28.3\\
\M 21&49$^{+1}_{-15}$&43.4$^{+0.9}_{-12.2}$&11.5$^{+0.4}_{-4.2}$&43.9$^{+0.8}_{-14.1}$&40.0$^{+0.6}_{-11.3}$&25$^{+1}_{-9}$&27.9&19.1&23.9\\
\M 22&8.4$^{+0.5}_{-4.9}$&13.4$^{+0.1}_{-9.2}$&1.7$^{+0.3}_{1.6}$&7.1$^{+0.4}_{-4.6}$&12.0$^{+0.1}_{-8.3}$&3.6$^{+0.8}_{-3.3}$&14.9&12.7&21.8\\
\M 23&6$^{+1}_{-4}$&7.9$^{+0.1}_{-4.9}$&1.9$^{+0.2}_{-1.8}$&5$^{+1}_{-3}$&7.1$^{+0.1}_{-4.7}$&4.3$^{+0.6}_{-4.0}$&11.1&10.7&22.3\\
\M 24&9.8$^{+0.2}_{-3.0}$&6.3$^{+0.7}_{-2.4}$&0.16$^{+0.03}_{-0.09}$&6.9$^{+0.2}_{-2.3}$&4.7$^{+0.5}_{-1.8}$&0.32$^{+0.06}_{-0.18}$&78.9&25.8&22.5\\
\M 25&8$^{+1}_{-5}$&7.6$^{+0.1}_{-3.7}$&1.8$^{+0.2}_{-1.6}$&6$^{+1}_{-4}$&6.7$^{+0.2}_{-3.3}$&4.0$^{+0.6}_{-3.5}$&8.4&9.4&22.1\\
\M 26&6.1$^{+0.7}_{-3.0}$&7.8$^{+0.2}_{-4.2}$&5.4$^{+0.1}_{-3.1}$&6.3$^{+0.8}_{-3.1}$&8.0$^{+0.1}_{-4.2}$&12.4$^{+0.2}_{-7.1}$&23.9&19.7&25.0\\
\M 27&3.0$^{+0.2}_{-2.0}$&10.1$^{+0.6}_{-5.7}$&6.2$^{+0.3}_{-4.2}$&2.4$^{+0.2}_{-1.6}$&9.5$^{+0.4}_{-5.2}$&14.3$^{+0.7}_{-9.6}$&8.1&9.4&23.2\\
\B\M 28$^*$&2.1$\pm$0.7&24.9$^{+0.1}_{-4.6}$&8.79$^{+0.06}_{-2.07}$&1.8$\pm$0.6&22.6$^{+0.1}_{-4.1}$&19.8$^{+0.1}_{-4.7}$&13.1&11.8&23.2\\
\hline \hline
\end{tabular}
\begin{tablenotes}[flushleft]
\item \T Notes: These fluxes and conversion factors are determined from spectral fitting.  The \textit{NuSTAR} fluxes represent the average of the FPMA and FPMB fluxes.  Errors provided are 1$\sigma$ confidence intervals.  

\item ($^*$) Due to the poor photon statistics of NNR 28, it was not possible to perform spectral fitting in the same way as for the other sources.  Adopting an absorbed power-law model for this source with $\Gamma=1.8$ and $N_{\mathrm{H}} = 10^{23}$~cm$^{-2}$, we determined the \textit{Chandra} and \textit{NuSTAR} fluxes using the C-statistic (see \S~\ref{sec:spectral} for more details).  
\end{tablenotes}
\label{tab:convfactors}
\end{threeparttable}
\end{adjustwidth}
\end{minipage}
\end{table}

\begin{table}
\begin{minipage}{\linewidth}
\begin{adjustwidth}{-0.9in}{-0.9in}
\centering
\footnotesize
\begin{threeparttable}
\caption{Classification of \textit{NuSTAR} Norma region sources}
\begin{tabular}{cl} \hline \hline
\T \B Source No. & Classification \\
\hline
\T &\textbf{Tier 1 - Confirmed}\\
\T 1 & BH LMXB \\
2 & NS HMXB \\
3 & pulsar/PWN \\
\hline
\T &\textbf{Tier 1 - Candidate} \\
4 & CV (IP) \\
5 & bow shock PWN \\
6 & CV (IP) \\
7 & CWB \\
8 & young PWN \\ 
9 & CV (IP) \\
10 & magnetar or AB \\
11 & CV (polar or non-magnetic) or AGN \\
12 & CV (IP) \\
13 & CV (IP) \\
14 & CWB \\
15 & BH LMXB \\
16 & CV or AGN \\
17 & CV or AGN \\
18 & CV (IP), SB, SyXB \\
19 & CV (polar) \\
20 & CV (polar or non-magnetic) or AGN \\
21 & PWN/SNR \\
22 & CV or AGN \\
23 & CV or AGN \\
25 & CV (polar) \\
26 & CV or AGN \\
27 & CV or AGN \\
28 & CV (polar or non-magnetic) or AGN \\
\hline
\T & \textbf{Tier 2 - Tentative} \\
29 & Galactic \\
30 & Galactic \\
31 & Galactic \\
32 & Galactic or AGN \\
33 & Galactic or AGN \\
34 & Galactic or AGN \\
35 & Galactic or AGN \\
36 & Galactic \\
37 & Galactic or AGN \\
\B 38 & Galactic \\
\hline \hline
\end{tabular}
\begin{tablenotes}[flushleft]
\item \T Notes: Classifications of NNR sources are discussed in \S~\ref{sec:classification}.  The classifications of NNR 1, 2, and 3 are robust, while all other classifications for tier 1 sources should be considered candidate identifications (see \S~\ref{sec:classification} for details).  For candidate CVs, we provide in parenthesis the most likely CV type when possible.  For tier 2 sources we provide only tentative classifications of these sources as Galactic or extragalactic AGN.
\end{tablenotes}
\label{tab:class}
\end{threeparttable}
\end{adjustwidth}
\end{minipage}
\end{table}

\end{document}